\documentclass[12pt]{article}
\pdfoutput=1
\usepackage{putex}
\usepackage{graphicx}
\usepackage{enumitem}
\usepackage{cite}
\usepackage{slashed}
\usepackage[margin=15pt,small]{caption}
\usepackage{booktabs}

\usepackage[
      colorlinks=false,
      linkcolor=blue,  
      urlcolor=blue,    
      filecolor=blue,     
      citecolor=red,
	  linktocpage=true,
      pdfstartview=FitV,
      bookmarksopen=true    
      ]{hyperref}

\numberwithin{equation}{section} 

\setlist[itemize]{itemsep=2 pt}
\setlist[enumerate]{noitemsep}

\let\oldbibliography\thebibliography
\renewcommand{\thebibliography}[1]{\oldbibliography{#1}
\setlength{\itemsep}{0pt}} 

\newcommand{\grp}[1]{\mathrm{#1}}

\newcommand{\grO}{\grp{O}}
\newcommand{\grU}{\grp{U}}
\newcommand{\grSU}{\grp{SU}}
\newcommand{\grSO}{\grp{SO}}

\newcommand{\grSL}{\grp{SL}}

\def\bb{{\bar\beta}}
\def\m{\mu}
\def\n{\nu}

\def\bz{{\bar z}}
\def\pa{\partial}

\def\cP{\cals P}
\def\a{\alpha}
\def\b{\beta}

\def\d{\delta}
\def\g{\gamma}
\def\ca{\cals A}
\def\eps{\epsilon}
\def\k{\kappa}
\def\D{\Delta}

\def\overleftarrow#1{\vbox{\ialign{##\crcr
     $\leftarrow$\crcr\noalign{\kern-0pt\nointerlineskip}
     $\hfil\displaystyle{#1}\hfil$\crcr}}}

\def\leftDs{\overleftarrow{\mathop{\slashed{\nabla}}}}

\def\Mmat{X}
\def\coeff#1#2{\relax{\textstyle {#1 \over #2}}\displaystyle}
\def\Neql#1{{\cal N}\!=\!{#1}}

\def\eql{~=~}
\def\eeql{~\equiv~}
\def\cals#1{\mathcal{#1}}
\def\bphi{\phi}
\def\cA{\cals A}
\def\cB{\cals B}
\def\cO{\cals O}
\def\co{\cals O}
\def\cl{\cals L}

\def\<{\langle}
\def\>{\rangle}

\def\lab{\label}
\def\fA{\mathfrak{A}}
\def\tB{\text{B}}

\newcommand{\abs}[1]{\left\lvert #1 \right\rvert}

\newcommand {\be} {\begin {equation}}
\newcommand {\ee} {\end {equation}}
\newcommand {\bea}{\begin{eqnarray}}
\newcommand {\eea}{\end{eqnarray}}

\newcommand {\bes} {\begin {equation*}}
\newcommand {\ees} {\end {equation*}}
\newcommand{\reef}[1]{\eqref{#1}}

\newcommand{\es}[2] {\begin{equation} \label{#1} \begin{split} #2 \end{split} \end{equation}}

\def\sA{A}

\def\cn{\cals N}
\def\vx{\vec x}
\def\vy{\vec y}
\def\vz{\vec z}
\def\vw{\vec w}

\begin{document}
\begin{titlepage}

\begin{flushright}
\texttt{MIT-CTP/4851}\\
\texttt{PUPT-2510}\\
\end{flushright}

\begin{center}
{\Large \bf Boundary Terms and Three-Point Functions:\\} 
{\Large \bf An AdS/CFT Puzzle Resolved\\} 

\vspace*{0.5 cm}

{\bf Daniel Z.~Freedman,$^{1,2}$ Krzysztof Pilch,$^{3}$\\[2 mm] Silviu S.~Pufu,$^{5}$ and Nicholas P.~Warner$^{3,4}$}
\bigskip

$^{1}$ SITP and Department of Physics\\
Stanford University\\ Stanford, CA 94305, USA
\bigskip

$^2$ Center for Theoretical Physics and Department of Mathematics\\
 Massachusetts Institute of Technology\\
Cambridge, MA 02139, USA
\bigskip

$^{3}$ Department of Physics and Astronomy\\
$^{4}$ Department of Mathematics\\
University of Southern California\\
Los Angeles, CA 90089, USA
\bigskip

$^{5}$ Joseph Henry Laboratories\\ Princeton University\\
Princeton, NJ 08544, USA
\bigskip

\texttt{dzfmit@gmail.com, pilch@usc.edu,  spufu@princeton.edu, warner@usc.edu} \\
\end{center}

\vspace*{0.25cm}

\begin{abstract} 
{}
\noindent{}${\cal N} = 8$ superconformal field theories, such as the ABJM theory at Chern-Simons level $k=1$ or $2$, contain 35 scalar operators  $\co_{IJ}$ with $\Delta=1$ in the ${\bf 35}_v$ representation of SO(8). The \hbox{3-point} correlation function of these operators is non-vanishing, and indeed can be calculated non-perturbatively in the field theory.   But its   AdS$_4$  gravity dual, obtained from  gauged $\cn=8$ supergravity, has no cubic $\sA^3$ couplings in its Lagrangian, where $\sA^{IJ}$ is the bulk dual of $\co_{IJ}$.  So conventional Witten diagrams  cannot furnish the field theory result.  We show  that the extension of bulk supersymmetry to the  AdS$_4$  boundary requires the introduction of a finite $\sA^3$ counterterm  that does provide a perfect match to the 3-point correlator.  Boundary supersymmetry also requires  infinite  counterterms which agree with the method of holographic renormalization.  The generating functional of correlation functions of the $\D=1$ operators is the  Legendre transform of the on-shell action, and the supersymmetry properties of this functional play a significant role in our treatment. 

\end{abstract}

\end{titlepage}

\tableofcontents

\section{Introduction}

The anti-de Sitter / conformal field theory (AdS/CFT) duality has passed many tests.  When precise comparisons of gravity and  field theory results can be made, the results generally agree.  This paper focuses on an aspect of the AdS$_4$/CFT$_3$ duality in which there is an apparent acute conflict between the two sides of the duality, but we find that the conflict is resolved through oft-neglected boundary terms.    
   
The conflict involves the holographic computation of the 3-point function of dimension-$1$ operators of the CFT$_3$\@.  For concreteness, let us describe it in the case of the maximally supersymmetric (${\cal N} = 8$) 3d superconformal field theories (SCFTs) whose holographic description includes four-dimensional, $\cn=8$ gauged supergravity \cite{deWit:1982bul}.  The representation theory of the ${\cal N} = 8$ superconformal algebra shows that any 3d, local ${\cal N} = 8$ SCFT must contain scalar operators $\co_{IJ}, ~1\leq I,J\leq 8$ transforming in the traceless symmetric tensor description of the ${\bf 35}_{v}$ representation of the $\grSO(8)$ global R-symmetry group with scale dimension $\Delta=1$.  These scalars are present in any local ${\cal N} = 8$ SCFT because they belong to the same superconformal multiplet as the stress tensor.  As we will explain,  superconformal Ward identities imply that the 3-point correlation function $\<\co_{IJ}(\vec x_1)\co_{JK}(\vec x_2)\co_{KI}(\vec x_3)\>$ for given $I$, $J$ and $K$ (no sum) must be non-vanishing and related to the 2-point function of the canonically normalized stress tensor. This 2-point function can be calculated exactly using supersymmetric localization \cite{Pestun:2007rz} whenever an explicit Lagrangian description is available.  The AdS/CFT correspondence requires  the  3-point functions $\<\co_{IJ}(\vec x_1)\co_{JK}(\vec x_2)\co_{KI}(\vec x_3)\>$ be matched by a calculation in the gravity bulk, where 3-point functions are usually calculated  by evaluating a Witten diagram containing a cubic   vertex from the bulk Lagrangian.  The problem is that the Lagrangian of   $\cn=8$ gauged supergravity in four dimensions does not contain any cubic scalar couplings!  Thus another way to obtain $\<\co_{IJ}(\vec x_1)\co_{JK}(\vec x_2)\co_{KI}(\vec x_3)\>$ must be found. 

Note that the foregoing description of the  conflict does not rely on a specific field theory realization of the ${\cal N} = 8$ SCFT dual to four-dimensional, ${\cal N} = 8$ gauged supergravity.   In fact, the four-dimensional, ${\cal N} = 8$ gauged supergravity theory does not correspond to a unique ${\cal N} = 8$ SCFT;  it corresponds instead to a universal sector describing the stress tensor multiplet of all known ${\cal N} = 8$ SCFTs with holographic duals. These are the large $N$ limits of three distinct families:   $\grU(N)_1 \times \grU(N)_{-1}$ ABJM theory \cite{Aharony:2008ug}, $\grU(N)_2 \times \grU(N)_{-2}$ ABJM theory, and $\grU(N)_2 \times \grU(N+1)_{-2}$ ABJ theory \cite{Aharony:2008gk}. (See also~\cite{Bagger:2006sk, Gustavsson:2007vu, Bagger:2007jr, Bagger:2007vi} for earlier work that was generalized in \cite{Aharony:2008ug,Aharony:2008gk}.)\footnote{The ABJ(M) theory  \cite{Aharony:2008ug, Aharony:2008gk} is a $\grU(N)_k \times \grU(M)_{-k}$ Chern-Simons matter theory in three dimensions that has only ${\cal N} = 6$ manifest  supersymmetry. It is the effective theory on $N$ coincident M2-branes placed at the singular point of a certain $\mathbb{C}^4 / \mathbb{Z}_k$ orbifold. When $k=1$ or $2$ and $M = N$ or $M = N+1$, the infrared limit  is believed to have enhanced ${\cal N} = 8$ supersymmetry \cite{Aharony:2008ug,Bashkirov:2010kz, Benna:2009xd, Gustavsson:2009pm, Benna:2008zy,Kwon:2009ar}.  The $\grU(N)_1 \times \grU(N+1)_{-1}$ theory is dual to the $\grU(N)_1 \times \grU(N)_{-1}$ one, so there are only three distinct families of ${\cal N} = 8$ SCFTs of this type.  A fourth family of ${\cal N} = 8$ SCFTs is given by the $SU(2)_k \times SU(2)_{-k}$ BLG theories \cite{Bagger:2006sk, Bagger:2007jr, Bagger:2007vi, Gustavsson:2007vu} but do not have classical supergravity duals.  The BLG theories have manifest ${\cal N} = 8$ supersymmetry.} 
These ${\cal N} = 8$ SCFTs are believed to be, respectively, the infrared limits of maximally supersymmetric 3d Yang-Mills theory with gauge group $\grU(N)$, $\grO(2N)$, and $\grO(2N+1)$.  At large $N$ they have a dual description in terms of eleven-dimensional supergravity, of which four-dimensional ${\cal N} = 8$ gauged supergravity of \cite{deWit:1982bul} is a consistent truncation.

An important clue to the resolution of the conflict appears in \cite{Freedman:2013ryh}, where an $\cn=2$ truncation of the $\cn =8$ supergravity theory was studied.\footnote{To our knowledge, this truncation was first given in \cite{Cvetic:1999xp}.} The truncation contains 3~complex scalars $z^\alpha= A^\alpha+iB^\alpha$,~$\alpha=1,2,3$. The goal of \cite{Freedman:2013ryh} was to match the field theory calculation of the $S^3$ free energy of an $\cn=2$-preserving mass deformation of the ABJM theory obtained in \cite{Jafferis:2011zi} by the method of supersymmetric localization developed in\cite{Kapustin:2009kz, Jafferis:2010un, Hama:2010av} (for recent reviews, see \cite{Willett:2016adv,Pufu:2016zxm}).  Obtaining the match is not straightforward.  First, the bulk scalars $A^\alpha$ dual to the three $\Delta=1$ field theory operators $\co_{\alpha}$ in the truncation\footnote{The 3 operators $\co_{\alpha}$ constitute the subset of the  35 $\co_{IJ}$ that is part of the truncated theory. This subset is defined in the next section.}  must be quantized by alternate quantization \cite{Breitenlohner:1982jf}.  Second, the infinite counterterms  obtained from the method of holographic renormalization  must be supplemented by a finite counterterm.  Both alternate quantization and the finite counterterm 
\cite{Bianchi:2001de, Bianchi:2001kw} are required by the supersymmetry of the Legendre transformed on-shell action which is the generator of correlation functions in the boundary field theory \cite{Klebanov:1999tb}.  We focus on the counterterm obtained in \cite{Freedman:2013ryh} by a Bogomolny factorization argument for the action of planar domain walls \cite{Skenderis:1999mm}.  The counterterm turns
out to be proportional to  $\int d^3x \sqrt{-h}\,A^1A^2A^3$ with a determined coefficient.\footnote{The induced metric at the boundary is  $h_{ab}$.} It turns out that this boundary term and its extension to the full $\cn=8$ theory are exactly what we need to compute $\<\co_1(\vec x_1)\co_2(\vec x_2)\co_3(\vec x_3)\>$ and $\<\co_{IJ}(\vec x_1)\co_{JK}(\vec x_2)\co_{KI}(\vec x_3)\>$.

The main effort in this paper is to obtain the essential cubic counterterm\footnote{It is well known that a boundary term quadratic in fermion fields must be added to the bulk action in order to obtain the 2-point correlator of fermion operators in the boundary theory,  \cite{Henningson:1998cd,Mueck:1998iz,Arutyunov:1998ve,Henneaux:1998ch}.  Also, a cubic boundary counterterm plays a role in the holographic story of extremal correlation functions in $\cn=4$ SYM theory
\cite{D'Hoker:1999ea}. See, also \cite{Bastianelli:1999en,Bianchi:2003bd,Bianchi:2003ug}.}
by modifying the bulk theory so that supersymmetry extends to the boundary.  The principle we employ is that the on-shell supergravity action, seen via the AdS/CFT dictionary as a functional of the sources for the field theory operators, should be supersymmetric.   We analyze this first at the level of a  limit of four-dimensional $\cn=1$ supergravity in which the back-reaction of matter fields on the AdS$_4$ background is consistently suppressed and  the resulting theory, similar to that of \cite{Festuccia:2011ws}, enjoys global AdS supersymmetry.  We also discuss the changes needed to extend the treatment to $\cn=1$ supergravity.   Then we move on to $\cn=8$ and show how the cubic counterterm emerges from an extended Bogomolny argument and finally how it is generated in the full  $\cn=8$ gauged supergravity.  In both the ${\cal N} = 1$ and ${\cal N} = 8$ analyses, the alternate quantization of \cite{Klebanov:1999tb}, implemented through a Legendre transform of the on-shell action, plays an important role.  It is worth noting that for  $\cn=1$ global supersymmetric theories with boundaries, the  boundary terms we find here (and their derivation) are in some ways very similar to those encountered  in lower dimensions \cite{Witten:1978mh,Fendley:1990zj, Warner:1995ay}.

We should emphasize that the framework developed here goes beyond the immediate application to the correspondence between the ${\cal N} = 8$ gauged supergravity and its maximally supersymmetric 3d SCFT dual.  Indeed, any holographic computation of a 3-point correlator of dimension-$1$ operators in a 3d CFT with a gravity dual must be similar to the present study in that the bulk cubic vertex must vanish\footnote{Suppose that the on-shell action did contain an $A^3$ or $A\pa_\m A\pa^\m A$ vertex.  It is curious to note that the results for the Witten diagrams given in \cite{Freedman:1998tz} are both infinite when $d=3$, and $\D_1=\D_2=\D_3 = 1$.}
 and the answer comes from a (super)gravity boundary term.\footnote{An interesting example of dimension-$1$ operators in a non-supersymmetric instance of AdS$_4$/CFT$_3$ is present in the higher spin / $O(N)$ vector model duality conjectured in \cite{Klebanov:2002ja}.  For this model, the dimension-$1$ scalar operators have $s=0$ for the higher spin currents of spin $s$.  The match of 3-point functions of higher-spin currents between field theory and holography was performed in \cite{Giombi:2009wh} for all $s$.  For $s=0$, the authors of \cite{Giombi:2009wh} argued for a match of the 3-point function of dimension-$1$ scalar operators somewhat indirectly by considering the analytic continuation of the result for arbitrary $s$, and not by explicitly computing a boundary term as we do here.  Perhaps one can provide a more direct argument by explicitly computing the required boundary term by imposing the condition that the higher spin symmetry should extend to the boundary.} Our claim is that in a four-dimensional  ${\cal N} \geq 1$ supergravity theory this boundary term can be determined from the requirement that the  theory is supersymmetric, including boundary terms.

It is worth contrasting the situation here to that of four-dimensional \hbox{$\cn=4$} supersymmetric Yang-Mills theory, where the 3-point correlator of the chiral primary operator $\co_{\Delta=2}$ in the same multiplet as the stress tensor is \emph{protected\/} \cite{Lee:1998bxa,D'Hoker:1998tz}. This means that it is independent of the gauge coupling constant, and so it can be computed at weak coupling by performing Wick contractions.  This is \emph{not} true for the scalars $\co_{IJ}$ of  $\cn=8$ SCFTs, where there are strong coupling effects. It is worth displaying the result for the supergravity limits of the 2- and 3-point function of the operators $\co_\alpha$ in the truncation of \cite{Freedman:2013ryh}: 
 \es{3pttrunc}{
   \langle {\cal O}_\alpha(\vec{x}_1) {\cal O}_\beta(\vec{x}_2) \rangle &= \frac{ L^2}{2 \pi^3 G_4} \frac{\delta_{\alpha\beta}}{\abs{\vec{x}_{12} }^2} = \frac{\sqrt2 N^{3/2} k^{1/2}}{3\pi^3} \frac{\delta_{\alpha\beta}}{\abs{\vec{x}_{12} }^2} \,, \\
  \<\co_1(\vec x_1)\co_2(\vec x_2)\co_3(\vec x_3)\> &= \frac{L^2}{4\pi^4G_4}\frac{1}{|\vec x_{12}|  |\vec x_{23}| |\vec x_{31}|}
   = \frac{\sqrt2 N^{3/2} k^{1/2}}{6\pi^4}\frac{1}{|\vec x_{12}|  |\vec x_{23}| |\vec x_{31}|} \,.
 } 
In these expressions, $L$ is the radius of the dual AdS$_4$ solution, $G_4$ is the effective four-dimensional Newton constant, $\vec{x}_{ij} \equiv \vec{x}_i - \vec{x}_j$, $N$ was defined above, and $k = 1$ or $2$ is the Chern-Simons level of the ABJ(M) theory.   We will first explain how to derive \eqref{3pttrunc}  in the $\cn=8$ ABJM theory based on previous results that use supersymmetric localization and then  derive it from  $\cn =8$ supergravity.   Equality of the  coefficients follows from the AdS/CFT dictionary.   

Therefore, in addition to uncovering the essential role of supergravity boundary terms in the computation of CFT three-point functions, the results presented in this paper also provide another precision test of holography: the equality in \eqref{3pttrunc}.

The rest of this paper is organized as follows.  In Section~\ref{sec:FTcals} we review the field theory computations of correlation functions of dimension-$1$ operators.  In Section~\ref{BOGO} we start with a toy example in ${\cal N} = 1$ supergravity in 4 dimensions, in which we derive the boundary terms needed to ensure supersymmetry.  In Section~\ref{sec:3pttoy} we use these boundary terms to calculate holographically the 3-point function of dimension-1 operators, thus resolving the puzzle mentioned above in an ${\cal N} = 1$ toy example.  In Sections~\ref{N8sugra}--\ref{sec:3ptfnct} we generalize this computation to ${\cal N} = 8$ gauged supergravity:  We start with a brief review in Section~\ref{N8sugra}, we develop a Bogomolny argument that motivates the presence of a boundary term in Section~\ref{Sect:BogN8}, we use this boundary term to verify  supersymmetry in Section~\ref{N8sugrsusy}, and we perform the holographic computation of the 3-point functions of dimension-1 scalar operators in Section~\ref{sec:3ptfnct}.  We end with concluding remarks in Section~\ref{CONCLUSIONS}.


\section{Field theory computations}
\label{sec:FTcals}

In this Section we discuss 3-point functions of dimension-$1$ scalar operators from a field theory perspective.  We start in Section~\ref{GENERALSCFT} with a general discussion of dimension-$1$ scalar operators in 3d SCFTs.  In Section~\ref{N8THREE} we then specialize to ${\cal N} = 8$ SCFTs, which are the main focus of this paper.

\subsection{Dimension-1 scalar operators in 3d SCFTs}
\label{GENERALSCFT}

In 3d SCFTs with at least ${\cal N} = 2$ supersymmetry, scalar operators of dimension $1$ are very common.  Indeed, these operators appear in one of two ways:  either as part of a chiral or anti-chiral multiplet, where they carry R-charge $1$ or $-1$, respectively, or as part of the same multiplet as a conserved flavor symmetry current, where they have vanishing R-charge.  There are no other multiplets of the ${\cal N} = 2$ superconformal algebra that contain dimension-$1$ scalar operators.  Of course, not every ${\cal N} = 2$ SCFT must have a chiral or anti-chiral operator of dimension $1$, but if there are any flavor symmetries present, then dimension-$1$ operators must be present as part of the conserved flavor current multiplets.  

When we consider extended supersymmetry, dimension-$1$ scalar operators can, of course, only arise in multiplets that upon reduction to ${\cal N} = 2$ contain either a flavor current multiplet, a chiral multiplet of R-charge $1$, or an anti-chiral multiplet of R-charge $-1$.  This always happens, for instance, in SCFTs with ${\cal N} \geq 4$ supersymmetry.  Indeed, in such SCFTs some of the R-symmetry currents (which are in the same ${\cal N} \geq 4$ supermultiplet as the stress tensor) can be interpreted as flavor currents upon reduction to ${\cal N} = 2$, and these flavor currents belong to ${\cal N} = 2$ supermultiplets also containing dimension-$1$ scalar operators.  Therefore, local ${\cal N} \geq 4$ SCFTs must always contain scalar operators of dimension $1$ that belong to the same ${\cal N} \geq 4$ supermultiplet as the stress energy tensor.  

In ${\cal N} = 2$ SCFTs, supersymmetry techniques allow for the computation of certain \hbox{3-point} functions of 
dimension-$1$ scalar operators exactly.  Without extended supersymmetry, the \hbox{3-point} functions that are calculable with existing supersymmetric localization techniques are those of precisely one chiral operator, one anti-chiral operator, and one operator in a conserved flavor current multiplet.  Such a 3-point function is non-vanishing only if the chiral and anti-chiral operators carry non-vanishing charges under the flavor symmetry corresponding to the third operator, and in this discussion we will assume this.  The other type of non-zero three point function, namely between three operators in conserved current multiplets, does not seem to be accessible through supersymmetric localization in theories with just ${\cal N} =2$ supersymmetry, but it can of course also be computed in theories with extended supersymmetry in which supersymmetry relates it to a chiral-anti-chiral-conserved current 3-point function.

To be precise,  consider a dimension-$1$ chiral operator ${\cal O}$, an anti-chiral operator $\overline {\cal O}$, and a dimension-$1$ real operator $J$ in the same multiplet as a conserved flavor current $j^\mu$.  Let the operators ${\cal O}$ and $\overline {\cal O}$ have charges $q$ and $-q$, respectively, under the symmetry generated by $j^\mu$.  It is important to be precise about the normalization of these operators.  For the chiral and anti-chiral operators, let us normalize them such that 
 \es{NormChiral}{
  \langle {\cal O}(\vec x) \overline {\cal O}(0) \rangle = \frac{1}{8 \pi^2 \abs{\vec{x}}^2} \,.
 }
It is convenient to normalize $J$ such that it is related to the canonically normalized $j^\mu$ in a canonical way.  Canonical normalization of $j^\mu$ means that the following OPE holds
 \es{CanJmu}{
   j^\mu(\vec{x}) {\cal O}(0) = q \frac{x^\mu}{4 \pi \abs{\vec{x}}^3} {\cal O}(0) + \ldots \,.
 }
We take the canonical normalization of $J$ to mean that if the conserved current $j^\mu$ is normalized as in \eqref{CanJmu}, then $J$ should be normalized such that it gives the OPE
 \es{CanJnorm}{
   J(\vec{x}) {\cal O}(0) = q \frac{1}{4 \pi \abs{\vec{x}}} {\cal O}(0) + \ldots \,.
 }
With this normalization, we have the following 2-point functions at separated points
 \es{TwoPointJ}{
  \langle J(\vec{x}) J(0)  \rangle = \frac{\tau}{16 \pi^2 \abs{\vec{x} }^2} \,, \qquad
    \langle j^\mu(\vec{x}) j^\nu( 0)  \rangle = \frac{\tau}{16 \pi^2} \left(\partial_\lambda \partial^\lambda \eta^{\mu\nu} - \partial^\mu \partial^\nu  \right) \frac{1}{\abs{\vec{x}}^2}    \,.
 }

The coefficient $\tau$ can be computed using supersymmetric localization of a certain deformation of the SCFT on $S^3$.  The deformation can be interpreted as a modification of the supersymmetry algebra where we change the R-charges of all chiral operators by adding to them the flavor charges under $j^\mu$ multiplied by a parameter $t$.  It is possible to compute the $S^3$ free energy $F(t)$ for this deformation of the theory exactly. Then one extracts \cite{Closset:2012vg} (for recent reviews, see \cite{Pufu:2016zxm, Dumitrescu:2016ltq})
 \es{Gottau}{
  \tau = -\frac{2}{\pi^2} \frac{d^2 F}{dt^2} \biggr|_{t=0} \,.
 }

The 3-point function $\langle {\cal O}(\vec x_1) \overline {\cal O}(\vec{x}_2) J(\vec{x}_3) \rangle$ can be computed using these results very easily. Indeed, by conformal invariance, it takes the form 
 \es{ThreePoint}{
  \langle {\cal O}(\vec x_1) \overline {\cal O}(\vec x_2) J(\vec x_3) \rangle
   = \frac{\lambda_{{\cal O} \overline {\cal O} J}}{\abs{\vec{x}_1 - \vec{x}_2} \abs{\vec{x}_1 - \vec{x}_3} \abs{\vec{x}_2 - \vec{x}_3}} \,.
 }
Using the OPE \eqref{CanJmu} and the 2-point function \eqref{NormChiral}, we obtain
 \es{ThreePointAgain}{
  \lambda_{{\cal O} \overline {\cal O} J} = \frac{q}{32 \pi^3} \,.
 }

The simplicity of \eqref{ThreePointAgain} is misleading, because it relies on the canonical normalization of $J$ as well as on the normalization of the chiral and anti-chiral operators in \eqref{NormChiral}.  The following ratio of three and 2-point functions is a constant that is independent of the normalization of these operators:
 \es{Three}{
  \frac{\langle {\cal O}(\vec x_1) \overline {\cal O}(\vec x_2) J(\vec x_3)  \rangle^2}{\langle {\cal O}(\vec x_1) \overline {\cal O}(\vec x_3) \rangle\, \langle {\cal O}(\vec x_3) \overline {\cal O}(\vec x_2) \rangle \, \langle J(\vec x_1)J(\vec x_2) \rangle} = \frac{q^2}{{\tau}} \,.
 }
It depends on both the charge $q$ as well as the coefficient $\tau$ obtained through \eqref{Gottau}.

\subsection{Application to ${\cal N} = 8$ SCFTs}
\label{N8THREE}

This framework can be applied to the computation of the 3-point function of dimension-$1$ operators in maximally supersymmetric ${\cal N} = 8$ SCFTs, as we now explain.   As described above, any SCFT with at least ${\cal N} = 4$ supersymmetry must have dimension-$1$ scalars in the same multiplet as the stress energy tensor.  In an interacting ${\cal N} = 8$ theory, these are the only dimension-$1$ operators that can exist.  They transform in a 35-dimensional representation of the $\grSO(8)$ R-symmetry that, by a choice of convention,  we take to be the ${\bf 35}_v$.   In addition to the stress tensor and the dimension-$1$ scalar operators transforming in the ${\bf 35}_v$, the ${\cal N} = 8$ stress tensor multiplet also contains an R-symmetry current transforming in the adjoint of $\grSO(8)$, the supercurrent of spin $3/2$ transforming (by a choice of conventions) in the ${\bf 8}_s$ of $\grSO(8)$, dimension-$2$ pseudoscalars transforming in the ${\bf 35}_c$, as well as dimension-$3/2$ operators of spin $1/2$ transforming in the ${\bf 56}_s$.

The 2-point function of the canonically normalized stress tensor is determined by conservation and conformal invariance to be
 \es{TT}{
  \langle T_{\mu\nu}(\vec{x}) T_{\rho\sigma}(0) \rangle 
   = \frac{c_T}{64} (P_{\mu\rho} P_{\nu\sigma} + P_{\nu \rho} P_{\mu\sigma} - P_{\mu\nu} P_{\rho\sigma}) \frac{1}{16 \pi^2 \abs{\vec{x}}^2} \,,
 }
where $P_{\mu\nu} \equiv \eta_{\mu\nu} \partial^\lambda \partial_{\lambda} - \partial_\mu \partial_\nu$, and $c_T$ is a constant that depends on the theory.  This definition means that one has  $c_T = 1$ in a non-supersymmetric theory of a free massless real scalar.   A straightforward computation then shows that one has $c_T = 1$ in a non-supersymmetric theory of a free massless Majorana fermion.  The free ${\cal N}=8$ theory contains 8 real scalars and 8 Majorana fermions and it thus has $c_T = 16$. 

The 2-point function of the canonically normalized $\grSO(8)$ R-symmetry current is also determined up to an overall constant by conformal invariance and conservation.  Moreover the superconformal algebra relates this constant to $c_T$, and the 2-point function takes the form
 \es{JJ}{
  \langle j^\mu_{IJ}(\vec x) j^\nu_{KL}(0)  \rangle = \frac{c_T}{64} (\delta_{IK} \delta_{JL} - \delta_{IL} \delta_{JK}) P^{\mu\nu} \frac{1}{16 \pi^2 \abs{\vec{x}}^2} \,,
 }
where $j^\mu_{IJ}$ is antisymmetric in the $IJ$ indices.  The constant, $c_T$, has been computed in many examples by considering  Abelian flavor currents and using the method described around \eqref{Gottau}.  We will provide a few explicit examples shortly.

We now focus on the dimension-$1$ scalar operators in the ${\bf 35}_v$ of $\grSO(8)$, which we will represent by a symmetric traceless tensor ${\cal O}_{IJ}(\vec{x})$.  To simplify the following formulas, it is convenient to pass to an index free notation by contracting ${\cal O}_{IJ}$ with a traceless symmetric matrix $M^{IJ}$, thus defining
 \es{OCompact}{
  {\cal O}(\vec{x}, M) = M^{IJ} {\cal O}_{IJ}(\vec{x}) \,.
 }
The two and 3-point functions of ${\cal O}(\vec{x}, M)$ are restricted by conformal and $\grSO(8)$ invariance to take the form
 \es{Two}{
  \langle {\cal O}(\vec{x}_1, M_1) {\cal O}(\vec{x}_2, M_2) \rangle  &= c_2 \frac{\tr (M_1 M_2)}{{\abs{\vec{x}_1 - \vec{x}_2}^2} } \,, \\
  \langle {\cal O}(\vec{x}_1, M_1) {\cal O}(\vec{x}_2, M_2)  {\cal O}(\vec{x}_3, M_3) \rangle  &= c_3 \frac{\tr (M_1 M_2 M_3 + M_1 M_3 M_2)}{{\abs{\vec{x}_1 - \vec{x}_2} \abs{\vec{x}_1 - \vec{x}_3} \abs{\vec{x}_2 - \vec{x}_3}} } \,,
 } 
for some constants $c_2$ and $c_3$.  Of course, $c_2$ can be changed by changing the normalization of the operators, so it may not  be meaningful, and one might want to consider instead a combination of two and three point functions that is invariant under rescalings of the operators:
  \es{TwoThreeRatio}{
  \frac{\langle {\cal O}(\vec{x}_1, M_1) {\cal O}(\vec{x}_2, M_2)  {\cal O}(\vec{x}_3, M_3) \rangle^2}
   {{\langle {\cal O}(\vec{x}_1, M_1) {\cal O}(\vec{x}_3, M_2) \rangle \langle {\cal O}(\vec{x}_2, M_1) {\cal O}(\vec{x}_3, M_2) \rangle \langle {\cal O}(\vec{x}_1, M_3) {\cal O}(\vec{x}_2, M_3) \rangle   }}  \\[6 pt]
   = \frac{c_3^2}{c_2^3} \frac{\left[ \tr (M_1 M_2 M_3 + M_1 M_3 M_2) \right]^2}{{\tr (M_1 M_2) \tr(M_1 M_2) \tr (M_3 M_3)}} \,.
 } 

In order to connect \eqref{Two}--\eqref{TwoThreeRatio} with the discussion of the previous section, which considered ${\cal N} = 2$ SCFTs, we should understand how the ${\bf 35}_v$ operators ${\cal O}_{IJ}$ transform under an ${\cal N} =2$ superconformal subalgebra of the ${\cal N} = 8$ algebra.  One can choose an embedding of the ${\cal N} = 2$ superconformal algebra $\mathfrak{osp}(2|4)$ into $\mathfrak{osp}(8|4)$ such that the ${\cal N} = 2$ $\grSO(2)_R$ R-symmetry is generated by the anti-Hermitian $8\times 8$ matrix
 \es{MatSO2}{
  R = \frac i2 \begin{pmatrix}
   \sigma_2 & 0 & 0 & 0  \\
   0 & \sigma_2 & 0 & 0 \\
   0 & 0 & \sigma_2 & 0 \\
   0 & 0 & 0 & \sigma_2
  \end{pmatrix}
 }
acting in the ${\bf 8}_v$ representation of $\grSO(8)$.  In other words, the ${\cal N} = 2$ R-symmetry current is $j^\mu \equiv R^{IJ} j^\mu_{IJ}$. It is not hard to see that the ${\bf 35}_v$ operators have the following ${\cal N} = 2$ R-charges:  ten of them have R-charge $1$ and are thus chiral operators from the ${\cal N} = 2$ point of view;  ten of them have R-charge $-1$ and are thus anti-chiral operators from the ${\cal N} = 2$ point of view;  and fifteen of them have vanishing R-charge and therefore belong to flavor current multiplets from the ${\cal N} = 2$ point of view.  Indeed, from an ${\cal N} =2$ perspective, the flavor symmetry is $\grSU(4)$, because this is the subgroup of $\grSO(8)$ that commutes with \eqref{MatSO2}.  Since $\grSU(4)$ has rank three, there are three commuting Abelian flavor currents that can be taken to correspond to the $\grSO(8)$ generators:
 \es{FlavorGen}{
  F_{(1)} &= \frac i2 \begin{pmatrix}
   \sigma_2 & 0 & 0 & 0  \\
   0 & \sigma_2 & 0 & 0 \\
   0 & 0 & -\sigma_2 & 0 \\
   0 & 0 & 0 & -\sigma_2
  \end{pmatrix} \,, \qquad
  F_{(2)} =  \frac i2 \begin{pmatrix}
   \sigma_2 & 0 & 0 & 0  \\
   0 & -\sigma_2 & 0 & 0 \\
   0 & 0 & \sigma_2 & 0 \\
   0 & 0 & 0 & -\sigma_2
  \end{pmatrix} \,, \\[6 pt]
&\hspace{1.1 in} F_{(3)} =   \frac i2 \begin{pmatrix}
   \sigma_2 & 0 & 0 & 0  \\
   0 & -\sigma_2 & 0 & 0 \\
   0 & 0 & -\sigma_2 & 0 \\
   0 & 0 & 0 & \sigma_2
  \end{pmatrix} \,.
 }
These flavor currents are thus $j^\mu_{(\alpha)} \equiv F_{(\alpha)}^{IJ} j^\mu_{IJ}$.  They are normalized so that
 \es{jFlavorNorm}{
   \langle j^\mu_{(\alpha)}(\vec{x}) j^\nu_{(\beta)}(0) \rangle =  \frac{c_T}{16}  P^{\mu\nu} \frac{\delta_{\alpha\beta}}{16 \pi^2 \abs{\vec{x}}^2} \,.
 }
The dimension-$1$ scalars that are part of ${\bf 35}_v$ and that belong to the same ${\cal N} = 2$ multiplet as these flavor currents in \eqref{FlavorGen} are $J_{(\alpha)} = M_{(\alpha)}^{IJ} {\cal O}_{IJ}$, where
 \es{MiDef}{
  M_{(1)} &= \frac 14 \diag\{ 1, 1, 1, 1, -1, -1, -1, -1 \} \,, \\[6 pt]
  M_{(2)} &= \frac 14 \diag\{ 1, 1, -1, -1, 1, 1, -1, -1 \} \,, \\[6 pt]
  M_{(3)} &= \frac 14 \diag\{ 1, 1, -1, -1, -1, -1, 1, 1 \} \,,
 } 
respectively.  From \eqref{Two}, we have
 \es{JJCorrSO8}{
  \langle J_{(\alpha)}(\vec{x}) J_{(\beta)}(0)  \rangle = \frac{c_2}{ 2\abs{\vec{x}}^2} \delta_{\alpha \beta}\,.
 }
Comparing \eqref{jFlavorNorm} and \eqref{JJCorrSO8} to \eqref{TwoPointJ}, we see that the real scalars $J_{(\alpha)}$ are canonically normalized in the sense of \eqref{TwoPointJ} provided that 
 \es{Gotc2}{
   c_2 = \frac{c_T}{8 (4\pi)^2}  \,, \qquad
    \tau = \frac{c_T}{16} \,.
 }
 
In order to find $c_3$ by using \eqref{Three}, we should identify linear combinations of the ${\bf 35}_v$ operators that reduce to chiral and anti-chiral operators from an ${\cal N} = 2$ point of view.  It can be checked that
 \es{ChiralFrom35}{
  {\cal O} = {\cal O}_{11} - {\cal O}_{22} + 2i {\cal O}_{12} \,, \qquad
   \overline {\cal O} = {\cal O}_{11} - {\cal O}_{22} - 2i {\cal O}_{12} \,, 
 }
are such operators because they have R-charges $1$ and $-1$ under \eqref{MatSO2}.  From \eqref{FlavorGen}, we see that they have flavor charges $1$ and $-1$, respectively, under each of the currents $j^\mu_{(\alpha)}$.  From \eqref{TwoThreeRatio}, we have
  \es{TwoThreeRatioExplicit}{
  \frac{\langle {\cal O}(\vec x_1) \overline {\cal O}(\vec{x}_2) J_{(\alpha)} (\vec x_3) \rangle \rangle^2}{{\langle {\cal O}(\vec x_1) \overline {\cal O}(\vec x_3) \rangle\, \langle {\cal O}(\vec x_3) \overline {\cal O}(\vec x_2) \rangle \, \langle J_{(\alpha)}(\vec x_1)J_{(\alpha)}(\vec x_2) \rangle}} =  \frac{c_3^2}{2 c_2^3}  \,.
 } 
Identifying $q=1$ and using $\tau = c_T / 16$ as in \eqref{Gotc2}, we have from \eqref{Three} that
  \es{TwoThreeRatioExplicit2}{
  \frac{\langle {\cal O}(\vec x_1) \overline {\cal O}(\vec x_2) J_{(\alpha)} (\vec x_3) \rangle \rangle^2}{{\langle {\cal O}(\vec x) \overline {\cal O}(\vec x_2) \rangle\, \langle {\cal O}(\vec x_3) \overline {\cal O}(\vec x_2) \rangle \, \langle J_{(\alpha)}(\vec x_1)J_{(\alpha)}(\vec x_2) \rangle}} =  \frac{16}{{c_T}}  \,.
 } 
A comparison of \eqref{TwoThreeRatioExplicit} and \eqref{TwoThreeRatioExplicit2} gives 
 \es{Gotc3}{
  \frac{c_3^2}{c_2^3} = \frac{32}{{c_T}} \,.
 }
For canonically normalized ${\cal O}_{IJ}$ for which $c_2$ is given by \eqref{Gotc2}, we have
 \es{c3Can}{
  c_3 = \frac{c_T}4 \frac{1}{(4 \pi)^3 } \,.
 }

\subsubsection{Summary}

To summarize, the 2- and 3-point functions of the canonically normalized ${\bf 35}_v$ operators in an ${\cal N} = 8$ SCFT are
  \es{TwoSummary}{
  \langle {\cal O}(\vec{x}_1, M_1) {\cal O}(\vec{x}_2, M_2) \rangle  &= \frac{c_T}{8} \frac{1}{(4\pi)^2} \frac{\tr (M_1 M_2)}{{\abs{\vec{x} - \vec{x}_2}^2} } \,, \\[6 pt]
  \langle {\cal O}(\vec{x}_1, M_1) {\cal O}(\vec{x}_2, M_2)  {\cal O}(\vec{x}_3, M_3) \rangle  &= \frac{c_T}4 \frac{1}{(4 \pi)^3 }  \frac{\tr (M_1 M_2 M_3 + M_1 M_3 M_2)}{{\abs{\vec{x}_1 - \vec{x}_2} \abs{\vec{x}_1 - \vec{x}_3} \abs{\vec{x}_2 - \vec{x}_3}} } \,,
 } 
where $c_T$ is defined in \eqref{TT}.

In general, the quantity, $c_T$, depends on the parameters and dynamics of the (S)CFT in question\@.  For an (S)CFT with a holographic dual, $c_T$ is a simple universal function of $L$ and $G_4$---it must be universal because the correlator $\<T_{\mu\nu}(\vec{x}) T_{\rho\sigma}(0)\>$ is unique and depends only on $L$ and $G_4$.  In the rest of this paper, we will be interested in theories with AdS$_4$ duals.  If $L$ is the radius of AdS$_4$ and $G_4$ is the effective Newton constant in four-dimensions, we have \cite{Chester:2014fya}:
 \es{cTLargeN}{
  c_T = \frac{32 L^2}{\pi G_4} \,.
 }
The correlation functions \eqref{TwoSummary} then become:
 \es{CorrFinal}{
  \langle {\cal O}(\vec{x}_1, M_1) {\cal O}(\vec{x}_2, M_2) \rangle  &= \frac{ L^2}{4 \pi^3 G_4} \frac{\tr (M_1 M_2)}{\abs{\vec{x}_1 - \vec{x}_2}^2} \,, \\
   \langle {\cal O}(\vec{x}_1, M_1) {\cal O}(\vec{x}_2, M_2) {\cal O}(\vec{x}_3, M_3) \rangle  &= \frac{L^2}{8  \pi^4 G_4}\frac{\tr (M_1 M_2 M_3 + M_1 M_3 M_2)}{{\abs{\vec{x}_1 - \vec{x}_2} \abs{\vec{x}_1 - \vec{x}_3} \abs{\vec{x}_2 - \vec{x}_3}} }\,. 
 }
One of our main goals in the remainder of this paper is to reproduce these formulas from a holographic computation.

\subsubsection{An example}

Ref.~\cite{Freedman:2013ryh} considered only three of the 35 operators, denoted ${\cal O}_{\alpha}$, with $\alpha=1, 2, 3$, corresponding to 
 \es{OBDef}{
  {\cal O}_{\alpha} = 2 J_{(\alpha)} 
 }
with $J_{(\alpha)}$ defined right above \eqref{MiDef}.  The 2-point function of ${\cal O}_\alpha$ is
 \es{ParticularTwoPoint}{
   \langle {\cal O}_\alpha(\vec{x}_1) {\cal O}_\beta(\vec{x}_2) \rangle = \frac{ L^2}{2 \pi^3 G_4} \frac{\delta_{\alpha\beta}}{\abs{\vec{x}_1 - \vec{x}_2}^2} \,.
 } 
 Using \eqref{CorrFinal}, one can check that all 3-point functions between ${\cal O}_{\alpha}$ vanish except for
  \es{ParticularThreePoint}{
    \langle {\cal O}_{1}(\vec{x}_1) {\cal O}_{2}(\vec{x}_2) {\cal O}_{3}(\vec{x}_3) \rangle  &= \frac{L^2}{4  \pi^4 G_4}\frac{1}{{\abs{\vec{x}_1 - \vec{x}_2} \abs{\vec{x}_1 - \vec{x}_3} \abs{\vec{x}_2 - \vec{x}_3}} } \,,
  }
as well as symmetric permutations of ${\cal O}_{\alpha}$.  For a different computation of these correlation functions, see Appendix~\ref{ALTERNATIVE}.\footnote{It should also be possible to calculate $c_2$ and $c_3$ directly in the SCFT using the gauged quantum mechanics obtained in \cite{Dedushenko:2016jxl}.}

\section{Boundary terms in $\cn =1$ truncations}\label{BOGO}

\subsection{Review of the Bogomolny argument in \cite{Freedman:2013ryh}}
\label{ss:N1bog}

The first (not so gentle) hint that a boundary counterterm may provide the answer to the puzzle of the vanishing $\<\co_1(\vx_1)\co_2(\vx_2)\co_3(\vx_3)\>$ correlator from bulk supergravity came from Appendix \ref{Appendix:Trunc} of \cite{Freedman:2013ryh}.  In this reference, a Bogomolny argument was used to generate the BPS equations for a general $\cn=1$ supergravity model with asymptotically AdS$_4$ solution. The model contains chiral multiplets with a K\"ahler target space with K\"ahler potential $K(z,\bar z)$ and a holomorphic superpotential $W_\text{SG}(z)$.  We now summarize the results.

When the domain wall Ansatz 

\begin{equation}\label{domwall}
ds^2 = e^{2 \cals A(r)}\eta_{ab}dx^a dx^b + dr^2 \,,\qquad    z^\a =z^\a(r)\,, \qquad \bar z^{\bb}= \bar z^{\bb}(r) 
\end{equation}
is inserted in the (Lorentzian signature) bosonic action
\begin{equation} \label{Sbps}
\begin{split}
S & \eql \frac{1}{8\pi G_4}\int d^4x\sqrt{-g}\,\Big[ \frac12 R - K_{\a\bb}\pa_\m z^\a \pa^\m \bz^{\bb} - V_\text{SG}\Big]\,,  \\[6 pt]
V_\text{SG} & \eql e^{K}\,\Big[g^{\a\bb}\nabla_\a W_\text{SG}\nabla_\bb\overline W_\text{SG} - 3W_\text{SG}\overline W_\text{SG}\Big]\,,\\[6 pt]
\nabla_\a W_\text{SG} & \eql (\pa_\a + K,_\a)W_\text{SG}\,,   \quad \nabla_\bb\overline W_\text{SG} =(\pa_\bb +K,_\bb)\overline W_\text{SG}\,, \end{split}
\end{equation}
the action can be manipulated by partial integration and turned into a sum of quadratic factors which are the BPS equations
\begin{equation}
\begin{split}
 \label{floweqs}
\pa_r z^\a &\eql - e^{K/2} \sqrt{W_\text{SG}/\overline{W}_\text{SG}}\, K^{\a\bar\g}\nabla_{\bar\g}\overline{W}_\text{SG}\,,\\[6 pt]
 \pa_r \bz^{\bb}&\eql - e^{K/2} \sqrt{\overline{W}_\text{SG}/W_\text{SG}}\, K^{\d\bb}\nabla_\d W_\text{SG}\,,\\[6 pt]
\pa_r \ca &\eql  e^{K/2} |W_\text{SG}| \,,
\end{split}
\end{equation}
plus the boundary term (at the cutoff $r_0$)
\be \label{cutoff}
S_{\rm cutoff} = \frac{1}{4\pi G_4}\int d^3x\, dr\, \frac{\pa}{\pa r}\Big(\sqrt{-g} e^{K/2} |W_\text{SG}|\Big) =\frac{1}{4\pi G_4}\int d^3x\, e^{3\cals A} e^{K/2} |W_\text{SG}| \,.
\ee
This surface term must be cancelled by adding an equal and opposite counterterm to the action, which we will do momentarily.

In the specific 3-scalar truncation studied in \cite{Freedman:2013ryh}, the superpotential and K\"ahler potential are
 \es{SuperpotKahler}{
  W_\text{SG}= \frac{1+z^1z^2z^3}{L} \,, \qquad K = - \sum_{\alpha=1}^3 \log \left[ 1- \abs{z^\alpha}^2 \right] \,.
 }
The constant term in $W_\text{SG}$ determines AdS scale. The warp factor of the domain wall solution tends to $e^{2\ca(r)}  \to e^{2r/L}$ at large $r$, 
and the scalars vanish at the rate $z^\a(r) \sim e^{-r/L}$.  The counterterm, which is K\"ahler invariant, is (at fixed large $r$)
\begin{equation}\label{bpsct}
\begin{split}
S_{\rm BPS} &\eql -\frac{1}{4\pi G_4}\int d^3x  e^{3A} e^{K/2} |W_\text{SG}|\\[6 pt]
&\eql -\frac{1}{4\pi G_4L}\int d^3x \, e^{3r/L}\,\Big[\,1  +\frac12 \delta_{\alpha\bar\beta} z^\alpha(r)\bar z^{\bar\beta}(r)\\[6 pt]
& \hspace{2.19in} +\frac12(z^1(r)z^2(r)z^3(r) + \text{c.c.})+\ldots\Big].
\end{split}
\end{equation}

The constant part of $|W_\text{SG}|$ gives a cubic divergence as $r\to\infty$, and the quadratic term from the K\"ahler potential gives a linear divergence. Both terms agree with standard  counterterms from holographic renormalization. The third term is finite, and it is this that provides the boundary cubic vertex which will be used to calculate $\<\co_1(\vec x_1)\co_2(\vec x_2)\co_3(\vec x_3)\>$ in Section~\ref{sec:corrO}.

It is important to point out that the precise agreement of the free energy found in \cite{Freedman:2013ryh} between the AdS/CFT result and that from supersymmetric localization in the dual ABJM field depended crucially on the added cubic counterterm.  Since BPS domain walls are supersymmetric, the new counterterm is a consequence of SUSY.

\subsection{Boundary terms required by supersymmetry}
\label{sec:Btfe}

In most studies of supergravity theories, boundary terms  generated in the process of checking local supersymmetry are  discarded, since the supersymmetry parameters, $\eps(r,\vx)$, are arbitrary functions and may be assumed to vanish rapidly at the boundary.   However, in AdS, the spinors $\eps(r,\vx)$, are required to  approach an AdS Killing spinor at the boundary and this leads to finite and even divergent boundary contributions.  Without the addition of appropriate boundary terms, as we will explain, the action is simply not supersymmetric.

Let us be more precise.  The most basic AdS/CFT setup involves the study of the states in a CFT\@.  This is to be contrasted with the study of relevant deformations and correlation functions of the CFT via holographic sources, which we will discuss in the next paragraph.  In general, the states of a CFT are described by bulk field configurations obeying boundary conditions that 1) provide a well-defined Euler-Lagrange principle, namely that the Euler-Lagrange equations follow from the vanishing of the variation of the action, without discarding any boundary terms.  If the CFT is supersymmetric, which is the case of interest here, the boundary conditions used to describe states of the CFT must also 2) be preserved under arbitrary supersymmetry variations; and 3) ensure that the action is supersymmetric, also without discarding any boundary terms.  The point we will make is that these conditions cannot be obeyed without the addition of certain boundary counterterms.  See also~\cite{Henningson:1998cd,Mueck:1998iz,Arutyunov:1998ve,Henneaux:1998ch,D'Hoker:1999ea,Belyaev:2008ex,vanNieuwenhuizen:2006pz,vanNieuwenhuizen:2005kg,Hawking:1983mx,Grumiller:2009dx,Belyaev:2008xk,Corley:1998qg,Volovich:1998tj,D'Hoker:1999ea, Rashkov:1999ji,Bianchi:2001de,Bianchi:2001kw,Amsel:2009rr,Andrianopoli:2014aqa}. 

As a more involved application of AdS/CFT to supersymmetric field theories, one generalizes the boundary conditions discussed above to allow for deformations of the CFT by introducing sources for relevant operators.  For a general given source configuration, the action will not be supersymmetric.\footnote{For certain special source configurations, the action may preserve a fraction of the supersymmetries preserved by the vacuum.}  Instead, supersymmetry relates various source configurations to one another.  So, the on-shell supergravity action, when viewed as a functional of the various field theory sources, should still be supersymmetric, provided that the sources are are transformed appropriately instead of being held fixed.  Indeed, it is \emph{usually} the on-shell action $S_\text{on-shell}$, viewed as a functional of various field theory sources, that is interpreted by the AdS/CFT dictionary as the generating functional of connected correlation functions, and this generating functional should be supersymmetric.   We will actually deal with \emph{a somewhat exceptional application of AdS/CFT}, because the three bulk scalars $A^\a = \Re z^\a$ in the $\cn=1$ truncation (and the 35 $\alpha^{ijkl} $ in the $\cn=8$ theory) are dual to $\D=1$ operators in the dual CFT.  In this case it is not $S_\text{on-shell}$ but rather its Legendre transform \cite{Klebanov:1999tb},  defined and called $\tilde S_\text{on-shell}$ in Section~\ref{sec:countN1} below, that is the generating functional. 
Supersymmetry requires that this generating functional is supersymmetric, provided that the field theory sources are assigned appropriate  transformation rules.\footnote{Alternate quantization and the Legendre transform are needed to describe CFT operators whose scale dimension is given by the lower sign in the AdS/CFT mass fomula $\D = (d \pm \sqrt{d^2 +4 m^2L^2})/2.$}

In the remainder of this section, we determine the boundary counterterms that ensure that $\tilde S_\text{on-shell}$ is supersymmetric.  In the limit in which the cutoff is removed, $r_0\to\infty$, we find a set of infinite and  finite boundary counterterms.  The infinite counterterms agree with those obtained by holographic renormalization and the finite ones include the finite term of $S_\text{BPS}$ in \eqref{bpsct}.

Since gauged $\cn=8$ supergravity is a rather complicated theory,  we first present a detailed illustration of the technique in a far simpler model, an $\cn=1$ model with global SUSY in AdS$_4$.  This model is obtained in a limit of $\cn =1$ 
supergravity, similar to that of \cite{Festuccia:2011ws}, in which the back-reaction of matter fields on the spacetime geometry is consistently suppressed.  We then outline the extension of the method to $\cn =1$ supergravity and finally proceed  to derive the analogous results in the $\cn=8$ theory.

\subsection{The global limit of $\cals N=1$, AdS$_4$ supergravity}

In this section we derive the action and transformation rules of  chiral multiplets of a global SUSY model on a fixed  AdS$_4$ background geometry. We derive this model from $\cn =1$ supergravity written in conventions very similar\footnote{Here we scale the SUSY parameter  $\eps$ of \cite{book} to $\sqrt2\eps.$} to those of Chapter 18 of \cite{book}.  The action is normalized as\footnote{To avoid potential confusion, we note that complex scalars in this section are canonical and have engineering dimension 1. They are related to the dimensionless scalars of \cite{Freedman:2013ryh} and previous sections of this paper by $z_{\rm here} = z_{\rm there}/\k$. When  this and the analogous scaling is made for spinors, the supergravity action acquires the overall factor $1/8\pi G_4$.\label{kappascaling}}\footnote{For clarity, we write $\slashed{\nabla} = \gamma^\mu \partial_\mu$ as an operator acting on the 4d fields, and $\slashed{\partial} = \gamma^a \partial_a$ as an operator acting only in the boundary directions.} 
 \es{normaction}{
S_\text{SG} &= \int d^4x\sqrt{-g}\,\Big[\,\frac{1}{2\k^2} \left( R - \bar \psi_\mu \gamma^{\mu\nu\rho} \nabla_\nu \psi_\rho  \right) \\
 &\qquad\qquad{}- g_{\a\bb}\biggl(\pa_\m z^\a\pa^\m\bz^{\bb} + \frac12 \,\bar\chi^\a\slashed{\nabla}P_R\chi^\bb + \frac12\,\bar\chi^\bb\slashed{\nabla}P_L\chi^\a\biggr) +  \cdots\,\Big]\,.
 }
Factors of $\k$ with $\k^2 =8\pi G_4$ are included in the non-linear terms indicated by $\cdots$.  

The dynamics of the supergravity model is specified by a K\"ahler potential $K(z,\bz)$ and a holomorphic superpotential of the form
\be\label{wsg}
W_\text{SG}(z) \eql  \frac{1}{\k^2 L} + W(z)\qquad \longrightarrow\qquad \frac{1}{\k^2 L} +\frac{\k}{L}z^1z^2z^3 \,.
\ee
The superpotential in the $\cn =1$ truncation of $\cn=8$ supergravity studied in \cite{Freedman:2013ryh} appears on the right.
The condition that the theory admit a supersymmetric   AdS$_4$ solution of scale $L$ is that
\be\label{bkgd}
\nabla_\a W_\text{SG} \eeql \big(\pa_\a +\kappa^2\pa_\a K \big)W_\text{SG}\eql 0\,,
\ee
is satisfied at $z_\a=0$.  This condition is fulfilled in the model of \cite{Freedman:2013ryh}.

The global limit of the supergravity action is obtained via the following procedure:  
\begin{enumerate}
\item Fix the  AdS$_4$ background and use coordinates $r,x^a,~a=0,1,2$ in which 
\be\label{adsmet}
ds^2 = e^{2r/L}\eta_{ab}dx^a dx^b + dr^2\,.
\ee
\item Set the gravitino field to $\psi_\m=0$.  This is consistent if we require that the SUSY parameters are Killing spinors of  AdS$_4$ and thus satisfy
\be \label{kspin}
\nabla_\mu \eps = - \frac{1}{2L}\g_\m\eps \qquad\implies \qquad  \slashed{\nabla}\eps = -\frac{2}{L}\eps \,.
\ee
\item  Use \eqref{wsg} to obtain the superpotential $W(z)$ of the global model.
\item  Keep the $\kappa$ factors in $K(z,\bz)$ and $W(z)$, but otherwise drop all terms in the supergravity action with positive powers of $\k$. 
 \end{enumerate}

When this procedure is applied to the scalar potential of $\cn=1$  supergravity, one obtains

\begin{equation}\label{}
\begin{split}
V_\text{SG} &\eeql e^{\k^2K}[g^{\a\bb}\nabla_\a W_\text{SG}\nabla_{\bb} \overline W_\text{SG} -3\k^2 W_\text{SG}\overline W_\text{SG} ]\\[6 pt]
&\eql g^{\a\bb}(\pa_\a W(z)+\frac{1}{L}\pa_\a K)(\pa_\bb\overline W(\bz) +\frac{1}{L}\pa_\bb K) -\frac{3}{L}(W+\overline W) -3K + O(\k^2)\,,
\end{split}
\end{equation}
which agrees with (3.8) of \cite{Festuccia:2011ws}.
An additional cosmological constant term $-3/\k^2L^2$ has been dropped since it is 
part of the gravitational sector whose solution is fixed.  

The entire action obtained from our procedure agrees with (3.5) of \cite{Festuccia:2011ws}.  However, we now make a further  assumption which simplifies the
analysis needed for our main purpose which is to determine the boundary terms in the variation of the action.  
Namely, we assume that the K\"ahler metric is flat.  This is 
justified because the K\"ahler potential of the $\cn=1$ truncation and the parent $\cn=8$ theory has the structure
\be
K(z,\bz) =  z\bz + a_2(z\bar z)^2 + a_3(z\bar z)^3+\cdots \,.
\ee
In models with cubic $W(z)$, scalar masses $m^2 = -  2/L^2$ are entirely determined by the conformal coupling, so the leading asymptotic behavior of scalar fields is $z(r,x)\sim e^{-r/L}$.  Thus the effects of target space curvature are suppressed by $e^{-2r/L}$ relative to the leading term, and they play no role in the determination of boundary terms.  

After the procedure above is implemented we make the further step of introducing auxiliary $F,~\bar F$ fields.  It is also sufficient to consider a single chiral mutiplet $(z,P_L\chi,F)$. This enables us to write the action  as\footnote{This action was studied in Section~3 of \cite{Burges:1985qq}.}
\begin{equation}\label{globLag}
S_\text{bulk} \eql S_\text{kin} +S_F + S_{\bar F}\,,
\end{equation}
where
\begin{equation}\label{}
\begin{split}
S_\text{kin}\eql \int d^4x\sqrt{-g}\,\Big[-\pa_\m z \pa^\m \bz & - \frac12\,\big(\bar\chi \slashed{\nabla} P_L\chi + \bar\chi\slashed{\nabla} P_R\chi\big)\\[6 pt]
&  +\Big(F+{z\over L}\Big)\Big(\overline F +{\bar z\over L}\Big) +{2\over L^2}\,z\bar z\Big]\,,
\end{split}
\end{equation}
\begin{equation}\label{}
S_F\eql \int d^4x\sqrt{-g}\,\Big[FW' -\frac12 W''\bar\chi P_L\chi + {3\over L}\,W\Big]\,,\qquad 
S_{\bar F} \eql (S_F)^\dagger \,.
\end{equation}
 It is very useful to have three terms which are separately invariant under the transformation rules:
\begin{equation}\label{trafos}
\begin{split}
\d z &\eql \bar\eps P_L\chi\,,\qquad \d P_L\chi =P_L(\slashed{\nabla}z +F)\eps\,,\qquad\d F=\bar\eps(\slashed{\nabla} -1/L)P_L\chi\,,\\[6 pt]
\d \bz & \eql \bar\eps P_R\chi\,,\qquad \d P_R\chi =P_R(\slashed{\nabla}\bz +\bar F)\eps\,,\qquad\d \bar F=\bar\eps(\slashed{\nabla} -1/L)P_R\chi\,.
\end{split}
\end{equation}

The proof of invariance is quite simple for $S_F$:
\begin{equation}\label{}
\begin{split}
\d S_F  \eql \int d^4x\sqrt{-g}\,\Big[ & F W''\,\bar\eps P_L\chi  - W''\,\bar\eps(-\slashed{\nabla}z+F)P_L\chi\\ &    +
W'\,\bar\eps\Big(\slashed{\nabla} -{1\over L}\Big)P_L\chi 
+ {3\over L}\,W'\, (\bar\eps P_L\chi) -W'''\, (\bar\eps P_L\chi)(\bar\chi P_L \chi)\Big] \,.
\end{split}
\end{equation}
Terms involving  $F$ cancel and the $W'''$ term vanishes by Fierz rearrangement. The remaining terms can be written as
\begin{equation}\label{dsf}
\begin{split}
\d S_F  &\eql \int d^4x\sqrt{-g}\,\Big[\,\bar\eps\,\slashed{\nabla}(W'P_L\chi) +{2\over L}\,W'\,\bar\eps  P_L\chi\Big]\\[6 pt]
&\eql  \int d^4x\sqrt{-g}\,\Big[\nabla_\m(W'\,\bar\eps\g^\m  P_L\chi) -\bar\eps\Big(\leftDs -{2\over L}\Big)\,W'P_L\chi\Big]\,.
\end{split}
\end{equation}
The last term vanishes by the (adjoint of the) Killing spinor equation \eqref{kspin}, and the first term is the total derivative which is the goal of the calculation. 

It is more difficult to show that $\d S_\text{kin}$ is invariant up to boundary terms.  Details are given in Appendix~\ref{appendixSkin}.  Here we simply write the final expression that contains the residual boundary terms
\be\label{dskinfinal}
\d S_\text{kin} = \frac12 \int d^4x\sqrt{-g} \,\nabla_\m\Big[\,\bar\eps\g^\m\Big(-\slashed{\nabla}(z P_R +\bz P_L) +{2\over L}\, (z P_R+\bz P_L) +(FP_R+\bar FP_L)\Big)\chi\Big] \,.
\ee 

The analysis above is valid for a general superpotential, $W(z)$.  However, we are specifically concerned with a cubic $W(z)$, which, for the purpose of providing a toy model, we take to be
 \es{WChoice}{
  W(z) = \frac{\k z^3}{3 L} \,.
 }
The consistent truncation of $\cn =8$ studied in \cite{Bobev:2013yra,Pilch:2015dwa,Pilch:2015vha} contains three identical chiral multiplets and is trivially related to ours, as is the truncation to three chiral multiplets with $W =\k z^1z^2z^3/L$ of \cite{Freedman:2013ryh}.

Finally we note that auxiliary fields are eliminated and real fields are introduced using
\begin{equation}\label{aux}
F=-{z\over L} -\overline W{}'\eql  -{z\over L}-\k\,{\bz^2\over L^2}\,,\qquad z=A+iB\,.
\end{equation}

\subsection{Further conventions and asymptotic behavior} 

Before determining the boundary counterterms that ensure 
supersymmetry it is useful to state our conventions more completely and to discuss the asymptotic behavior of the various actors in our drama.

In the natural Lorentz frame, $e^a =e^{r/L} dx^a$ and $e^3=dr$, $a=0,1,2$, for the metric \eqref{adsmet},  
$(\gamma^a,\gamma^3)$ are constant $\gamma$-matrices for signature $(-+++).$ As usual, the $\gamma$-matrices with a Greek index are defined by $\gamma^\mu=e_a{}^\mu \gamma^a+e_3{}^\mu\gamma^3$. 
In the language of the Cartan structure equations,   the connection 1-forms are $\omega^{  a 3} = e^{ a}/L,~~ \omega^{  a  b}=0.$ 

The Killing spinors of the Poincar\'e patch are Majorana spinors.  In AdS$_4$, the Killing spinor equation \eqref{kspin} has solutions of the form
\begin{align}
 \label{kspin1}
&\eps \eql e^{r/2L}\eta_- +e^{-r/2L}\eta_+ \,,
\end{align}
with coefficients $\eta_-(\vx)$ and $\eta_+(\vx)$ that obey $\slashed{\partial} \eta_+ = 0$ and $\slashed{\partial}\eta_- = -(3/L) \eta_+$ and have definite ``radiality'':
 \es{etarad}{
  \g^3\eta_{\pm} = \pm\eta_{\pm}\,,\qquad  \bar\eta_{\pm}\g^3=\mp\bar\eta_{\pm} \,.
 }
In particular, there are two linearly independent Poincar\'e supersymmetries that have $\eta_+ = 0$ and $\eta_-=  {\rm constant}$, as well as two superconformal supersymmetries that have $\eta_+ = {\rm constant}$ and $\eta_- = -\g_{ a}x^a \eta_+/L$.\footnote {The designations Poincar\'e and superconformal arise because the associated supercharges anti-commute to translations and, respectively, special conformal transformations of the isometry group SO(3,2).}

The behavior of solutions of the field equations as $r\to \infty$ is (with $z = A+iB$)
\begin{equation}\label{asympts}
\begin{split}
A(r,\vx) &\eql e^{-r/L}A_1(\vx) + e^{-2r/L} A_2(\vx)+\dots\,, \\[6 pt]
B(r,\vx) &\eql e^{-r/L} B_1(\vx)+ e^{-2r/L}B_2(\vx) +\dots\,,\\[6 pt]
\chi(r,\vx) &\eql e^{-3r/2L}\chi_{3/2}(\vx) +e^{-5r/2L}\chi_{5/2}(\vx)+\dots\,.
\end{split}
\end{equation}
The leading rates are  standard in AdS$_4$/CFT$_3$ for scalars of mass $m^2=-2/L^2$ and massless spinors.  In a free theory, i.e. $W(z)=0$, the asymptotic series for $A$ and $B$ would contain exponential rates $e^{-kr/L}$ with $k$ either even or odd \cite{Breitenlohner:1982jf}.  The presence of mixed even and odd integer rates occurs with interactions and is important in our analysis.

From the bulk supersymmetry variations \eqref{trafos} and the decomposition  \eqref{kspin1}  for the Killing spinors, we find  the supersymmetry transformations of the various coefficients appearing in the boundary expansion \eqref{asympts}:
 \es{InducedSUSY}{
  \delta A_1 &= \frac 12 \bar \eta_- \chi_{3/2 +} \,, \qquad \qquad  \delta A_2 = \frac 12 \left( \bar \eta_- \chi_{5/2+} + \bar \eta_+ \chi_{3/2-} \right) \\
  \delta B_1 &= -\frac i2 \bar \eta_- \gamma^5 \chi_{3/2-} \,, 
   \qquad 
   \delta B_2 = -\frac i2 \left(\bar \eta_- \gamma^5 \chi_{5/2-} + \bar \eta_+ \gamma^5 \chi_{3/2+} \right) \,, \\
  \delta \chi_{3/2-} &= \left( \frac{1}{L}A_2 - \frac{\kappa}{L} (A_1^2 - B_1^2) + i \gamma^5 \slashed{\partial} B_1 \right) \eta_- - \frac{2i}{L} B_1 \gamma^5 \eta_+ \,, \\
  \delta \chi_{3/2+} &= i \gamma^5 \left( \frac{1}{L}B_2 + \frac{2\kappa}{L} A_1 B_1 \right) \eta_- + \slashed{\partial} A_1 \eta_- - \frac{2}{L} A_1 \eta_+ \,. 
 }
Here and in the rest of this section we find it convenient to split the coefficient functions $\chi_k(\vx)$ appearing in the expansion of $\chi(r, \vx)$ into components of even and odd radiality,  denoted by an additional $\pm$ subscript:
 \es{FermRadiality}{
  \chi_k(\vx) = \chi_{k+}(\vx) + \chi_{k-}(\vx) \,, \qquad
   \gamma^{3} \chi_{k \pm} = \pm \chi_{k \pm} \,.
 }

\subsection{Counterterms and CFT states}
\label{sec:countStates}

We now turn to our goal of finding the appropriate boundary counterterms that ensure supersymmetry.  As already mentioned, the appropriate requirement in its most general form is that the Legendre transform of $S_\text{bulk} + S_\text{bdy}$, seen as a functional of the boundary theory sources, is supersymmetric.  As a particular simpler case that does not require a Legendre transform, we first study the case where the boundary sources vanish and find the boundary counterterms $S_\text{bdy}$ that ensure supersymmetry, as explained in Section~\ref{sec:Btfe}.  The counterterms $S_\text{bdy}$ are initially evaluated at the cutoff $r=r_0$; in the limit $r_0\to \infty$ they are expressed in terms of the asymptotic coefficients of \eqref{asympts}.

In determining the boundary conditions and boundary counterterms that ensure supersymmetry, we can take guidance from the fact that the pseudoscalar $B(r, \vec{x})$ is dual to a dimension $2$ operator in the dual CFT\@.  Consequently, the standard AdS/CFT dictionary identifies $B_1(\vec{x})$ as the field theory source for this operator.  The condition of vanishing sources should therefore include $B_1(\vec{x}) = 0$.  The supersymmetry variations \eqref{InducedSUSY} then identify a consistent set of boundary conditions on the other fields.  Indeed, by considering $\delta B_1(\vec{x})$, one also obtains $\chi_{3/2-}(\vec{x}) = 0$, and then from $\delta \chi_{3/2-}(\vec{x}) = 0$ one further obtains $A_2(\vec{x}) - \kappa A_1^2(\vec{x}) = 0$.  In summary, the conditions of vanishing sources are
 \es{VanishingSources}{
  B_1(\vec{x}) = 0 \,, \qquad \chi_{3/2-}(\vec{x}) = 0 \,, \qquad
   A_2(\vec{x}) - \kappa A_1^2(\vec{x}) = 0 \,,
 }
    and they represent our desired boundary conditions.\footnote{It is well known \cite{Henningson:1998cd, Mueck:1998iz, Arutyunov:1998ve,Henneaux:1998ch} that one should choose one of the two asymptotic projections $\chi_{3/2\,\pm}$ in \eqref{InducedSUSY} as the fermion source.}  

The boundary counterterms are then determined by ensuring that the boundary conditions \eqref{VanishingSources} are consistent with the Euler-Lagrange variational principle.  Let us examine the scalar part of the action first.  Integrating out the auxiliary fields and using the cubic superpotential \eqref{WChoice}, the scalar part of the bulk action becomes
 \es{ScalarAction}{
   S_\text{bulk} &= \int d^4x\sqrt{-g}\,\Big[-\pa_\m z \pa^\m \bz    +{2\over L^2}\,z\bar z - \frac{\kappa^2}{L^2}  (z \bar z)^2 \Big] \,.
 }
The Euler-Lagrange variation of the action reads 
 \es{EulerScalar}{
  \delta S_\text{bulk} = \int d^4 x \sqrt{-g} \, \Bigl[
    \delta z (\text{eom for $\bar z$}) +  \delta \bar z (\text{eom for $z$}) -\nabla_\m(\d z\pa^\m\bz + \d\bz\pa^\m z ) \Bigr] \,.
 } 
The variational principle implies the equations of motion provided that we add a boundary term whose variation cancels the second term in \eqref{EulerScalar}.  Using the asymptotic expansion \eqref{asympts} and the boundary conditions \eqref{VanishingSources}, we have 
 \es{VarPrinc2}{
  \delta S_\text{bdy} + \frac 1L \int d^3 x \, 
   \Bigl[ 2  e^{r_0/L} (A_1 \delta A_1) +
    8 \kappa A_1^2 \delta A_1 
     \Bigr] = 0 \,,
 }
where the second term in \eqref{VarPrinc2} comes from the last term in \eqref{EulerScalar}.  From this expression we deduce that the required boundary term is
 \es{SbdyQuadCubic}{
  S_\text{bdy} = -  {1\over L} \, \int d^3xe^{3r_0/L}\,\Big[A^2 +{2\kappa \over 3}\, A^3 \Big] \,,
 } 
because its variation gives \eqref{VarPrinc2}, again after using the boundary conditions \eqref{VanishingSources}.   A similar analysis for the fermionic part of the action shows that there are no fermionic boundary terms that do not vanish under \eqref{VanishingSources}, the boundary term $S_\text{bdy}$ being the only boundary term that is needed.  One can then check that the combined action $S_\text{bulk} + S_\text{bdy}$ is supersymmetric.  This calculation is a particular case of the calculation performed in the next section, and we will defer it until then.  

What we have done so far amounts to a ``minimal supersymmetric completion''  of the bulk action via the boundary term \eqref{SbdyQuadCubic}. Without this boundary term and the boundary conditions \eqref{VanishingSources}, the theory would not be supersymmetric.

\subsection{More general counterterms and the Legendre transform}
\label{sec:countN1}

We now proceed to an analysis that is not restricted to the CFT states but allows non-vanishing sources for relevant operators.  In particular, we relax the conditions \eqref{VanishingSources} by allowing arbitrary field theory sources, as we will explain.  We will determine a more general boundary action $S_\text{bdy}$ that reduces to \eqref{SbdyQuadCubic} when the sources are taken to vanish as in \eqref{VanishingSources}.

Recall that the supersymmetry variation of the bulk action, $\delta S_\text{bulk} = \delta S_\text{kin} + \delta S_F + \delta S_{\bar F}$, reduces to a boundary term given in \eqref{dsf}--\eqref{dskinfinal}.  It is straightforward to cancel various contributions to $\delta S_\text{bulk}$ against the variation of an appropriately chosen boundary counterterm $S_\text{bdy}$.  For instance, it is clear that $\d S_F$ (and its conjugate $\d S_{\bar F}$) are finite at the boundary and can be nicely cancelled by the variation of the finite counterterm
\be
\label{S3}
S_{3} =-\int d^3x \,e^{3r_0/L} [W(z)+ \bar  W(\bar z)] \,.
\ee
The remaining boundary term $\delta S_\text{kin}$ is ``linearly divergent.''  Its leading term grows as $e^{r_0/L}$ at the boundary when we include the factor $\sqrt{-h} = e^{3r_0/L}$, $h$ being the determinant of the boundary metric.  We expect that such divergences are cancelled by counterterms determined by holographic renormalization.  The relevant counterterm can be obtained from (6.5) of \cite{Freedman:2013ryh}.   With a sign change for Lorentzian signature and in the global limit and with current normalization,  it is given by 
\be\label{sb}
S_2 \eql -{1\over L}\, \int d^3x\,e^{3r_0/L}\,\bar z z\,.
\ee
Upon adding $S_2$, the supersymmetry variation of the kinetic term \eqref{dskinfinal} is finite at the boundary.  After adjusting the normalization to that of Section~\ref{BOGO} and for cubic $W(z)$, $S_{3}$ and $S_2$
agree perfectly with the cubic and quadratic terms of \eqref{bpsct}. In the rest of this section we will work with the cubic $W(z)= \k z^3/3L$ introduced in \eqref{WChoice}.

The remaining finite terms of $\d(S_\text{kin} + S_2)$ (to be displayed in the next section) must still be cancelled, and two further modifications are needed.  The first is to add another finite counterterm
\be\label{schi}
S_\chi \eql \frac{c}{4} \int d^3x\,e^{3r_0/L}\, \bar\chi\chi\,.
\ee
This was proposed in the earliest papers on fermions in AdS/CFT  \cite{Henningson:1998cd,Mueck:1998iz,Arutyunov:1998ve,Henneaux:1998ch} in order to obtain non-trivial 2-point correlators of fermionic operators in the boundary theory.  The coefficient $c$ will be fixed at the value $c=1$ below.\footnote{ For a Dirac fermion, the coefficient was fixed in \cite{Arutyunov:1998ve}  and \cite{Henneaux:1998ch}. }

The second modification involves the Legendre transform that was mentioned in Section~\ref{sec:Btfe}.  It is a more subtle issue that we now discuss in detail.   We know that the scalar field $A$ is dual to a field theory operator of dimension $1$, and hence obeys ``alternate boundary conditions'' as explained in \cite{Klebanov:1999tb}.  Let us explain what this means by comparison to the pseudoscalar $B$, which is dual to a dimension-$2$ operator and obeys standard boundary conditions.  For $B$, the leading coefficient in the boundary expansion \eqref{asympts}, $B_1(\vx)$, is interpreted as a source for the dual operator.  The Euler-Lagrange equations of motion are solved with the boundary condition of a prescribed value for $B_1(\vx)$, and the on-shell action is naturally thought of as a functional of $B_1(\vx)$.  For $A$, it is not the leading coefficient, $A_1(\vx)$, that should be interpreted as the source for the field theory operator, but instead its canonically conjugate quantity \cite{Klebanov:1999tb}
 \es{calADefSUSY}{
  \fA(\vx) = -\frac{\delta S_\text{on-shell}[A_1, \ldots]}{\delta A_1(\vx)} \,.
 }
Here, the ellipsis stands for other boundary data, such as $B_1(\vx)$, that can be interpreted as sources for field theory operators.  The source $\fA(\vx)$ is sometimes loosely referred to as $A_2(\vx)$, because a simple calculation, 
 \es{calAAgainSUSY}{
  \fA(\vx) = -\lim_{r_0\to \infty} e^{-r_0/L} \frac{\delta S_\text{on-shell}}{\delta A(r_0, \vx)}
   &=  - \lim_{r_0 \to \infty} e^{-r_0/L} \Pi_A(r_0, \vx) \\
   &= -  \left( \frac{2}{L} A_2(\vx)  - \frac{2\kappa}{L} (A_1(\vx)^2 - B_1(\vx)^2) \right) \,,
 }
shows that, up to a normalization factor, it is equal to $A_2(\vx)$ plus non-linear corrections coming from the boundary terms \eqref{S3}--\eqref{sb}.  
Note that $\fA (\vx)$ is the boundary limit of canonical momentum for the field $A(r,\vx)$, namely\footnote{Here $\cal L $ is the Lagrangian obtained from the  action \eqref{globLag},  augmented by conversion of the boundary actions \eqref{S3}--\eqref{sb} into total $\pa_r$ derivatives.}
\be\lab{canp}
\Pi_A(r, \vx) = e^{3r/L}\frac{\pa\cal L} {\pa(\partial_r A(r,\vx))} = -2 e^{3 r/L} \left(\partial_r A + \frac{1}{L}A + \frac{\kappa}{L} (A^2 - B^2) \right) \,,
\ee 
and that  the second equality in \eqref{calAAgainSUSY} follows from the Hamilton-Jacobi equation.

The generating functional for connected correlators is the Legendre transform
 \es{LegTransfSUSY}{
  \tilde S_\text{on-shell}[\fA, \ldots] = S_\text{on-shell}[A_1, \ldots] + \int d^3x \,  \fA(\vx) A_1(\vx) 
 }
evaluated after extremizing the RHS with respect to $A_1(\vx)$.  This extremization yields precisely \eqref{calADefSUSY}.  It is $\tilde S_\text{on-shell}$, and not $S_\text{on-shell}$, that is required to be supersymmetric when sources are present.

To ensure that $\tilde S_\text{on-shell}$ is supersymmetric, we need the supersymmetry variation $\d \fA(\vx)$, and this must be chosen as the variation of \eqref{calAAgainSUSY} when the equations of motion are used.  In particular, the fermion equation of motion implies
 \es{chi52}{
  \chi_{5/2+} =L \slashed{\partial} \chi_{3/2-} + 2\k( A_1 \chi_{3/2+} +  i B_1 \gamma^5 \chi_{3/2-}) \,.
 }
When combined with \eqref{calAAgainSUSY} and \eqref{InducedSUSY}, this yields 
 \es{onshell}{
  \delta \fA  =  -  2 \delta \left( \frac{A_2}{L} - \frac{\kappa}{L} ( A_1^2 - B_1^2) \right) 
   =  -  \left(  \bar \eta_- \slashed{\partial} \chi_{3/2-} + \bar \eta_+\frac{1}{L} \chi_{3/2-}  \right) \,.
 }

To summarize, we have added boundary terms to the bulk action of \reef{globLag} to obtain the renormalized action 
 \es{SrenDef}{
S_\text{ren} = S_\text{bulk} + S_\text{bdy} \,, 
 }
where the bulk and boundary terms are
 \es{SbulkSbdy}{ 
 S_\text{bulk} \equiv S_\text{kin} +S_F+S_{\bar F} \,, \qquad
 S_\text{bdy} \equiv S_2 +S_\chi+S_3 \,.
}
The renormalized action $S_\text{ren}$ is denoted by $S_\text{on-shell}[A_1, \ldots]$ when equations of motion are satisfied.  We identified the boundary limit $\fA (\vx)$ of the canonical momentum.   
We then defined the Legendre transform in \reef{LegTransfSUSY} which a functional of $\fA $.  This is the generating functional for correlation functions and will be used for this purpose in Section~\ref{sec:3pttoy}.
In the next subsection we show that 
 \es{SrenSL}{
  \delta (S_\text{ren} +S_L) = 0 \,, \qquad S_L \equiv \int d^3x\, \fA (\vx) A_1(\vx)\,,
 }
on-shell.

Before checking supersymmetry, let us make a comment about the field theory sources, which we have identified as $B_1,~\chi_{3/2\,-}$ and $\fA$.  As  argued in Section~\ref{sec:countStates}, the three sources should then transform among themselves under SUSY\@.  
It is worth writing the SUSY variations of the sources that result from these assignments:
\bea  \label{susyforsources}
\delta B_1 &=& -\frac i2 \bar \eta_- \gamma^5 \chi_{3/2-}\,,\\
\delta \chi_{3/2-} &=& \left( -i  \slashed{\partial} B_1 \gamma^5+ \frac12 \fA \right) \eta_- - \frac{2i}{L} B_1 \gamma^5 \eta_+ \,, \\
\d \fA &=&  - \left( \bar \eta_- \slashed{\partial} \chi_{3/2-} + \bar \eta_+ \frac{1}{L}\chi_{3/2-} \right) \,.
\eea
These transformations resemble the standard superconformal transformations of an $\cn=1, d=3$ scalar multiplet, albeit with artefacts of their origin as the boundary limits of the bulk theory.  It is straightforward
to compute the commutator of two Poincar\'e supersymmetry transformations, those with $\eta_+=0$ and $\pa_a\eta_-=0$, as described below \eqref{etarad}.  In terms of the effectively two-component spinor parameters $\eps =i\g^5\eta_-$, the result is 
\be\label{susyalg}
[\d_1, \d_2] \Phi(\vx) = - (\bar\eps_1\gamma^a\eps_2)\pa_a\Phi(\vx)\,,
\ee
for all components $\Phi = B_1,~\chi_{3/2\,-},~\fA$ of the multiplet.\footnote{We suggest that faithful readers try the Fierz rearrangement needed for the fermion.}

\subsection{Cancellation of the supersymmetry variation of the on-shell action}

Let us now show that \eqref{SrenSL} holds.  We have already argued that
 \es{SFCanc}{
  \delta (S_F + S_{\bar F} + S_{3}) = 0 \,.
 }
Our remaining task is to show that
 \es{deltaSRemaining}{
  \delta \left(S_\text{kin} + S_2 + S_\chi + S_L \right) = 0 \,,
 }
which we now proceed to do.

The variation of $S_2$ in \eqref{sb} is
\be\label{sbvar}
\d S_2 \eql  - {1\over L}\,\int d^3x\sqrt{-g}\,\bar\eps\,\big( z P_R +\bar z P_L\big)\chi\,.
\ee
By adding it to $\delta S_\text{kin}$ in \eqref{dskinfinal} we obtain
\be\label{sum2}
\begin{split}
\d(S_\text{kin} + S_2) \eql \frac12\int d^3x\sqrt{-g}\,\Big[\bar\eps\,\Big(-\g^3\slashed{\pa} +{2\over L}\,(\g^3-I)\Big)&\big(zP_R +\bz P_L\big)\chi \\[6 pt] &+\bar\epsilon\,\g^3\big(FP_R +\bar F P_L\big)\chi\Big]\,.
\end{split}
\ee
Using \eqref{aux}
as well as the boundary asymptotics \eqref{asympts}, we obtain
 \es{dSkSb}{
  \delta(S_\text{kin} + S_2) 
   &= \frac 1{2} \int d^3 x \biggl[
    \bar \eta_- \left[ \frac{A_2}{L} - \frac{\kappa}{L} (A_1^2 - B_1^2) -  \slashed{\partial} A_1 \right] \chi_{3/2} \\
    &{}+ \bar \eta_- \left[-\frac{B_2}{L} - \frac{2\kappa}{L} A_1 B_1 +  \slashed{\partial} B_1 \right]i \gamma^5 \chi_{3/2} 
    + \frac{1}{L} \bar \eta_+ \left[ - 2 A_1 + 2 B_1 i \gamma^5 \right] \chi_{3/2}  \biggr] \,,
 }
where we took the limit $r_0 \to \infty$. 
 
Next, we have the variation $\delta S_\chi$:
 \es{dSchi}{
  \delta S_\chi &= \frac c2 \int d^3 x \biggl[
    -\bar \eta_- \left[ - \frac{1}{L} (A_2 + i \gamma^5 B_2) + \frac{\kappa}{L} \left( A_1^2 - B_1^2 - 2 i A_1 B_1 \gamma^5 \right) \right] \chi_{3/2} \\
    &{}- \bar \eta_- \left[ \slashed{\partial} A_1 - i \gamma^5 \slashed{\partial} B_1 \right] \chi_{3/2} 
    -  \frac{1}{L} \bar \eta_+ \left[ 2 A_1 + 2 i \gamma^5 B_1 \right] \chi_{3/2} 
   \biggr] \,,
 } 
as well as the variation of $S_L$ computed after using \eqref{calAAgainSUSY} and \eqref{onshell}
 \es{deltaSL}{
  \delta S_L = -\int d^3 x \, \biggl[ \bar \eta_- A_1 \slashed{\partial} \chi_{3/2-} + \frac{1}{L}  \bar \eta_+ A_1 \chi_{3/2-} 
   +  \bar \eta_-  \left(\frac{A_2}{L} -  \frac{\kappa}{L} (A_1^2 - B_1^2) \right)\chi_{3/2 +}  \biggr]\,.
 }

We see that $\delta (S_\chi + S_L)$ can cancel $\delta (S_\text{kin} + S_2)$ in \eqref{dSkSb} only if $c=1$.  With this choice, the sum of \eqref{dSkSb}--\eqref{deltaSL} is
  \es{Diff}{
   \delta \left(S_\text{kin} + S_2 + S_\chi  + S_L \right)
   =  \int d^3 x \biggl[
   - \bar \eta_- (\slashed{\partial} A_1) \chi_{3/2-} - \frac{3}{L} \bar \eta_+ A_1 \chi_{3/2-}
     -  \bar \eta_- A_1 \slashed{\partial} \chi_{3/2-}  \biggr]  \,.
 }
Finally,  using $\slashed{\partial} \eta_- = -(3/L) \eta_+$ as explained below \eqref{kspin1}, we see that the integrand in this expression is a total derivative.  Thus \eqref{deltaSRemaining} follows.

\subsection{The $AB^2$ boundary term: a minor puzzle resolved}
\label{AB2counter}

While we have found a boundary term $S_\text{bdy}$ defined in \eqref{SbulkSbdy} that ensured supersymmetry, we have not mentioned whether it is unique.  In fact, if it were unique, then the following puzzle could be raised.  The cubic boundary term \reef{S3} decomposes as
\begin{align}
S_3 = -\frac{2\k}{L}\int d^3x\, (A^3 -3AB^2) \,,    
\end{align} 
where we have indicated its form for the case $W = \kappa z^3/(3 L)$ as in \eqref{WChoice}.\footnote{When $W = \kappa z^1z^2z^3 / L$ the RHS of this expression contains the combination $A^1A^2A^3 - A^1B^2B^3-A^2B^3B^1 - A^3B^1B^2$.}  The $A^3$ term will be used to calculate the 3-point correlator of three $\D=1$ scalar operators in the next section while the $AB^2$ term would generate a correlator of one scalar and two $\D=2$ pseudoscalars.  The puzzle arises because both correlators are non-vanishing in the $\cn=1$ models, but $\grSO(8)$ symmetry forces\footnote{The subscripts here indicate scale dimension.} $\<\cO_1(\vx_1)\cO_2(\vx_2)\cO_2(\vx_3)\>$ to vanish in $\cn=8$ supergravity.  This is suspicious because both the $z^3$ and the $z^1z^2z^3$ models are supposed to be consistent $\cn=1$ truncations of $\cn=8$.  

The resolution of this issue is that the finite boundary counterterm 
\begin{align}
S' = c' \int d^3x A_1B_1^2\,,
\end{align} 
can be added to the $z^3$ model with arbitrary constant $c'$ and maintains supersymmetry of the Legendre transform $\tilde S$.  Further, the more general cubic polynomial 
\begin{align}
\hat S= \int d^3x \left[{c_3A_1^3 + c_2 A_1^2B_1+ c_1 A_1B_1^2+c_0 B_1^3}\right]\,,
\end{align} 
 violates  supersymmetry unless $c_0=c_2=c_3 =0$.  This is quite fortunate. One can choose $c'=c_1=-2\k/L$ and cancel the  $\<\cO_1(\vx_1)\cO_2(\vx_2)\cO_2(\vx_3)\>$ correlator which must vanish in a consistent truncation of $\cn=8$,
 while the coefficient of $\<\cO_1(\vx_1)\cO_1(\vx_2)\cO_1(\vx_3)\>$ retains the value which matches the non-perturbative physics of the boundary $\cn =8$ SCFT\@.
 
It is easy to establish the facts mentioned above.  In particular:
 \begin{enumerate}
 \item  The addition of the boundary term $\hat S$ requires that we recompute the extremal point of $\tilde S$.
We find that $\fA$ shifts as $\fA \to \fA + \hat\fA$ with
 \es{GotAHat}{
  \hat\fA = -\left ( 3 c_3A_1^2 +2c_2A_1B_1 + c_1B_1^2 \right) \,.
 }
The boundary term $S_L$ shifts as  $S_L \to S_L + \hat S_L$   with  $\hat S_L=\int d^3x\,\hat\fA \,A_1$.
 \item  These changes are compatible with supersymmetry if
 \es{hatsusy}{
   \d (\hat S +\hat S_L) &= 
   \int d^3x\left[ \d(c_3A_1^3 + c_2 A_1^2B_1+ c_1 A_1B_1^2+c_0 B_1^3) +\d\hat \fA A_1 +\hat \fA \d A_1\right] \\
   &= \int d^3x\left[ -2 A_1 (c_2 B_1 + 3 c_3 A_1) \delta A_1 + (3 B_1^2 c_0 - A_1^2 c_2) \delta B_1 \right] =0 \,.
 }
where we used \eqref{GotAHat}, and where $\delta A_1$ and $\delta B_1$ are understood to be computed from \eqref{InducedSUSY}.   It is then straightforward to determine the integrand of \reef{hatsusy} and observe that it vanishes if and only if $c_0=c_2=c_3 =0$, while $c_1$ is  arbitrary.
 \end{enumerate}

\subsection{Boundary SUSY for $\cn=1$ truncations of supergravity.}
\label{sec:bsusytr}

In this section we discuss, qualitatively, the steps that are needed to show that the boundary terms obtained above in the global limit are not changed by reanalysis at the level of $\cn=1$ supergravity.  In supergravity we must use $\eps(r,\vx)$ parameters with arbitrary dependence on the coordinates of the bulk theory.  The terms in the general $\cn=1$ Lagrangian (as presented in (18.6) of \cite{book}) that must be considered are the chiral multiplet terms
that have obvious limits to the global Lagrangian in  \cite{Festuccia:2011ws}.  These include, respectively,  the $m_{3/2}$ and $m_{\a\b}$ terms in (18.15) and (18.16) of \cite{book}.  We must also include the Noether current term (written for a single multiplet)
\be\lab{lnoether}
\cl_{\rm Noether} = \frac{1}{\sqrt2}\bar\psi_\m\Big[\big(\slashed{\nabla}\bar z\g^\m +\g^\m\nabla W_\text{SG}\big) P_L\chi+ \text{c.c.}\Big]\,,
\ee
and use its gravitino variation $\d\psi_\m= \sqrt2 (\nabla_\mu \eps +\frac{1}{2L}\g_\mu)\eps$.
We can drop terms in the supergravity Lagrangian, 
such as the quartic fermion terms, and in transformation rules, whose contribution to  possible boundary terms vanishes when the AdS/CFT asymptotic conditions of \reef{asympts} are used.\footnote{Boundary conditions on the gravitino are not needed for our purposes. They are discussed in \cite{Corley:1998qg,Volovich:1998tj,Rashkov:1999ji,Amsel:2009rr,Andrianopoli:2014aqa}.}

With the action and transformation rules limited in this manner, one finds that the $\nabla_\mu\eps$ terms from $\delta \cl_{\rm Noether}$ combine with others elsewhere in $\delta S$ to produce the same set of boundary  terms found in \reef{dsf} and \reef{dskinfinal}, but with  general spinor parameters $\eps(r,\vx)$. The assumption that they approach Killing spinors as $r \to \infty$  is then used to study the boundary terms in more detail.

\section{2- and 3-point correlators from ${\cal N} = 1$ supergravity}
\label{sec:3pttoy}

In this section we present a  holographic calculation of 2- and 3-point functions of a $\D= 1$ CFT operator ${\cal O}_1$ in the example \eqref{globLag} from ${\cal N} = 1$ supergravity with the cubic superpotential \eqref{WChoice}.  The computation in this toy model will be generalized to ${\cal N} = 8$ supergravity in Section~\ref{sec:3ptfnct}.  

\subsection{2- and 3-point correlators}

As we explained in the previous section, the operator ${\cal O}_1$ is dual to the bulk scalar $A = \Re z$.  The pseudoscalar field $B$ and fermion $\chi$ play no role in the calculation of correlators of ${\cal O}_1$, so we set them to  zero. The part of the action \eqref{globLag} involving $A$ and the boundary term \eqref{SbdyQuadCubic} that we need is 
\be   \label{actionA}
S =  \frac 12 \int d^4x \sqrt{g}\,\Big[\pa_\m A\pa^\m A - \frac{2}{L^2} A^2 \Big] +\frac{1}{2L}  \int d^3x\, e^{3r_0/L}\,\Big[\,A^2 +\frac{2\k}{3} A^3\Big]\,,
\ee
where we Wick rotated to Euclidean signature and multiplied \eqref{globLag} by an overall factor of $1/2$ for a more conventional normalization.   We set $\kappa=L=1$ in this section.  

The field $A$ obeys the equation of motion
 \es{Aeom}{
  \left( \square + 2\right) A = 0 \,.
 }
As in \eqref{asympts}, the solution of this equation of motion can be expanded at large $r$ as 
 \es{AExpansion}{
  A(r, \vx) = e^{-r} A_1(\vx) + e^{-2r} A_2(\vx) + \cdots \,.
 }
In fact, the equation of motion \eqref{Aeom} implies that the entire bulk field can be reconstructed in terms of $A_1(\vx)$ with the help of the bulk-to-boundary propagator
 \es{bulkToBdry}{
   A(r, \vx) = \int d^3y\, K_2 (r, \vx; \vy) A_1(\vy) \,, \qquad K_2(r, \vx; \vy) \equiv \frac{1}{\pi^2} \frac{e^{-2r}}{\left( e^{-2r} + \abs{\vx-\vy}^2 \right)^2} \,.
 } 
Plugging this expression into \eqref{actionA} one obtains the on-shell action written as a functional of the boundary coefficient $A_1(\vx)$:\footnote{Here and in the following formulas, the integration kernel $\frac{1}{\abs{\vx-\vy}^4}$ is understood to be regularized by replacing it with $\frac{1}{(\epsilon^2 + \abs{\vx-\vy}^2)^2}$, where $\epsilon = e^{-r_0}$ is a holographic UV cutoff, and discarding the power divergences in $\epsilon$.  Discarding such power divergences can be unambiguously done in a CFT\@.} 
 \es{actionAOnShell}{
  S_\text{on-shell}[A_1] = - \frac 12 \int d^3 x \, d^3 y \frac{A_1(\vx) A_1(\vy)}{\pi^2\abs{\vx-\vy}^4} + \frac 13 \int d^3 x \, A_1(\vx)^3 + O(A_1^4) \,.
 }

This expression would be the  goal of our computation if $A$ were dual to a dimension $2$ field theory operator $\co_2$. In that case $A_1(\vx)$ would be interpreted as the source of $\co_2$, and, 
by the AdS/CFT dictionary, $-S_\text{on-shell}[A_1]$ becomes  the generating functional of its connected correlators,  certainly not what we want.   

Indeed, in our case of interest, the field $A$ is dual to a dimension $1$ operator ${\cal O}_1$, but $A_1(\vx)$ is not the field theory source for ${\cal O}_1$.  Instead, the field theory source, denoted $\fA (\vx)$, is the canonically conjugate variable to $A_1$ \cite{Klebanov:1999tb} and the generating functional is the Legendre transform  $\tilde S_\text{on-shell}[\fA]$ defined as in \eqref{LegTransfSUSY} by 
 \es{LegTransfSUSYAgain}{
  \tilde S_\text{on-shell}[\fA] = S_\text{on-shell}[A_1] + \int d^3x \,  \fA(\vx) A_1(\vx)  \,.
 }
These ideas were introduced in  Section~\ref{sec:countN1}, where our main purpose was to demonstrate that the generating functional $\tilde S_\text{on-shell}[B_1,\chi_{3/2\,-},\fA]$ is a supersymmetric functional of its sources.  In this section our purpose is more pragmatic; we wish to express $\tilde S_\text{on-shell}[\fA]$ in a form in which  functional derivatives with respect to $\fA (\vx)$ can be applied to produce correlators of $\co_1$.  

Toward that end, we  proceed to extremize the RHS of \eqref{LegTransfSUSYAgain} with respect to $A_1(\vx)$ after inserting the toy model expression \eqref{actionAOnShell}.  Extremization yields the result
 \es{calAfromA1}{
  \fA (\vx) = -\frac{\delta S_\text{on-shell}[A_1]}{\delta A_1(\vx)} = \frac{1}{\pi^2} \int d^3 y\, \frac{A_1(\vy)}{\abs{\vx-\vy}^4} -   A_1(\vx)^2  + O(A_1^3) \,.
 }
This expression can be inverted by taking its convolution with $1/(2 \pi^2 \abs{\vz-\vx}^2)$ and using the relation
 \es{RelationInverse}{
  \int d^3 x\, \frac{1}{2 \pi^2 \abs{\vz-\vx}^2} \frac{1}{\pi^2 \abs{\vx-\vy}^4} = - \delta^{(3)} (\vz -\vy) \,,
 } 
which can be derived, for instance, by passing to Fourier space.\footnote{A more careful regulated analysis gives $$\int d^3 x\, \frac{1}{2 \pi^2 \abs{\vz-\vx}^2} \frac{1}{\pi^2 (\epsilon^2 + \abs{\vx-\vy}^2)^2} = \frac{1}{\epsilon} \frac{1}{2 \pi^2 \abs{\vz-\vy}^2}- \delta^{(3)} (\vz - \vy) \,.$$    To derive \eqref{RelationInverse} one must discard the linear UV divergence. The regulated expression above is simply a combination of \eqref{K1Def} and \eqref{K1Expansion} below multiplied by $\eps^2=e^{-2r_0}$.}   The expression for $A_1(\vx)$ in terms of $\fA (\vx)$ is finally
 \es{A1FromcalA}{
  A_1(\vx) = -\int d^3 y \, \frac{\fA (\vy)}{2 \pi^2 \abs{\vx-\vy}^2} - \frac{1}{(2 \pi^2)^3} \int d^3 y\, d^3 z\, \fA (\vy) \fA (\vz) I(\vx, \vy, \vz)  + O(\fA ^3) \,,
 }
where
 \es{IDef}{
  I(\vx, \vy, \vz) = \int d^3 w \, \frac{1}{\abs{\vx-\vw}^2 \abs{\vy-\vw}^2 \abs{\vz-\vw}^2} = \frac{\pi^3}{\abs{\vx-\vy} \abs{\vy-\vz} \abs{\vx-\vz}} \,.
 } 
Plugging this into \eqref{LegTransfSUSYAgain} and using \eqref{IDef} again gives 
 \es{Stilde}{
  \tilde S_\text{on-shell}[\fA ] = -\frac 1{4 \pi^2} \int d^3x \, d^3 y\, \frac{\fA (\vx) \fA (\vy)}{ \abs{\vx-\vy}^2}
   - \frac{1}{24 \pi^3} \int d^3 x\, d^3 y\, d^3 z \frac{\fA (\vx) \fA (\vy) \fA (\vz)}{ \abs{\vx-\vy} \abs{\vy-\vz} \abs{\vx-\vz}}  + O(\fA ^4) \,.
 }
The first term in this expression agrees with the result of \cite{Klebanov:1999tb} in a free bulk theory.  The expression \eqref{Stilde} thus generalizes this result to include a cubic boundary interaction.

Since $-\tilde S_\text{on-shell}[\fA ]$ is interpreted as the generating function of connected correlators for the operator ${\cal O}_1$, we obtain the 2- and 3- point functions 
 \es{2pt3ptO}{
  \langle {\cal O}_1(\vx_1) {\cal O}_1(\vx_2) \rangle &= \frac{1}{2 \pi^2 \abs{\vx_1-\vx_2}^2}  \,, \\
  \langle {\cal O}_1(\vx_1) {\cal O}_1(\vx_2) {\cal O}_1(\vx_3) \rangle &= \frac{1}{4 \pi^3} \frac{1}{ \abs{\vx_1-\vx_2} \abs{\vx_2-\vx_3} \abs{\vx_1-\vx_3}} \,.
 }
In Section~\ref{sec:3ptfnct}, a similar computation is used to obtain the 2- and 3-point functions of the dimension $1$ operators of an ${\cal N}  = 8$ SCFT transforming in the ${\bf 35}_v$ representation of $\grSO(8)$ R-symmetry.

\subsection{On the nonlinear boundary condition  for $A(r,\vx)$}

The toy model provides the opportunity to explore the Legendre transform further and hopefully gain further insight into its workings.  Toward that end we express the bulk field $A(r,\vx)$ in terms of boundary data for a source $\fA (\vx)$ of compact support.   We then study its boundary limit in a region where the source vanishes and show explicitly that the boundary condition 
  \be \label{NonLinear}
   \fA(\vx) = A_2(\vx) -A_1(\vx)^2 =0\,, \qquad \vx \in (\text{supp($\fA )$})^c
\ee
is satisfied.

The bulk field $A(r, \vx)$ can be expressed in terms of the boundary data $\fA (\vx)$ by combining \eqref{bulkToBdry} and \eqref{A1FromcalA}.  Performing the required integrals, one can write the resulting expression as
 \es{AFromcalA}{
  A(r, \vx) = \int d^3 y \, K_1(r, \vx; \vy) \left(\fA (\vy) + \frac{1}{(2 \pi^2)^2} \int d^3 z \, d^3 w\, \frac{\fA (\vz) \fA (\vw)}
   {\abs{\vy-\vz}^2 \abs{\vy-\vw}^2} \right) + O(\fA ^3)\,,
 }
where
 \es{K1Def}{
  K_1(r, \vx; \vy) \equiv \int d^3 z\, K_2 (r, \vx; \vz) \frac{-1}{2 \pi^2 \abs{\vz-\vy}^2} = -\frac{1}{2 \pi^2} \frac{e^{-r}}{e^{-2r} + \abs{\vx-\vy}^2} \,.
 }
 To check \eqref{NonLinear}, we should expand $A(r, \vx)$ at large $r$ and assume that $\vx$ lies outside the support of $\fA $.  Since $r$ only appears in $K_1$, we can expand $K_1$ at large $r$ first.  To leading order in $e^{-r}$, we have that $K_1(r, \vx; \vy)$ approaches $-\frac{e^{-r}}{2 \pi^2\abs{\vx-\vy}^2}$.  The first subleading correction can be computed as
  \es{K1Sub}{
   e^{2r} \left( K_1(r, \vx; \vy) + \frac{e^{-r}}{2 \pi^2\abs{\vx-\vy}^2}\right) = \frac{1}{2 \pi^2} \frac{e^{-r}}{\abs{\vx-\vy}^2 \left(e^{-2r} + \abs{\vx-\vy}^2 \right)} \to \delta^{(3)} (\vx-\vy) \,.
  }
So
 \es{K1Expansion}{
  K_1(r, \vx; \vy) \to -\frac{e^{-r}}{2 \pi^2\abs{\vx-\vy}^2} + e^{-2r} \delta^{(3)} (\vx-\vy) \,.
 }
Using this large $r$ expansion in \eqref{AFromcalA} and comparing with \eqref{AExpansion}, we identify 
 \es{GotA1A2}{
  A_1(\vx) &= -\frac{1}{2 \pi^2} \int d^3 y \frac{\fA (\vy)}{ \abs{\vx-\vy}^2} + O(\fA ^2) \,, \\
  A_2(\vx) &= \frac{1}{(2 \pi^2)^2} \int d^3 z \, d^3 w\, \frac{\fA (\vz) \fA (\vw)}
   {\abs{\vx-\vz}^2 \abs{\vx-\vw}^2} + O(\fA ^3)
 }
By examining \eqref{GotA1A2} it is easy to see that, indeed, the non-linear boundary condition \eqref{NonLinear} is obeyed.

\section{The  $\mathbf{ \cals N}\bf =  8$ supergravity}
\label{N8sugra}

We begin with a brief summary of the $\cals N=8$ gauged supergravity in four dimensions  \cite{deWit:1981sst,deWit:1982bul} with the $\grSO(8)$ gauge fields  set to zero. In the bosonic sector one is then left with  the  metric, $g_{\mu\nu}$, and  the scalar/pseudoscalar fields parametrizing the non-compact coset space $\rm E_{7(7)}/SU(8)$.   In the symmetric gauge \cite{Cremmer:1979up,deWit:1982bul},  the scalar 56-bein, $\cals V$, is explicitly given by
\begin{equation}\label{56bein}
\cals V\eeql \left(\begin{matrix}
u_{ij}{}^{IJ} & v_{ijIJ} \\ v^{klIJ} & u^{kl}{}_{KL}
\end{matrix}\right)\eql  \exp \left(\begin{matrix}
0 & -\coeff 1 4 \sqrt 2\,\phi_{ijkl}\\
 -\coeff 1 4 \sqrt 2\,\bphi^{ijkl} & 0
\end{matrix}\right)\in {\rm E_{7(7)}}\,,
\end{equation}
where 
\begin{equation}\label{selfphi}
\phi_{ijkl}\eql {1\over 24}\epsilon_{ijklmnpq}\phi^{mnpq}\,,\qquad \phi^{ijkl}\eql(\phi_{ijkl})^*\,,
\end{equation}
are complex self-dual fields, whose real and imaginary parts,
$\phi_{ijkl}=\alpha^{ijkl}+i\,\beta^{ijkl}$,
 are the $35_v$ scalars, $\alpha^{ijkl}$,  and $35_c$ pseudoscalars, $\beta^{ijkl}$, respectively, where the labels $s$ and $c$ indicate the assignment of $\grSO(8)$ representations.\footnote{See, e.g., Table~7 in \cite{Duff:1986hr}.}  In the fermionic sector, there are $8_s$ left/right-handed gravitini, $\psi_\mu{}^i$/$\psi_{\mu\,i}$, and $56_s$ left/right-handed gauginos $\chi^{ijk}/\chi_{ijk}$. 
 As for  complex scalars, see \eqref{selfphi},   complex conjugation of the fermions amounts simply to raising/lowering of the $\grSU(8)$ indices, for example  $(\chi_{ijk})^*=\chi^{ijk}$,  with the corresponding change of chirality. 
 
The scalar fields enter the action and the supersymmetry transformations through  the composite $\grSU(8)$ connection, $\cals B_\mu{}^i{}_j$, and the self-dual tensor, $\cals A_\mu{}^{ijkl}$, defined by\cite{Cremmer:1979up}:
\begin{equation}\label{Bmufield}
\cals B_\mu{}^i{}_j\eql {2\over 3}(u^{ik}{}_{IJ}\partial_\mu u_{jk}{}^{IJ}-v^{ikIJ}\partial_\mu v_{jkIJ})\,,
\end{equation}
\begin{equation}\label{Amufield}
\cals A_\mu{}^{ijkl}\eql -2\sqrt 2(u^{ij}{}_{IJ}\partial_\mu v^{klIJ}-v^{ijIJ}\partial_\mu u^{kl}{}_{IJ})\,,
\end{equation} 
and the two $A$-tensors \cite{deWit:1982bul}: 
\begin{equation}\label{Atensors}
A_1{}^{ij}\eql{4\over 21} \,T_k{}^{ikj}\,,\qquad A_{2i}{}^{jkl}\eql -{4\over 3}\, T_i{}^{[jkl]}\,,
\end{equation}
defined in terms of the $T$-tensor,
\begin{equation}\label{ttensor}
T_i{}^{jkl}~\equiv~\left(u^{kl}_{IJ}+v^{klIJ}\right)\left(u_{im}{}^{JK}u^{jm}{}_{KI}-v_{imJK}v^{jmKI}\right)\,.
\end{equation}
Note that $A_1{}^{ij}=A_1{}^{ji}$, while $A_{2\,i}{}^{jkl}=A_{2\,i}{}^{[jkl]}$. 

The bosonic action of the $\cals N=8$ supergravity in the gravity plus scalar sector is \cite{deWit:1982bul}\footnote{In this section we set $\kappa^2=1/8\pi G_4=1$.}
\begin{equation}\label{N8bosac}
S_\tB\eql \int d^4x \,\sqrt{-g}\,\Big[\, {1\over 2}\,R -{1\over 96}\,\cals A_\mu{}^{ijkl}\cals A^{\mu}{}_{ijkl}-g^2\,\cals P\Big]\,,
\end{equation}
where
\begin{equation}\label{N8pot}
\cals P  \eql -{3\over 4}\,|A_1{}^{ij}|^2+{1\over 24}\,|A_{2\,l}{}^{ijk}|^2\,,
\end{equation}
is the scalar potential. The maximally supersymmetric solution is given by the  AdS$_4$ metric \eqref{adsmet} of radius 
\begin{equation}\label{lrad}
L\eql {1\over\sqrt 2\,g}\,,
\end{equation}
 and vanishing scalar fields, $\phi_{ijkl}=0$. 
 
 For a general solution, the  asymptotic expansion of the scalar fields is similar to that in \eqref{asympts}, namely
\begin{equation}\label{scasymp}
\begin{split}
\phi^{ijkl}(r,\vx) & \eql e^{-r/L}\phi_{(1)}{}^{ijkl}(\vx)+e^{-2r/L}\phi_{(2)}{}^{ijkl}(\vx)+\ldots\,, \\[6 pt] \phi_{(n)}{}^{ijkl}(\vx) & \eql \alpha_{(n)}{}^{ijkl}(\vx)-i\beta_{(n)}{}^{ijkl}(\vx)\,.
\end{split}
\end{equation}

Using the symmetric gauge \eqref{56bein}  and the definitions \eqref{Bmufield}--\eqref{Atensors}, one can verify by a somewhat tedious calculation  the following expansions of the composite fields \cite{Cremmer:1979up}:
\begin{equation}\label{ABexp1}
\begin{split}
\cals B_\mu{}^i{}_j & \eql -{1\over 24}\Big(\phi^{ipqr}\partial_\mu\phi_{jpqr}-\phi_{jpqr}\partial_\mu\phi^{ipqr}\Big)+O(\phi^4)\,,\\[6 pt]
\cals A_\mu{}^{ijkl} & \eql \partial_\mu\phi^{ijkl}+{1\over 24}\phi_{pqrs}\phi^{pq[ij}\partial_\mu\phi^{kl]rs}
-{1\over 24}\phi^{pq[ij}\phi^{kl]rs}\partial_\mu\phi_{pqrs}+O(\phi^5)\,,
\end{split}
\end{equation}
and of the $A$-tensors
\cite{deWit:1981sst},\footnote{We correct the sign in the first bracket on the right hand side in \eqref{Aexp2}.}
\begin{align}\label{Aexp1}
A_1{}^{ij} & \eql \big(1+{1\over 192}\,|\phi|^2\big)\,\delta^{ij}+{\sqrt 2\over 96}\,\bphi^{ikmn}\phi_{mnpq}\bphi^{pqkj}+O(\phi^4)\,,\\[6 pt] 
\label{Aexp2}
A_{2\,l}{}^{ijk}& \eql -{\sqrt 2\over 2}\big(1+{1\over 144}\,|\phi|^2\big)\,\bphi^{ijkl}-{3\over 8}\,\phi_{mnl[i}\bphi^{jk]mn}+{\sqrt 2\over 16}\,\phi_{lpqr}\bphi^{pqs[i}\bphi^{jk]rs}+O(\phi^4)\,,
\end{align}
where
$
|\phi|^2\eql \phi_{ijkl}\bphi^{ijkl}$. 
In particular, it follows from \eqref{Aexp1} and \eqref{Aexp2} that 
\begin{equation}\label{}
\begin{split}
|A_1{}^{ij}|^2 & \eql 8+{1\over 12}\,|\phi^2|-{\sqrt 2\over 96}\,\big(\bphi^{ijkl}\phi_{klmn}\bphi^{mnij}+\text {c.c.}\big)+O(\phi^4)\,,\\[6 pt]
|A_{2\,l}{}^{ijk}|^2 & \eql {1\over 2}\,|\phi|^2-{3\sqrt 2\over 16}\big(\bphi^{ijkl}\phi_{klmn}\bphi^{mnij}+\text {c.c.}\big)+O(\phi^4)\,.
\end{split}
\end{equation}
Hence the scalar potential \eqref{N8pot},
\begin{equation}\label{}
\begin{split}
\cals P\eql  -6-{1\over 24}\,|\phi|^2+O(\phi^4)\,,
\end{split}
\end{equation}
has no cubic terms in its expansion! This is the source of the puzzle  we resolve in this paper.

In the following we will also need the action for the spin-1/2 fields:
\begin{equation}\label{SN8f12}
\begin{split}
S_\text{$\chi$-bulk}\eql \int d^4 x\sqrt{-g}\,\Big[  -{1\over 12}\big(\bar \chi^{ijk} & \gamma^\mu D_\mu\chi_{ijk}  + \bar \chi_{ijk}\gamma^\mu D_\mu\chi^{ijk} \big) \\[3 pt] &  +{\sqrt 2\over 144}\,g\,\big(\epsilon^{ijkpqrlm}A_2{}^n{}_{pqr}\bar\chi_{ijk}\chi_{lmn}+\text{c.c.}\big)\Big]\,,
\end{split}
\end{equation}
and their Noether coupling to the gravitini:
\begin{equation}\label{}
\begin{split}
S_\text{Noether}\eql \int d^4x\sqrt{-g} \Big[-{1\over 12}\,\cals A_{\mu}{}^{ijkl}\bar\chi_{ijk}\gamma^\nu\gamma^\mu\psi_{\nu l}+{g\over 6} \,A_2{}^i{}_{jkl}\bar\psi_{\mu\,i}\gamma^\mu\chi^{jkl}+\text{c.c.}\Big]\,.
\end{split}
\end{equation}

The supersymmetry variation of the scalar fields is \cite{deWit:1982bul}
\begin{equation}\label{delsV}
(\delta \cals V\cals V^{-1})_{ijkl}\equiv -2\sqrt 2 \,\Sigma_{ijkl}\,,
\end{equation}
where
\begin{equation}\label{Sigma}
\Sigma_{ijkl}\eql \bar\epsilon_{[i}\chi_{jkl]}+{1\over 24}\eta_{ijklmnpq}\bar\epsilon^m\chi^{npq}\,,
\end{equation}
is self-dual. The expansion of \eqref{delsV} yields the result similar to \eqref{ABexp1} \cite{Cremmer:1979up,deWit:1982bul}, namely,
\begin{equation}\label{delphi}
\delta\phi_{ijkl}\eql 8\,\Sigma_{ijkl}(1+O(\phi^2))\,.
\end{equation}
Finally, the  supersymmetry variations of the  left-handed gravitinos and  gauginos in the $\cals N=8$ theory are given by \cite{deWit:1982bul}
\begin{align}
\delta\psi_\mu{}^i & \eql 2 D_\mu\epsilon^i ~+~ \sqrt {2} \, g\,A_1{}^{ij}\gamma_\mu\epsilon_j\,, 
\label{deltagravitino}\\[6 pt]
\delta\chi^{ijk} & \eql -\cals A_\mu{}^{ijkl}\,\gamma^\mu\,\epsilon_l  ~-~  2\, g\, A_{2\,l}{}^{ijk}\epsilon^l\,,
\label{deltagaugino}
\end{align}
with the corresponding complex conjugate variations of the right-handed fields, $\psi_{\mu i}$ and $\chi_{ijk}$.

To conclude this summary we note  that  the bosonic action \eqref{N8bosac} expanded about its maximally supersymmetric solution is 
\begin{equation}\label{bulkexp}
S_{\tB}\eql \int d^4x \,\Big[{1\over 2}\,R-{1\over 96}\,\partial _\mu\phi_{ijkl}\partial^\mu\phi^{ijkl}+g^2\Big(6+{1\over 24}|\phi|^2\Big)+\ldots \Big]\,.
\end{equation}
and has the same structure as the corresponding $\cals N=1$ action in \eqref{ScalarAction}. This suggests that we should find the boundary counterterms with   the same structure as those found in Section~\ref{BOGO}. To determine them we first consider the supersymmetry transformations of the fermions and the corresponding Bogomolny factorization as in Section~\ref{ss:N1bog} and then confirm the result by a direct supergravity calculation. 

\section{Bogomolny argument in $\cals N=8$ supergravity}
\label{Sect:BogN8}
\renewcommand{\coeff}[2]{\frac{#1}{#2}}

\subsection{Motivation}

It is useful to recall the form of the original BPS arguments \cite{Bogomolny:1975de,Prasad:1975kr}.  These are computations in field theories in flat backgrounds and, at least for monopoles, involve completing the square in the Hamiltonian. This completion of the square  requires boundary terms that bound  the energy from below. The bound is saturated precisely when the perfect square in the bulk action vanishes and this condition leads to the BPS equations.   Apart from time-independence, there  were no special assumptions about how the fields depended on coordinates and the original treatment involved flat space and did not incorporate gravitational back-reaction.

In this section, we will make a ``BPS-inspired'' argument by making a similar completion of squares, but there will be several important differences with the standard BPS story.  First, our metric will not be flat but will be that of the ``kink Ansatz,'' \eqref{domwall}, the most general metric that preserves Poincar\'e invariance in the boundary directions.  However, unlike  \eqref{domwall}, we will consider completely general scalar fields.   We use this metric Ansatz because we wish to consider fields in AdS and in asymptotically-AdS backgrounds. 

Exactly as in the BPS story, we will  complete the square in the bulk action and collect the essential boundary terms that are needed to achieve this. Since we are allowing a non-trivial scale factor, ${\cals A(r)}$, in our metric Ansatz, one should anticipate that the energy will not be bounded below.  Indeed, one finds  that the bulk action produces a signed sum of squares. 
Thus, unlike the BPS story, we cannot obtain a lower bound on the energy.  What is important here is that we show that the action with the completed  squares in the bulk has much better fall-off behavior at infinity in an asymptotically-AdS background.  The result is that the boundary terms obtained  from the ``BPS-inspired'' completion of squares are  precisely the boundary terms that one needs to regulate the  action in an asymptotically-AdS background. 
\subsection{The BPS equations in the ``kink Ansatz''}

As in Section~\ref{BOGO}, we start by assuming that the metric has the Poincar\'e-invariant form   \eqref{domwall}  
and  that the scalar fields depend only on the radial coordinate,~$r$.  Setting the spacetime components ($\mu=0,1,2$) of the gravitino variations \eqref{deltagravitino} and the gaugino variations \eqref{deltagaugino} to zero, we  obtain the following 
 equations: 
\begin{equation}\label{N8BPSeqs}
\begin{split}
\cals A' \,\gamma^3 \,\epsilon^i+\sqrt 2 \,g\,A_1{}^{ij}\epsilon_j & \eql 0\,,\\[6 pt]
-\cals A_r{}^{ijkl}\,\gamma^3\,\epsilon_l  ~-~  2\, g\, A_{2\,l}{}^{ijk}\epsilon^l & \eql 0\,,
\end{split}
\end{equation}
which, together with the  the complex conjugate equations, constitute a linear system 
for the Killing spinors, $\epsilon_i$ and $\epsilon^i$. 

Motivated by the known solutions to \eqref{N8BPSeqs} from   RG-flows in various truncations of the $\cals N=8$ theory (see for example \cite{Ahn:2000mf,Corrado:2001nv,Ahn:2001by,Ahn:2001kw,Ahn:2002qga,Pope:2003jp,Bobev:2013yra}), let us set
\begin{equation}\label{Xmatrix}
\gamma^3\epsilon^i\eql \Mmat^{ij}\epsilon_j\,,\qquad \gamma^3\epsilon_i\eql \Mmat_{ij}\epsilon^j\,, \qquad \Mmat^{ij}\eql (\Mmat_{ij})^*\,,
\end{equation}
where $X_{ij}$ is a symmetric matrix, which  by  consistency with  $(\gamma^3)^2=I$  must also be unitary. Then, substituting \eqref{Xmatrix} in \eqref{N8BPSeqs}, we find the following equations:
\begin{equation}\label{BPSoprs}
\begin{split}
\big(\cals A'\,X^{ij}+\sqrt 2\,g\,A_1{}^{ij}\big)\,\epsilon_j\eql 0\,,\qquad 
\big(\cals A_r{}^{ijkl}+2\,g\, X^{lm}A_{2m}{}^{ijk}\big)\,\epsilon_l\eql 0\,,
\end{split}
\end{equation}
where the matrices acting on the Killing spinors, $\epsilon_i$, are the BPS operators we are looking for.

We refer the reader to Appendices~\ref{Appendix:Trunc} and \ref{app:FP} for further discussion of truncations and flows. Here let us note that for   known RG flows the components of the BPS operators that act on the nonvanishing $\epsilon_i$'s reduce to the usual BPS equations for the metric function and the scalar fields, respectively. In particular, in the truncation discussed in \cite{Freedman:2013ryh} (see also Appendix~\ref{app:FP}) they yield the BPS equations  \eqref{floweqs}. 

In the following we will show that the BPS operators defined in \eqref{BPSoprs} provide  natural factors for the  $\cals N=8$ analogue of the Bogomolny argument in Section~\ref{BOGO}. At the same time one should keep in mind that the discussion below is completely general and independent of any solution of \eqref{BPSoprs}. In particular, the factorization in Section~\ref{ss:compsq} holds for scalar fields with arbitrary space-time dependence. All that we use is  that the metric has the form   \eqref{domwall}, the general form of the BPS operators and identities satisfied by  the $A$-tensors in $\cals N=8$,~$d=4$ supergravity. 

\subsection{Completing the square}
\label{ss:compsq}

We now generalize the result: the metric will still be required to be of the form \eqref{domwall}, but the scalar fields will be allowed to have arbitrary dependence on all coordinates.  With these choices, the  bosonic action \eqref{N8bosac}  reduces to the following effective action for the scalars and gravitational field:\footnote{The reduction of \eqref{N8bosac} to \eqref{effact1} introduces boundary terms that arise from the integration by parts of the second order derivatives of the metric inside the Ricci scalar. Those terms are then cancelled by the usual Gibbons-Hawking boundary counterterm. } 
\begin{align}
S_\tB\eql \int d^4 x\, e^{3A} \, \Big[ 3 (\cals A')^2  ~+~   \coeff{3}{4}\,g^2 \,\left|A_1{}^{ij}\right|^2~-~ \coeff{1}{96}\, \cals A_\mu{}^{ijkl}\cals A^\mu{}_{\,ijkl}  ~-~ \coeff{1}{24}\,g^2 \,\left|A_{2i}{}^{jkl}\right|^2 \,\Big] \,.
\label{effact1}
\end{align}

At first sight, it may seem inconsistent to employ such an action because scalars depending on the boundary directions will have an energy momentum tensor that sources metric components that violate the metric Anstaz in \eqref{domwall}.  There are two, essentially equivalent, ways to think about this.  First, we  want to work about a gravitational background that preserves Poincar\'e invariance in the boundary directions and, as in Section \ref{BOGO}, we want to ``consistently suppress'' all gravitational back-reaction that breaks the Poincar\'e invariance.  This can be reduced to a prescription in terms of powers of the gravitational coupling, $\kappa$, but we can simply take the view that we use (\ref{effact1}) and drop all the Einstein equations involving components of the energy-momentum tensor that break Poincar\'e invariance.

The second, and more practical perspective, is that our goal now is to examine the behavior of the action in asymptotically-AdS space and consider the asymptotic behavior of the bulk action as it approaches the boundary.  To that end, we note that (\ref{effact1}) contains precisely the degrees of freedom that remain non-trivial as the metric asymptotes that of AdS at infinity. We we discuss this more in Section \ref{ss:asymp1}. 

The supersymmetry means that this action can be written in terms of squares of the BPS operators introduced above.  Indeed, the 
first two terms in \eqref{effact1} may be written as 
\begin{equation}
\label{compsq1}
\begin{split}
 e^{3A} \, \Big[ 3 (\cals A')^2 ~+~   \coeff{3}{4}\,g^2 \,\left|A_1{}^{ij}\right|^2 \Big] ~=~ & \coeff{3}{8}\,  e^{3A} \Big | \cals A' \, \Mmat_{ij}  ~\mp~   \sqrt{2} \, g \, A_1{}_{ij} \Big|^2 \\[6 pt]
 & \qquad ~\pm~ \coeff{3}{4  \sqrt{2}}\, g \, \cals A' \,  e^{3A} \Big[ \Mmat_{ij} A_1{}^{ij}~+~  \Mmat^{ij} A_1{}_{ij}~ \Big]
\,,
\end{split}
\end{equation}
where $\Mmat_{ij}$ is any unitary matrix.  Similarly, the $A_2$-term and the radial component of the scalar kinetic term in  (\ref{effact1}) can be written as:
\begin{equation}\label{compsq2}
\begin{split}
 e^{3A} \,  \Big[  -\coeff{1}{96}\, \cA_r{}^{ijkl}\cA_r{}_{\,ijkl}   ~-~ & \coeff{g^2 }{24}\left|A_{2i}{}^{jkl}\right|^2 \Big]  \eql  - \coeff{1}{96}\, e^{3A} \, \Big| \cA_r{}^{ijkl} \pm  2\, g \, \Mmat^{i m} \,A_{2m}{}^{jkl} \Big |^2   \\[6 pt]  & \hspace{30 pt}
~\pm~   \coeff{g}{48}\,   e^{3A} \,\Big[ \cA_r{}_{\,ijkl}  \, \Mmat^{i m} \,A_{2m}{}^{jkl} + \cA_r{}^{\,ijkl}  \, \Mmat_{i m} \, {A_{2}}^m{}_{jkl} \Big]  \,.
\end{split}
\end{equation}
The role of the dynamical matrix $\Mmat_{ij}$ here is to preserve the $\grSU(8)$ covariance of the factors inside the squares.  In principle, we could choose $\Mmat_{ij}$ in any convenient manner and one could even choose these matrices to be different in \eqref{compsq1} and \eqref{compsq2}.  

There is, however, a very natural and canonical choice that  is motivated by flows and superpotentials of truncated theories.  We will also see that this choice also leads to a very simple boundary action.   Autonne-Takagi factorization \cite{Horn:1985} allows one to write the symmetric, complex matrix, $A_{1}{}^{ij}$, as 
\begin{equation}
A_{1}{}^{ij} ~=~  (S \, D \, S^T)^{ij} \,,
\label{diag1}
\end{equation}
where $S^i{}_j$ is a unitary matrix and $D^{ij}$ is real and diagonal with non-negative eigenvalues.   Indeed, multiplying this by its complex conjugate gives 
\begin{equation}
\, {A_{1}}^{ik} \, A_{1\, kj} \,   ~=~  (S \, D^2 \, S^\dagger)^i{}_j   \,.
\label{diag2}
\end{equation}
and so the eigenvalues of $D$ are the square-roots of the real eigenvalues of the hermitian matrix ${A_{1}}^{ik} A_{1\, kj}$.
Choose 
\begin{equation}
\Mmat_{ij}   ~=~  (S^* \,  S^\dagger)_{ji} \,,\qquad \Mmat^{ij}\eql (\Mmat_{ij})^*\,.
\label{Mmatdefn}
\end{equation}
Note that $X_{ij}=X_{ji}$ is a symmetric matrix.  Furthermore one has 
\begin{equation}
\Mmat_{ij} A_{1}{}^{ij}    ~=~  {\rm Tr} \big (S^* \,  S^\dagger \, S \, D \, S^T\big) ~=~  \mathop{\rm Tr} D   \,,
\label{MmatA1trace}
\end{equation}
which means that the squared term in \eqref{compsq1} may be written as 
\begin{equation}
\label{compsq1a}
\coeff{3}{8}\,  e^{3A} \Big | \cals A' \, \Mmat^{ij}  ~\mp~   \sqrt{2} \, g \, A_1{}^{ij} \Big|^2  ~=~\coeff{3}{8}\,  e^{3A} \Big | \cals A' \, \delta_{ij}  ~\mp~   \sqrt{2} \, g \, D_{ij} \Big|^2 \,.
\end{equation}
%

\subsection{Collecting the boundary terms}

Observe that using the identity \cite{deWit:1982bul}
\begin{equation}
D_\mu A_1{}^{ij} ~=~ \coeff{1}{12 \sqrt{2}} \,  \big( {A_{2}}^i {}_{klm}   \cA_\mu{}^{jklm} + {A_{2}}^j{}_{klm}   \cA_\mu{}^{iklm} \big)   \,,
\label{derA1}
\end{equation}
the extra terms in (\ref{compsq1}) and (\ref{compsq2}) can be combined to 
\begin{equation}
\pm  \coeff{g}{4\sqrt{2}}\,   \Big[\Mmat^{i j}  D_r (e^{3A} \,A_{1\, ij})  ~+~    \Mmat_{i j} \, D_r (e^{3A} \, A_1{}^{ij}) \Big]  \,.
\label{bdryterm2}
\end{equation}
Using the cyclic properties of the trace, one finds: 
\begin{equation}
\begin{aligned}
\Mmat_{ij} \partial_\mu A_{1}{}^{ij}    ~=~  & \mathop{\rm Tr} \big [S^* \,  S^\dagger \,  \partial_\mu (S \, D \, S^T) \big] ~=~  \mathop{\rm Tr} \big [(S^\dagger \,  \partial_\mu S) D ~+~ D \, (\partial_\mu S^T) \, S^* ~+~  \partial_\mu D  \big]  \\[6 pt]
 ~=~   & {\rm Tr} \big [ \partial_\mu D  ~+~( (S^\dagger \,  \partial_\mu S) + (S^\dagger \,  \partial_\mu S)^T) \,  D \big]  \,.
\end{aligned}
\label{MmatdA1trace}
\end{equation}
Note that $(S^\dagger \,  \partial_\mu S) + (S^\dagger \,  \partial_\mu S)^T$ is symmetric and in the Lie algebra of $\grSU(8)$.  It is therefore purely imaginary and so cancels when added to the complex conjugate:
\begin{equation}
\Mmat^{ij} \partial_\mu A_{1\, ij}   + \Mmat_{ij} \partial_\mu A_{1}^{ij}   ~=~ 2 \,  \partial_\mu \mathop{\rm Tr} D    ~=~ 2 \,  \partial_\mu \mathop{\rm Tr}  \sqrt{\smash[b]{ A_{1}A_{1}^\dagger}}    \,.
\label{ReMatdA1trace}
\end{equation}
Finally, there are the connection terms in the covariant derivative: 
\begin{equation}
 D_\mu  A_{1}{}^{ij}    ~=~  \partial_\mu   A_{1}{}^{ij} ~-~ \coeff{1}{2} \, \cB_\mu{}^{i}{}_{k} A_{1}{}^{kj} ~-~  \coeff{1}{2} \,  \cB_\mu{}^{j}{}_{k} A_{1}{}^{ik}   ~=~  \partial_\mu    A_{1}{}^{ij}  ~-~ \cB_\mu{}^{(i}{}_{k}  A_{1}{}^{j)k}  \,,
\label{covDA1}
\end{equation}
where we have used the symmetry of $A_{1}{}^{ij}$.  These connection terms yield a contribution:
\begin{equation}
- \Mmat_{ij}  \cB_\mu{}^{i}{}_{k} A_{1}{}^{kj}    ~=~  - {\rm Tr} \big [S^* \,  S^\dagger \, \cB_\mu\, S \, D \, S^T]  
 ~=~    - {\rm Tr} \big [ \cB_\mu\, (S \, D \, S^\dagger)]  \,.
\label{connterms}
\end{equation}
However, $(S \, D \, S^\dagger)$ is hermitian while $ \cB_\mu$ is anti-hermitian and so this trace is purely imaginary and therefore also cancels out when one adds the complex conjugate.  This means that with our choice of $\Mmat^{ij}$, the $\grSU(8)$ connection terms make no contribution to (\ref{bdryterm2}) and so the complete boundary term may be written as 
\begin{equation}
\pm  \coeff{g}{2 \sqrt{2}}\,   \partial_r   {\rm Tr}  \left[ e^{3A} \, D  \right] ~=~ \pm  \coeff{g}{2 \sqrt{2}}\,   \partial_r   {\rm Tr} \Big [e^{3A}  \, \sqrt{\smash[b]{ A_{1}A_{1}^\dagger}} \,  \Big]  \,.
\label{bdryterm3}
\end{equation}
Putting this all together, we see that the effective action (\ref{effact1}) can be written as
\begin{equation}\label{effact2}
\begin{split}
S_\tB\eql & \int d^3 x \, dr \, e^{3A} \, \Big[\, \coeff{3}{8}\,   \big | \cals A' \, \Mmat_{ij} ~-~   \coeff{1}{L} \, A_1{}_{ij} \big|^2 ~-~ \coeff{1}{96}\, \big| \cA_r{}^{ijkl} ~+~  \coeff{ \sqrt{2}}{L} \, \, \Mmat^{i p} \,A_{2p}{}^{jkl} \big |^2   \\[6 pt]
& \qquad\qquad\qquad~-~ \coeff{1}{96}\,  g^{ab} \,\cA_a {}^{ijkl}   \cA_b {}_{ijkl} \Big]  ~+~\coeff{1}{4 \, L}\, \int  d^3 x  \, e^{3A}\, \mathop{\rm Tr}  
\sqrt{\smash[b]{ A_{1}A_{1}^\dagger}} \,  \Big|_{r = r_{0}}   \,,
\end{split}
\end{equation}
where we have explicitly restored the spacetime components, $\cA_a$ and $\cA_b$ of $\cA_\mu$, $a,b=0,1,2$. 

In Appendix~\ref{Appendix:Trunc} we perform similar computations for consistent truncations of the $\Neql8$ theory to reduced levels of supersymmetry.  It particular we obtain  the analogous form of (\ref{effact2}) for such truncations.

%
\subsection{Asymptotics and counterterms}
\label{ss:asymp1}

We will now argue that for  the solutions of interest, namely with 
\begin{equation}\label{Aphiasymp}
\cals A(r) ~=~ \frac{r}{L}+O(e^{-2r/L}) \,,
\end{equation}
and \eqref{scasymp} for the scalars, 
the two squared terms in the first line in \eqref{effact2} obtained by choosing the upper signs in \eqref{compsq1} and \eqref{compsq2} vanish at the boundary and that the last term in the bulk integral vanishes as well. This makes the last term in  \eqref{effact2}  a natural candidate for the counterterm.  We will further confirm  that in Section~\ref{N8sugrsusy} by showing that this boundary  counterterm is  consistent with the local supersymmetry of the Legendre transformation of the renormalized on-shell action.

Observe that  \eqref{Aexp1} implies that the matrix, $(A_1{}_{ij})$, is diagonal to quadratic order in the fields. Using the asymptotic expansion \eqref{scasymp}, we thus have 
\begin{equation}\label{eigest}
A_{1\,ij}\eql \delta_{ij}+O(e^{-2r/L})\,,\qquad (A^\dagger_1A_1)_{ij}\eql \delta_{ij}+O(e^{-2r/L})\,,\qquad D_{ij}\eql \delta_{ij}+O(e^{-2r/L})\,.
\end{equation}
Together with \eqref{Aphiasymp} and \eqref{compsq1a}, where we choose the upper sign, this implies the estimate
\begin{equation}\label{asympf1}
\cals A' \, \delta_{ij} -   \sqrt{2} \, g \, D_{ij} ~\sim~ O(e^{-2r/L})\,.
\end{equation}
Similarly, from \eqref{eigest} and the definition \eqref{diag1}, we find the asymptotic expansion
\begin{equation}\label{}
S^i{}_j\eql S_{0}{}^i{}_j+O(e^{-3r/L})\,,
\end{equation}
where $S_0$ is a (complex) orthogonal matrix, $S_0^TS_0=1$. Then, cf.\ \eqref{Mmatdefn}, 
\begin{equation}\label{}
\Mmat^{ij}\eql \delta^{ij}+O(e^{-3r/L})\,.
\end{equation}
Once more choosing the upper sign in \eqref{compsq2} and using \eqref{ABexp1} and \eqref{Aexp2} we find:\footnote{Note the order of indices in (\ref{ABexp1}).}
\begin{equation}\label{asympf2}
\cA_r{}^{ijkl} +  2\, g \, \Mmat^{i m} \,A_{2m}{}^{jkl}  ~\sim~ O(e^{-2r/L})\,.
\end{equation}

Given the asymptotic expansions \eqref{asympf1} and \eqref{asympf2},  we see that 
\begin{equation}
\label{termasymp1}
\big | \cals A' \, \Mmat_{ij} ~-~   \coeff{1}{L} \, A_1{}_{ij} \big|^2   \,, \qquad   \big| \cA_r{}^{ijkl} ~+~  \coeff{ \sqrt{2}}{L} \, \, \Mmat^{i m} \,A_{2m}{}^{jkl}\big|^2   ~\sim~ \cO(e^{-4r/L}) \,.
\end{equation}
This shows that  the terms in the square bracket in \eqref{effact2} vanish at the boundary.

The  metric in the boundary directions is $g^{ab} = e^{-2r/L}\delta^{ab} $ and, from (\ref{scasymp}) and (\ref{ABexp1}), one has $\cA_a{}^{ijkl} \sim \cO(e^{-r/L})$ and so the third bulk term in (\ref{effact2}), including the factor of $e^{3 A}$, vanishes at infinity.   Thus even though we allowed scalar fields to depend on the boundary directions, the exponential fall-off of the components of the metric in the boundary directions means that such scalar fluctuations consistently decouples in our effective action near the boundary.

Thus we are led to the boundary scalar counterterm action: 
\begin{equation}\label{bctr1}
\begin{split}
S_\text{s-ct} & \eql -{1\over 4L}\int d^3x\,e^{3r_0/L}\,\mathop{\rm Tr} \sqrt{\smash[b]{A_1A_1^\dagger}}\\[6 pt]
&\eql \int d^3x\,e^{3r_0/L}\,\Big[ -{2\over L}-{1\over 96\,L}\,\phi_{ijkl}\bphi^{ijkl}\\
& \hspace{1.655 in}+{1\over 384\sqrt 2\,L}\,\big(\phi_{ijkl}\phi_{ijmn}\bphi^{klmn}+\text{c.c.}\big) +\ldots\Big]\,.
\end{split}
\end{equation}
The combined bulk (\ref{effact1}) and boundary  (\ref{bctr1}) action can be rewitten as:
\begin{equation}
\label{bogact}
\begin{split}
S_\text{B}+S_\text{s-ct} =
 \int d^3 x \, dr \, e^{3A} \, \Big[ \,\coeff{3}{8}\,   \big | \cals A' \, \Mmat_{ij} -   \coeff{1}{L} \, A_1{}_{ij} \big|^2 & -\coeff{1}{96}\, \big| \cA_r{}^{ijkl} +  \coeff{ \sqrt{2}}{L} \, \, \Mmat^{i m} \,A_{2m}{}^{jkl} \big |^2  \\
&- \coeff{1}{96}\,  g^{ab} \,\cA_a {}^{ijkl}   \cA_b {}_{ijkl}   \, \Big]\,.
 \end{split}
\end{equation}
This has a {\it vanishing} contribution in the asymptotic region.  Put differently, the original bulk action, $S_\tB$, has divergent and finite pieces at infinity but adding the boundary action, $S_\text{s-ct}$, precisely cancels these boundary terms.

Finally, we note that the cubic counterterm in \eqref{bctr1} depends only on the scalar fields,~$\alpha^{ijkl}$. Indeed, it is straightforward to check that for self-dual scalars, $\alpha^{ijkl}$, and   anti-self-dual pseudoscalars, $\beta^{ijkl}$, 
\begin{equation}\label{}
\alpha^{mn[ij}\alpha^{kl]mn}\qquad \text{and}\qquad \beta^{mn[ij}\beta^{kl]mn}\,,
\end{equation}
are also, respectively,  self-dual and anti-selfdual, see Appendix~\ref{app:so8tens}. Thus expanding the cubic counterterm we find
\begin{equation}\label{cubicct1}
\begin{split}
{1\over 384\sqrt 2\,L}(\phi_{ijkl}\phi^{klmn}\phi_{mnij}+\text{c.c.} )& \eql {\sqrt 2\over 384 \,L}\,\alpha^{ijkl}\alpha^{klmn}\alpha^{mnij}\,.
\end{split}
\end{equation}
This is of course in agreement with the branching rules for the $\rm SO(8)$ tensor products \cite{Slansky:1981yr}:
\begin{equation}\label{tenprod}
{\bf 35}_i\otimes {\bf 35}_i~\longrightarrow ~{\bf 1}+{\bf 35}_i+\ldots\,,\qquad {\bf 35}_i\otimes {\bf 35}_j~\longrightarrow ~{\bf 35}_k+\ldots\,,
\end{equation}
and the assignment of ${\bf 35}_v$ and ${\bf 35}_c$ to the scalars and the pseudoscalars, respectively. Hence the absence of a cubic coupling between the scalars and the pseudoscalars is a consequence of the $\grSO(8)$ symmetry.

We should finish this section by emphasizing that while the Bogomolny type argument uses the standard completion of the square that can be used to derive the BPS equations, the latter are not relevant to our focus here.   The Bogomolny type argument simply leads to a bulk action with stronger (vanishing) convergence properties at infinity and so can be used to derive the boundary counterterms needed to achieve this outcome.

\section{Boundary sources and $\cn=8$ supersymmetry}
\label{N8sugrsusy}

\subsection{Preliminaries}

In this section,  using methods similar to  Section~\ref{sec:countStates}, we show that boundary terms in the supersymmetry variation of the Legendre transformed on-shell action of $\cals N=8$ supergravity  are cancelled by the variation  of the boundary counterterms, 
\begin{equation}\label{}
S_\text{bdy}\eql S_\text{s-ct}+S_\text{$\chi$-ct}\,,
\end{equation}
where $S_\text{s-ct}$ is  the scalar counterterm   \eqref{bctr1} and $S_\text{$\chi$-ct}$ is the gaugino counterterm
 \begin{equation}\label{spinch}
S_\text{$\chi$-ct}\eql 
\int d^3x\,e^{-3r_0}\,\Big[ {1\over 24}\,\bar\chi^{ijk}\chi^{ijk}+\text{c.c.}\Big]\,.
\end{equation}
This fermionic counterterm  may, at first, seem surprising in that it breaks the $\grSU(8)$ symmetry of $\Neql8$ supergravity down to $\grSO(8)$.  Such a symmetry breaking is expected because  scalars and pseudoscalars in supergravity are quantized differently. 

As in Section~\ref{sec:countStates}, we consider only those variations that involve the scalar and spin-1/2 fields and work in the fixed  AdS$_4$  metric background with the corresponding Killing spinors,
\begin{equation}\label{Killspc}
\epsilon^i(r, \vec x) \eql e^{r/2L}\zeta_+{}^i(\vec x)+e^{-r/2L}\zeta_-{}^i\,,\qquad \epsilon_i(r, \vec x)\eql e^{r/2L}\zeta_+{}_i(\vec x)+e^{-r/2L}\zeta_-{}_i\,,
\end{equation}
\begin{equation}\label{radial8}
\gamma^3\zeta_\pm{}^i\eql\mp\zeta_{\pm\,i}\,, \qquad
\bar\zeta_\pm{}^i\gamma^3\eql\pm\bar\zeta_{\pm\,i}\,,
\end{equation}
\begin{equation}
 \label{slashzet}
\slashed\partial\zeta_+{}^i\eql -{3\over L}\,\zeta_{-}{}_i  \,,\qquad \slashed\partial\zeta_+{}_i\eql -{3\over L}\,\zeta_{-}{}^i\,,
\end{equation}
obtained by solving the BPS equations \eqref{deltagravitino},  $\delta\psi_\mu{}^i=\delta\psi_{\mu\,i}=0$,     with  vanishing scalar fields. However, unlike in Section~\ref{BOGO}, we will use the left- and right-handed spinors rather than the underlying Majorana spinors. This explains why the radiality conditions \eqref{radial8} look different from those in \eqref{etarad}.  The two are of course equivalent.

To take advantage of the radiality constraints \eqref{radial8} of the Killing spinors, it is convenient to introduce analogous projections of the spin-1/2 fields. To this end we define
\begin{equation}\label{}
\begin{split}
\Xi_{ijk} &\eql {1\over 2}\left(\chi_{ijk}-\gamma^3\chi^{ijk}\right)\,,\qquad \Upsilon_{ijk}\eql {1\over 2}\left(\chi_{ijk}+\gamma^3\chi^{ijk}\right)\,,\\[6 pt]
\Xi^{ijk} &\eql {1\over 2}\left(\chi^{ijk}-\gamma^3\chi_{ijk}\right)\,,\qquad \Upsilon^{ijk}\eql {1\over 2}\left(\chi^{ijk}+\gamma^3\chi_{ijk}\right)\,,
\end{split}
\end{equation}
where the level of indices indicates the $\gamma^5$-chirality.%
\footnote{In terms of the underlying Majorana spinors, $\chi_M^{ijk}$, the new fields are given by 
\begin{equation*}
\Xi^{ijk}\eql {1\over 4}(1+\gamma^5)(1-\gamma^3)\chi_M^{ijk}\,,\qquad \Upsilon^{ijk}\eql {1\over 4}(1+\gamma^5)(1+\gamma^3)\chi_M^{ijk}\,,\qquad \text{etc.}\,,
\end{equation*}
and hence are chiral projections of the fields with negative/positive radiality, respectively.
}
 Then
\begin{equation}\label{radXi}
\gamma^3\,\Xi^{ijk}\eql -\Xi_{ijk}\,,\qquad \gamma^3\,\Upsilon^{ijk}\eql \Upsilon_{ijk}\,.
\end{equation}

The asymptotic expansions of these fields are given by 
\begin{equation}\label{fermodes}
\begin{split}
\Xi^{ijk} & \eql e^{-3r/2L}\,\Xi_{(3/2)}{}^{ijk}+e^{-5r/2L}\,\Xi_{(5/2)}{}^{ijk}+\ldots\,,\\[6 pt]
\Upsilon^{ijk} & \eql e^{-3r/2L}\,\Upsilon_{(3/2)}{}^{ijk}+e^{-5r/2L}\,\Upsilon_{(5/2)}{}^{ijk}+\ldots\,,
\end{split}
\end{equation}
and similarly for the complex conjugate fields. In terms of the leading asymptotic  coefficients, the  supersymmetry variations \eqref{delphi} of the scalar fields become
\begin{equation}\label{varss}
\begin{split}
\delta\alpha_{(1)}{}^{ijkl} & \eql 8\,\bar\zeta_+{}^{[i}\Upsilon_{(3/2)}{}^{jkl]}+{1\over 3}\eta^{ijklmnpq}\bar\zeta_+{}^{m}\Upsilon_{(3/2)}{}^{npq}\,,\\[6 pt]
\delta\beta_{(1)}{}^{ijkl} & \eql -8i\,\bar\zeta_+{}^{[i}\Xi_{(3/2)}{}^{jkl]}+{i\over 3}\eta^{ijklmnpq}\bar\zeta_+{}^{m}\Xi_{(3/2)}{}^{npq}\,,
\end{split}
\end{equation}
\begin{equation}\label{varsss}
\begin{split}
\delta\alpha_{(2)}{}^{ijkl} & \eql
 8\,\left(\bar\zeta_-{}^{[i}\Xi_{(3/2)}{}^{jkl]}+\bar\zeta_+{}^{[i}\Upsilon_{(5/2)}{}^{jkl]}\right)\\
& \hspace{60 pt}+{1\over 3}\eta^{ijklmnpq}\left(\bar\zeta_-{}^{m}\Xi_{(3/2)}{}^{npq}+\bar\zeta_+{}^{m}\Upsilon_{(5/2)}{}^{npq}\right)\,,\\[6 pt]
 \delta\beta_{(2)}{}^{ijkl} & \eql  -8i\,\left(\bar\zeta_-{}^{[i}\Upsilon_{(3/2)}{}^{jkl]}+\zeta_+{}^{[i}\Xi_{(5/2)}{}^{jkl]}\right)\\
& \hspace{60 pt}+{i\over 3}\eta^{ijklmnpq}\left(\bar\zeta_-{}^{m}\Upsilon_{(3/2)}{}^{npq}+\bar\zeta_+{}^{m}\Xi_{(5/2)}{}^{npq}\right)\,,
\end{split}
\end{equation}
while the  supersymmetry variations \eqref{deltagaugino} for the leading modes of the gauginos are 
\begin{equation}\label{varXi}
\begin{split}
\delta \Xi_{(3/2)}{}^{ijk}  \eql -{2i\over L}\,\beta_{(1)}{}^{ijkl}\,\zeta_-{}^l
-{1\over L} { }\,\Big[ &\alpha_{(2)}{}^{ijkl}+{3\over 4\sqrt 2}\,\alpha_{(1)}{}^{mn [i j}\alpha_{(1)}{}^{k]lmn}\\[6 pt]
&+{3\over 4\sqrt 2}\,\beta_{(1)}{}^{mn [i j}\beta_{(1)}{}^{k]lmn}
-iL\gamma^3\slashed\partial \beta_{(1)}{}^{ijkl}\Big]\,\zeta_+{}^l\,,
\end{split}
\end{equation}
and
\begin{equation}\label{varUps}
\begin{split}
\delta \Upsilon_{(3/2)}{}^{ijk} \eql  {2\over L}\,\alpha_{(1)}{}^{ijkl}\,\zeta_-{}^{l}
-{i\over L}\,
\Big[&-\beta_{(2)}{}^{ijkl}+{3\over 4\sqrt 2}\,\alpha_{(1)}{}^{mn [i j}\beta_{(1)}{}^{k]lmn}\\[6 pt]
&-{3\over 4\sqrt 2}\,\beta_{(1)}{}^{mn [i j}\alpha_{(1)}{}^{k]lmn}
-iL\gamma^3\slashed\partial \alpha_{(1)}{}^{ijkl}\Big]\,\zeta_+{}^{l}\,.\end{split}
\end{equation}

The structure of the supersymmetry  variations \eqref{varss}--\eqref{varUps}, modulo the SO(8) indices, is exactly the same as in \eqref{InducedSUSY}. In particular, we can set the  sources:
\begin{equation}\label{N8zerosr}
\beta_{(1)}{}^{ijkl}(\vec x)\eql 0\,, \qquad \Xi_{(3/2)}{}^{ijk}(\vec x)\eql 0\,,  
\end{equation}
and
\begin{equation}\label{LegA}
\fA^{ijkl}(\vec x)~\equiv~-{1\over L}\,\Big[\alpha_{(2)}{}^{ijkl}(\vec x)+{3\over 4\sqrt 2}\,\alpha_{(1)}{}^{mn [i j}(\vec x)\alpha_{(1)}{}^{k]lmn} (\vec x)\Big]\,,
\end{equation}
to zero (cf.\ \eqref{VanishingSources}) consistent with supersymmetry. It follows from \eqref{cid5} that $\fA^{ijkl}$ is totally antisymmetric and self-dual.    The same calculation as in Section~\ref{sec:countN1} shows that 
\begin{equation}\label{}
\fA^{ijkl}\eql -\lim_{r\to\infty}e^{-r/L}\,\Pi^{ijkl}\,,
\end{equation}
where $\Pi^{ijkl}$ is the conjugate momentum 
\begin{equation}\label{}
\Pi^{ijkl}\eql -e^{3r/L}\left[\partial_r\alpha^{ijkl}+{1\over L}\alpha^{ijkl}-{3\over 4\sqrt 2L}\alpha^{mn[ij}\alpha^{kl]mn}\right]\,,
\end{equation}
obtained by varying the bulk plus boundary bosonic action \eqref {bogact}. Performing the Legendre transform on the scalars amounts then to the addition of  
\begin{equation}\label{theSLL}
S_L\eql  {1\over 48}\,\int d^3x\,\,\fA^{ijkl}(\vec x)\alpha_{(1)}{}^{ijkl}(\vec x)
\end{equation}
to the  action and then extremizing with respect to  $\alpha_{(1)}{}^{ijkl}$.

The equations of motion for the spin-1/2 fields that follow from the bulk action  \eqref{SN8f12} are:
\begin{equation}\label{N8feoms}
\begin{split}
\gamma^\mu D_\mu\chi_{ijk}-{1\over 12L}\,\eta_{ijkpqrlm}\,A_2{}_{n}{}^{pqr}\,\chi^{lmn}  \eql 0\,.
\end{split}\end{equation}
In the AdS$_4$ background, 
\begin{equation}\label{}
\gamma^\mu D_\mu\eql e^{-r/L}\slashed \partial+\gamma^3{\partial\over \partial r}+{3\over 2L}\,\gamma^3\,,
\end{equation}
where $\slashed{\partial}$ is the 3d Dirac operator along the boundary, and the  asymptotic expansion of  \eqref{N8feoms} and its complex conjugate equation yield
\begin{equation}\label{fereqs}
\begin{split}
\Xi_{(5/2)}{}^{ijk} &= -L\slashed\partial \Upsilon_{(3/2)}{}_{ijk}+{1\over 12\sqrt 2}\,\eta_{ijkpqrlm}\,\left(\alpha_{(1)}{}^{npqr}\,\Xi_{(3/2)}\,^{lmn}-i\,\beta_{(1)}{}^{npqr}\,\Upsilon_{(3/2)}\,^{lmn}\right)\,,\\
\Upsilon_{(5/2)}{}^{ijk} & = L\slashed\partial \Xi_{(3/2)}{}_{ijk}-{1\over 12\sqrt 2}\,\eta_{ijkpqrlm}\,\left(\alpha_{(1)}{}^{npqr}\,\Upsilon_{(3/2)}\,^{lmn}-i\,\beta_{(1)}{}^{npqr}\,\Xi_{(3/2)}\,^{lmn}\right)\,,
\end{split}
\end{equation}
with arbitrary $\Xi_{(3/2)}{}^{ijk}$ and $\Upsilon_{(3/2)}{}^{ijk}$. This 
is the $\cals N=8$ analogue of \eqref{chi52}.

\subsection{The boundary variation from the bulk $\cals N=8$ action}
\label{bryvarN8}

We will now demonstrate explicitly the invariance of the Legendre transformed on-shell renormalized action
\begin{equation}\label{Ltract}
\widetilde S\eql S_\text{bulk}+S_\text{bdy}+S_L\,,
\end{equation}
under the $\cals N=8$ superconformal symmetry \eqref{Killspc}. Since the calculation turns out somewhat lengthy, it is helpful to split it into several steps as to make various cancellations more transparent.

We start with the bulk action of the $\cals N=8$ supergravity, which is known to be invariant---in the bulk---under local supersymmetry variations \cite{deWit:1982bul}. However, the proof involves integration by parts which   on a space-time with a boundary gives rise to    nontrivial  boundary terms. A convenient way to identify them is to  group individual  terms in the variation of the bulk supergravity  Lagrangian, $\cals L_\text{bulk}$, according to whether they contain derivatives of the supersymmetry parameters or not. Schematically, we may write this as
\begin{equation}\label{}
\delta\cals L_\text{bulk}\eql \bar V_i\epsilon^i+\bar X^{\mu}{}_iD_\mu\epsilon^i+\text{c.c.}\,.
\end{equation}
The  invariance of the action in the bulk means that after integration by parts,
\begin{equation}\label{}
(\bar V_i-D_\mu \bar X^\mu{}_{i})\epsilon^i+\text{c.c.}\eql 0\,.
\end{equation}
In the AdS$_4$ background that we are considering, the  remaining boundary terms are thus given by 
\begin{equation}\label{bryX}
\delta S_\text{bulk}\eql \int d^3x \,e^{3r_0/L}\,\big[\,\bar X^3{}_i\epsilon^i+\text{c.c.}\,\big]\,.
\end{equation}
In practice, this means that in order to extract the boundary terms of interest, we must look only at those terms in the bulk action that upon the supersymmetry variation give rise to  derivatives of the supersymmetry parameters. There are three sources of such terms: the kinetic terms with derivatives of the varied fields, the Noether coupling of the gravitino to the supersymmetry current that is independent of $g$, and the additional Noether coupling  due to gauging. 

Since we only are interested   in the variations that contain the scalar fields and the gauginos,  we must  consider  only the following terms in the bulk action:
\begin{equation}\label{acts}
\begin{split}
 S_\text{bulk}\eql  \int d^4x\,&\sqrt{-g}\Big[-{1\over 96}\cals A_\mu{}^{ijkl}\cals A^\mu{}_{ijkl} 
-{1\over 12}\Big(\bar \chi^{ijk}\gamma^\mu D_\mu\chi_{ijk}+\bar\chi_{ijk}\gamma^\mu D_\mu\bar\chi^{ijk}\Big)\\[6 pt]
& 
-{1\over 12}\Big(\cals A_{\mu}{}^{ijkl}\bar\chi_{ijk}\gamma^\nu\gamma^\mu\psi_{\nu l}+\text{c.c.}\Big)+{g\over 6} \Big(A_2{}^i{}_{jkl}\bar\psi_{\mu\,i}\gamma^\mu\chi^{jkl}+\text{c.c.}\Big)\Big]\,.
\end{split}
\end{equation}
Then, using the supersymmetry variations \eqref{deltagravitino}, \eqref{deltagaugino} and \eqref{delsV}, we find essentially by inspection that the boundary terms in the variation of the bulk action are given by\begin{equation}\label{bdryvar}
\begin{split}
\bar\epsilon_j X^{\mu\,j} &\eql  -{1\over 6}\cals A^{\mu\,ijkl}\bar\epsilon_i\chi_{jkl}
+ {1\over 12}\cals  \cals A_\nu{}^{ijkl}\bar\epsilon_l\gamma^\nu\gamma^\mu\chi_{ijk}-{g\over 6}A_2{}^l{}_{ijk}\bar\epsilon_l\gamma^\mu\chi^{ijk} 
\\[6 pt]
&\qquad +
{1\over 6}\cals A_\nu{}^{ijkl}\bar\epsilon_i\gamma^\nu\gamma^\mu\chi_{jkl} +{g\over 3} A_2{}^i{}_{jkl}\bar\epsilon_i\gamma^\mu\chi^{jkl}\,,
\end{split}
\end{equation}
plus its complex conjugate, $\bar\epsilon^i X^\mu{}_i$. We have kept here the same  order of terms as in \eqref{acts} to indicate the origin of each term.  Note that the variation of the gaugino kinetic action (the second and third terms   in \eqref{bdryvar}) is proportional to the variation of the Noether coupling in the last two terms in  \eqref{bdryvar}.  Hence\footnote{To properly understand the signs, note the positions of the contracted indices $ijkl$  in the second a fourth terms in \eqref{bdryvar}: these two terms differ by a factor of $-2$ just like the third and fifth terms.}
\begin{equation}\label{simpvar}
\begin{split}
\bar\epsilon_j X^{\mu\,j} +\text{c.c.} & \eql   -{1\over 6}\cals A^{\mu\,ijkl}\bar\epsilon_i\chi_{jkl}
+{1\over 12}\cals A_\nu{}^{ijkl}\bar\epsilon_i\gamma^\nu\gamma^\mu\chi_{jkl}
+{g\over 6}A_2{}^i{}_{jkl}\bar\epsilon_i\gamma^\mu\chi^{jkl}+\text{c.c.}\,, \\[6 pt]
& \eql -{1\over 6}\cals A^{\mu\,ijkl}\bar\epsilon_i\chi_{jkl}-{1\over 12}\delta\bar\chi_{jkl}\gamma^\mu\chi^{jkl}+\text{c.c.}\,,
\end{split}
\end{equation}
where in the second line we used \eqref{deltagaugino}.

\subsection{Adding counterterms}
\label{sec:verinv}

The first term in the second line of \eqref{simpvar}
 is the only boundary contribution to \eqref{bryX} from the variation of the bulk bosonic action \eqref{N8bosac}, that is
 \begin{equation}\label{deltaSB}
\delta S_\tB\eql \int d^3x\, e^{3r_0/L}\,\Big[ -{1\over 6}\cals A^{\mu\,ijkl}\bar\epsilon_i\chi_{jkl}+\text{c.c.}\Big]\,.
\end{equation}
We will first show that modulo source terms on the boundary, 
 both the infinite and finite terms in \eqref{deltaSB} are cancelled by the variation of the bosonic counterterms in \eqref{bctr1}:
\begin{equation}\label{deltaSbtrm}
\begin{split}
\delta S_\text{s-ct} \eql \int d^3x e^{3r_0/L}\Big[-{c_2\over 48L}\,\big(\alpha^{ijkl}\delta\alpha^{ijkl} & +\beta^{ijkl}\delta\beta^{ijkl}\big)\\ & +{c_3\over 128\sqrt 2L}\,\alpha^{ijkl}\alpha^{klmn}\delta\alpha^{ijmn}
\Big]\,.
\end{split}
\end{equation}
To keep track of the origin of various terms, 
we have introduced here constants $c_2$ and $c_3$, to be set $c_2=c_3=1$ afterwards. 

 The expansion of \eqref{deltaSB} and \eqref{deltaSbtrm} yields the following result:%
\footnote{Note that we consider here only a subset of terms from the full supersymmetry variation of the bosonic action and hence there is no contradiction with the result of the asymptotic analysis in Section~\ref{ss:asymp1}, namely that the renormalized bosonic action vanishes at the boundary.     }
\begin{equation}\label{}
\begin{split}
\eqref{deltaSB}+\eqref{deltaSbtrm} = & {1\over 3L}\int d^3 x\,e^{r_0/L}\,   {(1-c_2) }\Big[   \alpha_{(1)}{}^{ijkl}\,\bar\zeta_+{}^i\Upsilon_{(3/2)}{}^{jkl}+i \beta_{(1)}{}^{ijkl}\,\bar\zeta_+{}^i\Xi_{(3/2)}{}^{jkl}\Big]\\[6 pt]
 & + {1\over 3L}\int d^3 x\,{(1-c_2) }\Big[  \alpha_{(1)}{}^{ijkl}\,\bar\zeta_+{}^i   \Upsilon_{5/2}{}^{jkl}+i \beta_{(1)}{}^{ijkl}\,\bar\zeta_+{}^i\Xi_{5/2}{}^{jkl}\\[6 pt]
 & \hspace{120 pt}+\alpha_{(1)}{}^{ijkl}\bar\zeta_-{}^i\Xi_{(3/2)}{}^{jkl}+i \beta_{(1)}{}^{ijkl}\bar\zeta_-{}^i\Upsilon_{(3/2)}{}^{jkl}\Big] \\[6 pt]
 & + {1\over 3L}\int d^3x\,\Big[\Big((2-c_2)\,\alpha_{(2)}{}^{ijkl}+{3c_3\over 4 \sqrt 2}\,\alpha_{(1)}^{ijmn}\alpha_{(1)}{}^{klmn}\Big)\,\zeta_+{}^i\Upsilon_{(3/2)}{}^{jkl}\\[6 pt]
& \hspace{180 pt} +(2-c_2) i \beta_{(2)}{}^{ijkl}\,\zeta_+{}^i\Xi_{(3/2)}{}^{jkl}\Big]\,.
\end{split}
\end{equation}
This shows that the quadratic counterterm removes the divergence at the boundary as well as it cancels a number of finite terms given by the second integral. After using \eqref{varss}, \eqref{LegA} and \eqref{theSLL}, the  remaining terms are
\begin{equation}\label{totdelbos}
\delta S_\tB+\delta S_\text{s-ct} \eql \int d^3x \Big[-{1\over 48} \fA^{ijkl}\delta\alpha_{(1)}{}^{ijkl}+ {i\over 3L} \beta_{(2)}{}^{ijkl}\,\zeta_+{}^i\Xi_{(3/2)}{}^{jkl}\Big]\,,
\end{equation}
and they indeed vanish in the absence of sources, cf.\ \eqref{N8zerosr}  and \eqref{LegA}.

Next consider the second boundary term in the variation of the bulk action \eqref{simpvar} and combine it with the variation of the gaugino counterterm \eqref{spinch} multiplied by an overall constant, $c_\chi$. In terms of the modes \eqref{fermodes}, we then have
\begin{equation}\label{varfer}
\begin{split}
& \int d^3x\,  e^{3r_0/L}\,\Big[ -{1\over 12}\,\delta\bar\chi^{ijk}\gamma^3\chi^{ijk} +{c_\chi\over 12}\,\delta\bar\chi^{ijk}\chi^{ijk}+\text{c.c.}\Big]\\[6 pt]
 & \eql \int d^3x  \Big[-{1\over 12}(1-c_\chi)\delta\bar\Xi_{(3/2)}{}^{ijk}\Upsilon_{(3/2)}{}^{ijk}+{1\over 12}(1+c_\chi)\delta\bar\Upsilon_{(3/2)}{}^{ijk}\Xi_{(3/2)}{}^{ijk}+\text{c.c.}\Big]\,.
\end{split}
\end{equation}
Examining the variations \eqref{varXi} and \eqref{varUps}, it is clear that we can cancel the second term in \eqref{totdelbos} only by setting $c_\chi=1$. Substituting \eqref{varUps} in \eqref{varfer} and then  using \eqref{cid7} and the radialities \eqref{radial8} and \eqref{radXi}, we find
\begin{equation}\label{varUPS}
\begin{split}
{1\over 6}\,\delta\bar\Upsilon_{(3/2)}{}^{ijk}\Xi_{(3/2)}{}^{ijk}+\text{c.c.} \eql &-{2\over 3L}\,\alpha_{(1)}{}^{ijkl}\,\bar\zeta_-{}^i\Xi_{(3/2)}{}^{jkl}-{1\over 3}\zeta_+{}_i\slashed \partial \alpha_{(1)}{}^{ijkl}\,\Xi_{(3/2)}{}^{jkl}
\\[6 pt]
&-\Big({i\over 2\sqrt 2}\,\alpha_{(1)}{}^{ijmn}\beta_{(1)}{}^{klmn}+{i\over 3L} \,\beta_{(2)}{}^{ijkl}\Big)\,\zeta_+{}^i\Xi_{(3/2)}{}^{jkl}\,.
\end{split}
\end{equation}
 
The variation of the last term in \eqref{Ltract} is
\begin{equation}\label{varSLL}
\delta S_L\eql \int d^3x \,\Big[{1\over 48}\,\fA^{ijkl}\delta \alpha_{(1)}{}^{ijkl}+{1\over 48}\,\alpha_{(1)}{}^{ijkl}\,\delta\fA^{ijkl}\,\Big]\,,
\end{equation}
where the first term cancels against the first term in  \eqref{totdelbos}. To complete the proof of invariance, we must show that the second term in \eqref{varSLL} combines with the first three terms in  \eqref{varUPS} into a total derivative {\it along the boundary}.

 From \eqref{LegA}, \eqref {varss}, \eqref{varsss}  and using the identities \eqref{cid5}, \eqref{cid7}, \eqref{epsalal} and \eqref {epsalbe}
in Appendix~\ref{app:so8tens}, we have
\begin{equation}\label{varSLLL}
\begin{split}
{1\over 48}\,\alpha_{(1)}{}^{ijkl}\,\delta\fA^{ijkl}   \eql & - {1\over 3L}\,\alpha_{(1)}{}^{ijkl}
\left(\bar\zeta_-{}^{i}\Xi_{(3/2)}{}^{jkl}+\bar\zeta_+{}^{i}\Upsilon_{(5/2)}{}^{jkl}\right)\\[6 pt]
& - {1\over 2\sqrt 2L}\,\alpha_{(1)}{}^{ijmn}\alpha_{(1)}{}^{klmn}\,\bar\zeta_-{}^{i}\Upsilon_{(3/2)}{}^{jkl}\\[6 pt]
\eql &  - {1\over 3L}\,\alpha_{(1)}{}^{ijkl}\,\bar\zeta_-{}^{i}\Xi_{(3/2)}{}^{jkl}+{i\over 2\sqrt 2}\,\alpha_{(1)}{}^{ijmn}\beta_{(1)}{}^{klmn}\,\zeta_+{}^i\Xi_{(3/2)}{}^{jkl}\\[6 pt]
& -{1\over 3}\,\alpha_{(1)}{}^{ijkl}\,\bar\zeta_+{}^{i}\slashed \partial \Xi_{(3/2)}{}_{jkl}\,.
\end{split}
\end{equation}
where in the second step we have also used the fermion equations of motion  \eqref{fereqs} to eliminte $\Upsilon_{(5/2)}{}^{jkl}$. Adding the variations in \eqref{totdelbos}, \eqref{varUPS} and \eqref{varSLLL} we are left with
\begin{align}\nonumber
\delta \widetilde S = &  \int d^3x \,\Big[-{1\over L}\,\alpha_{(1)}{}^{ijkl}\,\bar\zeta_-{}^i\Xi_{(3/2)}{}^{jkl}-{1\over 3}\zeta_+{}_i\slashed \partial \alpha_{(1)}{}^{ijkl}\,\Xi_{(3/2)}{}^{jkl}-{1\over 3}\,\alpha_{(1)}{}^{ijkl}\,\bar\zeta_+{}^{i}\slashed \partial \Xi_{(3/2)}{}_{jkl}
\Big]\\[6 pt]
\eql & \int d^3x \Big [ -{1\over 3}\,{\partial\over\partial {x^a}}(\alpha_{(1)}{}^{ijkl}\zeta_+{}_i\gamma^a\Xi_{(3/2)}{}^{jkl})\Big] \,,
\end{align}
which vanishes. This concludes the proof of invariance.

\section{2- and 3-point correlators  from $\cals N=8$ supergravity}
\label{sec:3ptfnct}

\subsection{The counterterms in the $\rm SL(8,\mathbb{R})$ basis}

To calculate the three point functions of the operators, $\cals O_{IJ}$, we  first transform the scalar fields from  the $\grSU(8)$ to the $\grSL(8,\mathbb{R})$ basis \cite{Cremmer:1979up}. This replaces the antisymmetric self-dual tensor, $\alpha^{ijkl}$, by the symmetric traceless tensor, $A^{IJ}$,
\begin{equation}\label{altoA}
\alpha^{ijkl}\eql {1\over 4} (\Gamma_{IK})^{ij}(\Gamma_{JK})^{kl}A^{IJ}\,,\qquad
A^{IJ}\eql {1\over 96}\,  (\Gamma_{IK})^{ij}(\Gamma_{JK})^{kl}\,\alpha^{ijkl}\,,
\end{equation}
and the anti-self-dual tensor, $\beta^{ijkl}$,  by the self-dual tensor, $B^{IJKL}$, 
\begin{equation}\label{betoB}
\beta^{ijkl}\eql {1\over 16}\, (\Gamma_{IJ})^{ij}(\Gamma_{KL})^{kl}\,B^{IJKL}\,,\qquad B^{IJKL}\eql {1\over 16}\,(\Gamma_{IJ})^{ij}(\Gamma_{KL})^{kl}\,\beta^{ijkl}\,,
\end{equation}
where $I,J,\ldots$ indices lie in ${\bf 8}_v$ and $\Gamma_{IJ}$ are chiral $\grSO(8)$  generators. In terms of the new fields, the bulk action \eqref{bulkexp} and the boundary counterterms \eqref{bctr1} read:
\begin{equation}\label{SBA}
\begin{split}
S_\tB \eql \int d^4x\,\sqrt{-g}\,\Big[\,{1\over 2}\,R-{1\over 4} \,\partial_\mu A^{IJ} \partial^\mu A^{IJ} & -{1\over 96}\,\partial_\mu B^{IJKL}\partial^\mu B^{IJKL}\\[6 pt] &+{1\over 2L^2}\Big(6+A^{IJ}A^{IJ}+{1\over 24}B^{IJKL}B^{IJKL}\Big)\Big]\,,
\end{split}
\end{equation}
\begin{equation}\label{SbA}
S_\text{s-ct}\eql \int d^3x\,e^{3r_0/L}\Big[-{2\over L}-{1\over 4L}\,A^{IJ}A^{IJ}+{1\over 6\sqrt 2\,L}\,A^{IJ}A^{JK}A^{KI}\Big]\,.
\end{equation}
The source field  \eqref {LegA} becomes
\begin{equation}\label{Aconst}
\fA^{IJ}\eql-{1\over L}\Big[ A_{(2)}{}^{IJ} +{1\over \sqrt 2}\, \Big(A_{(1)}{}^{IK}A_{(1)}{}^{JK}-{1\over 8}\,\delta^{IJ}A_{(1)}{}^{MN}A_{(1)}{}^{MN}\Big)\,\Big]\,,
\end{equation}
as can be verified by  calculating the momentum $\Pi^{IJ}$ from the action \eqref{SBA}--\eqref{SbA}.  

\subsection{The correlators of the operators $\co_{IJ}(\vx)$}
\label{sec:corrO}

Finally we calculate the 2- and 3-point functions of the 35 operators $\co_{IJ}(\vx)$ with scale dimension $\D=1$.  This computation parallels the one in Section~\ref{sec:3pttoy}, the only difference being that here we need to carefully keep track of the $\grSO(8)$ vector indices.  Just as in Section~\ref{sec:3pttoy}, let us set $L=1$.  We will reinstate $L$ by dimensional analysis at the end.

The starting point is the action \eqref{SBA}--\eqref{SbA} with the pseudoscalars set to zero.  In Euclidean signature, it reads
 \es{SAIJ}{
  S &= \frac{1}{\kappa^2} \int d^4 x \, \sqrt{g} \left[ \frac 14 \partial_\mu A^{IJ} \partial^\mu A^{IJ} - \frac 12 A^{IJ} A^{IJ} \right] \\[6 pt]
   &{}\qquad + \frac{1}{\kappa^2} \int d^3 x\, e^{3 r_0} \left[\frac 14 A^{IJ} A^{IJ}  - \frac{1}{6 \sqrt{2}} A^{IJ} A^{JK} A^{KI} \right] + O(A^4)\,,
 }
where we restored the factor of $1/\kappa^2$ that accounts for a proper normalization  of the Einstein-Hilbert term in \eqref{SBA}.
 As in \eqref{AExpansion}, we expand $A^{IJ}$ as 
 \es{AIJExpansion}{
  A^{IJ}(r, \vx) = e^{-r} A^{IJ}_{(1)}(\vx) + e^{-2r} A^{IJ}_{(2)}(\vx) + \cdots \,,
 }
and we can write the on-shell action as a simple generalization of \eqref{actionAOnShell}:
 \es{SOnShellN8}{
  S_\text{on-shell}[A_{(1)}^{IJ}] &= 
   -\frac 1{4\kappa^2} \int d^3 x\, d^3 y\, \frac{A_{(1)}^{IJ}(\vx) A_{(1)}^{IJ}(\vy)}{\pi^2 \abs{\vx-\vy}^4} \\[6 pt]
    &{}\qquad - \frac{1}{6 \sqrt{2} \kappa^2} \int d^3 x\, A^{IJ}_{(1)} (\vx) A_{(1)}^{JK}(\vx) A_{(1)}^{KI}(\vx) + O(A_{(1)}^4)\,. 
 }

To obtain the generating functional of connected correlators of $\cO_{IJ}(\vx)$, we should pass to the Legendre transform of \eqref{SOnShellN8}: 
 \es{LegTransfN8}{
  \tilde S_\text{on-shell}[\fA^{IJ}] = S_\text{on-shell}[A_{(1)}^{IJ}] 
   + \frac 1{2\kappa^2} \int d^3 x\, \fA^{IJ}(\vx) A_{(1)}^{IJ}(\vx) \,,
 } 
computed after extremizing the right-hand side with respect to $A_{(1)}^{IJ}(\vx)$.  By analogy with \eqref{calAfromA1}, this extremization gives
 \es{calAFromA1N8}{
  \fA^{IJ}(\vx) = \frac{1}{\pi^2 } \int d^3 y\, \frac{A_{(1)}^{IJ}(\vy)}{\abs{\vx-\vy}^4} - 
   \frac{1}{\sqrt{2}}  \biggl[ A^{JK}_{(1)} (\vx) A_{(1)}^{KI}(\vx) - \frac{1}{8} \delta^{IJ} A^{KL}_{(1)} (\vx) A_{(1)}^{KL}(\vx)\biggr]  + O(A_{(1)}^3) \,.
 }
Repeating the steps that led to \eqref{Stilde}, we obtain 
 \es{StildeN8}{
  \tilde S_\text{on-shell}[\fA^{IJ}] &= -\frac{1}{8 \pi^2 \kappa^2} \int d^3x \, d^3 y\, 
   \frac{\fA^{IJ}(\vx) \fA^{IJ}(\vy)}{\abs{\vx-\vy}^2} \\
    &+ \frac{1}{6 \sqrt{2} L \kappa^2}
     \int d^3x \, d^3 y\, d^3 z \frac{\fA^{IJ}(\vx) \fA^{JK}(\vy) \fA^{KI}(\vz)}
      {8 \pi^3 \abs{\vx-\vy} \abs{\vy-\vz} \abs{\vx-\vz}} + O(\fA^4) \,.
 }
where we restored the appropriate factors of $L$.

To identify $-\tilde S_\text{on-shell}[\fA^{IJ}]$ with the generating functional of connected correlators of ${\cal O}_{IJ}(\vx)$, we should account for the fact that $\fA^{IJ}(\vec{x})$ may not be precisely the field theory source for ${\cal O}_{IJ}(\vx)$, but it might differ from it by a constant,
\be \lab{Source}
  \text{Source for ${\cal O}_{IJ}(\vec{x})$ }= \frac{{\cal C}}{L} \fA^{IJ}(\vec{x}) \,,
\ee
with ${\cal C}$ being a dimensionless constant, and a factor of $1/L$ being required by dimensional analysis.  Adjusting for the proportionality constant in \eqref{Source}, we have from \eqref{StildeN8} that 
 \es{23N8}{
   \<\co_{IJ}(\vec x_1)\co_{IJ}(\vec x_2)\> &=\frac{L^2}{16\pi^3 G_4 {\cal C}^2}\frac{1}{|\vec x_1-\vec x_2|^2} \,, \\
   \langle {\cal O}_{IJ}(\vec{x}_1)  {\cal O}_{JK}(\vec{x}_2)  {\cal O}_{KI}(\vec{x}_3) \rangle  &= -\frac{L^2}{64 \sqrt{2} \pi^4 G_4 {\cal C}^3}\frac{1}{\abs{\vec{x}_1 - \vec{x}_2} \abs{\vec{x}_1 - \vec{x}_2} \abs{\vec{x}_2 - \vec{x}_3} }  \,,
 }
with no sum over $I$, $J$, and $K$. In the index free ``$M$ notation,'' these expressions become 
 \es{23N8IndexFree}{
   \<\co(\vec x_1,M_1)\co(\vec x_2,M_2)\> &=\frac{L^2}{8\pi^3 G_4 {\cal C}^2}\frac{\tr(M_1M_2)}{|\vec x_1-\vec x_2|^2}\,, \\
  \langle {\cal O}(\vec{x}_1,M_1)  {\cal O}(\vec{x}_2,M_2)  {\cal O}(\vec{x}_3,M_3) \rangle 
    &= -\frac{L^2}{16 \sqrt{2} \pi^4 G_4 {\cal C}^3}\frac{\tr (M_1 M_2 M_3 + M_1 M_3 M_2)}{\abs{\vec{x}_1 - \vec{x}_2} \abs{\vec{x}_1 - \vec{x}_2} \abs{\vec{x}_2 - \vec{x}_3} } \,.
 }

The relations \eqref{23N8IndexFree} are in complete agreement with the field theory results of Section~\ref{sec:FTcals}!  Indeed, these relations imply
 \es{c23FromHol}{
  c_2 = \frac{L^2}{8\pi^3 G_4 {\cal C}^2} \,, \qquad c_3 =- \frac{L^2}{16 \sqrt{2} \pi^4 G_4 {\cal C}^3} \,,
 }
where $c_2$ and $c_3$ are as in \eqref{Two}.  It is straightforward to see that the ratio $c_3^2 / c_2^3$, which is independent of the normalization constant ${\cal C}$ agrees with the result \eqref{Gotc3} provided that we use $c_T = 32 L^2 / (\pi G_4)$ as in \eqref{cTLargeN}.  Moreover, we see that if we work with operators ${\cal O}_{IJ}$ that are canonically normalized in the sense explained in Section~\ref{sec:FTcals}, for which the 2- and 3-point functions are given in \eqref{CorrFinal}, we have 
 \es{GotcalC}{
  {\cal C} = -\frac{1}{\sqrt{2}} \,.
 }
Up to an overall sign, this normalization constant could have also been inferred from \cite{Freedman:2013ryh}.

\vfill\eject

\section{Conclusions}
\label{CONCLUSIONS}

The goal of this paper was to resolve a puzzle concerning the 3-point functions of \hbox{dimension-1} scalar operators in 3d supersymmetric CFTs with gravity duals. In the case of the \hbox{${\cal N} = 8$} ABJM theory at Chern-Simons level $k=1, 2$, one can calculate this 3-point function exactly using the method of supersymmetric localization. It does not vanish.  When $k=1$, the gravity dual  of ${\cal N} = 8$ ABJM is 11d supergravity on $\text{AdS}_4 \times S^7$.  The 4d maximally supersymmetric  gauged $\grSO(8)$ supergravity theory captures the dynamics of the gravity multiplet in which the  superconformal primaries are a  ${\bf 35}_v$ of  scalar fields $A^{IJ}$ dual to the field theory operators $\co_{IJ}$ of dimension 1.  However the bulk action contains no cubic couplings of the $A^{IJ}$, so the traditional calculation of holographic 3-point functions is not applicable.  

The resolution is that the supergravity theory requires cubic boundary terms that provide precisely the right interactions  to reproduce the boundary 3-point functions.  Our main result \eqref{c23FromHol} obtained from holography agrees precisely with the field theory expectation \eqref{Gotc3}.  The boundary terms were first motivated by a Bogomolny argument for BPS domain walls. They were then  derived more rigorously by requiring that the total derivatives usually neglected in supersymmetry variations of an action are cancelled by boundary counter terms that include the necessary cubic.

Bulk fields dual to dimension-1 scalar operators in a 3d CFT enjoy alternate quantization as prescribed in \cite{Klebanov:1999tb}.
The generating functional for their correlators is the Legendre transform of the renormalized on-shell
action that includes the new cubic boundary term. The supersymmetry properties of the renormalized on-shell action and its Legendre transform are as follows:
\begin{enumerate}
 \item When sources are absent the on-shell action is invariant and the effect  of the cubic term is to produce  nonlinear boundary conditions on the bulk fields. Naive boundary conditions would break supersymmetry.
 \item   Sources are needed to calculate correlation functions. The sources and their supersymmetry transformations are determined from the near-boundary asymptotics  of the bulk fields. When sources are included,  only the Legendre transform is invariant.
\end{enumerate}

Independent of supersymmetry, the Legendre transform plays a crucial role in the calculation of the 3-point function $ \langle {\cal O}_{IJ}(\vec{x}_1)  {\cal O}_{JK}(\vec{x}_2)  {\cal O}_{KI}(\vec{x}_3) \rangle$.  This is developed in an $\cn=1$ toy model in Section 4 and extended to $\cn=8$ supergravity in Section 8. The argument is both intricate and elegant, and gives considerable insight into the working of the Legendre transform.

In the general framework of field theories with boundaries, the condition for a boundary to preserve a conserved charge of the bulk theory is very simple: In the absence of boundary sources, there must be no net flux of the conserved charge across the boundary.  In particular for supersymmetric theories, if there are no boundary sources, then flux of the supercurrent across the boundary should be zero.  The supersymmetric Noether currents of the $\Neql8$  (global) supersymmetries are:
\begin{equation}\label{}
\cals J_\mu{}^{i} ~\equiv~ {1\over 6}\cals A_\nu{}^{ijkl}  \gamma^\nu\gamma_\mu\chi_{jkl} +{g\over 3} A_2{}^i{}_{jkl} \gamma_\mu\chi^{jkl} \,.
\end{equation}
and so supersymmetric boundary conditions should imply no leakage of supercharge at infinity:
\begin{equation}\label{schgflux}
\int d^3 x \, e^{3 A} \, \left(\,\bar \epsilon_i \, \cals J^{i}_r +\text{c.c.}\right)\quad \longrightarrow\quad  0 \,, \qquad r \to \infty \,.
\end{equation}
It is relatively straightforward to establish that this indeed is a consequence of the vanishing of (\ref{N8zerosr}) and  (\ref{LegA}) and similarly, for the $\Neql 1$ theory, with the boundary conditions (\ref{VanishingSources}).

\section*{Acknowledgments}

DZF thanks Massimo Bianchi and Emery Sokatchev for useful suggestions and discussions.
KP and NPW would also like to thank Nikolay Bobev, Sergei Gukov and Vyacheslav Lysov for discussions that led to KP's and  NPW's interest in the issues considered in this paper. The interest of DZF, SSP and NPW was stimulated by discussions at a workshop organized by the Michigan Center for Theoretical Physics. DZF and KP  thank the Galileo Galilei Institute for Theoretical Physics for the hospitality and the INFN for partial support during the completion of this work. SSP thanks Simone Giombi, Zohar Komargodski, Andy Stergiou, and Ran Yacoby for useful discussions.  DZF was supported in part by the US NSF Grant No.~PHY 1620045.  
The work of KP and  NPW was supported in part by DOE grant DE-SC0011687.   SSP was supported by the US NSF under Grant No.~PHY-1418069.

\appendix

\section{An alternative computation of 3-point functions of dimension-$1$ operators}
\label{ALTERNATIVE}

Here we present an alternative method of computing 3-point functions of dimension-$1$ scalar operators that can be used in SCFTs with extended supersymmetry.  As mentioned in Section~\ref{sec:FTcals}, in ${\cal N} = 2$ SCFTs, 2-point functions of dimension-$1$ scalars in flavor current multiplets can be computed via supersymmetric localization by taking two derivatives of the $S^3$ free energy.  Indeed, given a flat space ${\cal N} = 2$ SCFT with R-symmetry current $j^\mu_R$ and Abelian flavor symmetries generated by $j^\mu_{(\alpha)}$, one can construct \cite{Jafferis:2010un} a unique supersymmetric theory on $S^3$ that is invariant under $\grSU(2|1)_\ell \times \grSU(2)_r$, whose bosonic part consists of the $\grSU(2)_\ell \times \grSU(2)_r$ isometry group of $S^3$ as well as a $U(1)$ symmetry generated by 
 \es{jRMod}{
   j^\mu_R + \sum_{\alpha} t_\alpha j^\mu_{(\alpha)} \,,
}
where $t_\alpha$ are parameters.  Using the technique of supersymmetric localization, one can moreover compute the $S^3$ free energy $F(t_\alpha)$ of this theory \cite{Jafferis:2010un}.  One then has \cite{Closset:2012vg}
 \es{JJTwo}{
  \langle J_{(\alpha)}(\vec{x}_1) J_{(\beta)}(\vec{x}_2) \rangle 
   &= -\frac{2}{\pi^2} \left( \frac{\partial^2 F}{\partial t^\alpha \partial t^\beta} \bigg|_{t_\alpha = 0} \right) \frac{1}{(4 \pi)^2 \abs{\vec{x}_1 - \vec{x}_2}^2} \,,
 }
where $J_{(\alpha)}$ are the dimension-1 scalars in the conserved current multiplets, normalized as in \eqref{TwoPointJ}. 

As we now argue, the 3-point function of $J_{(\alpha)}$ can also be computed from $F(t_\alpha)$ via
 \es{JJJAlternative}{
  \langle J_{(\alpha)}(\vec{x}_1) J_{(\beta)}(\vec{x}_2) J_{(\gamma)}(\vec{x}_3) \rangle 
   &= \frac{1}{\pi^2} \left( \frac{\partial^3 F}{\partial t^\alpha \partial t^\beta \partial t^\gamma} \bigg|_{t_\alpha = 0} \right) \frac{1}{(4 \pi)^3 \abs{\vec{x}_1 - \vec{x}_2}
    \abs{\vec{x}_1 - \vec{x}_3} \abs{\vec{x}_2 - \vec{x}_3}}   \,,
 } 
but only in SCFTs that have at least ${\cal N} = 4$  supersymmetry,  and where at least two of these ${\cal N} = 2$ flavor current multiplets descend from half-BPS multiplets of the extended supersymmetry.   Indeed,  $ \frac{\partial^3 F}{\partial t^\alpha \partial t^\beta \partial t^\gamma} \big|_{t_\alpha = 0} $ is proportional to the 3-point function $\langle J_{(\alpha)}(\vec{x}_1) J_{(\beta)}(\vec{x}_2) J_{(\gamma)}(\vec{x}_3) \rangle$ whenever all 3-point functions of the operators multiplying $t_\alpha$ in the $S^3$ action of \cite{Jafferis:2010un} are proportional to $\langle J_{(\alpha)}(\vec{x}_1) J_{(\beta)}(\vec{x}_2) J_{(\gamma)}(\vec{x}_3) \rangle$. This is true when at least two of the ${\cal N} = 2$ flavor current multiplets descend from half-BPS multiplets of the extended supersymmetry, because in this case there is only one superspace invariant that gives the 3-point function of the extended supersymmetry multiplets.   A free theory computation then gives the proportionality constant in \eqref{JJJAlternative}.

We now show how \eqref{JJTwo} and \eqref{JJJAlternative} work in  $\grU(N)_k \times \grU(N)_{-k}$ ABJM theory \cite{Aharony:2008ug}, first when $N=1$ where the theory is free, and afterwards in the large-$N$ limit where the theory has a holographic dual.  We will be primarily interested in taking $k=1$ or $2$ where supersymmetry is enhanced to ${\cal N} = 8$.  Recall that in ${\cal N} =2$ notation, ABJM theory has 2 vector multiplets with Chern-Simons levels $(k, -k)$, two bi-fundamental chiral multiplets ${\cal Z}_a$, $a=1, 2$ transforming in $({\bf N}, \overline{\bf N})$ of $\grU(N) \times \grU(N)$, and two bi-fundamental chiral multiplets ${\cal W}_a$, $a=1, 2$ transforming in the conjugate representation of the gauge group.  Due to the extended supersymmetry, the R-charges of these chiral multiplets take the free field \hbox{value $1/2$}.

There are 3 Abelian flavor symmetries with conserved currents $j^\mu_{(\alpha)}$, $\alpha = 1, 2, 3$, corresponding to the flavor charges of $({\cal Z}_1, {\cal Z}_2, {\cal W}_1, {\cal W}_2)$ being\footnote{This normalization of the $\grU(1)^3$ charges was chosen such that it agrees with the normalization in Section~\ref{N8THREE}.} $(\frac 12, \frac 12, -\frac 12, -\frac 12)$, $(\frac 12, -\frac 12, \frac 12, -\frac 12)$, and $(\frac 12, -\frac 12, -\frac 12, \frac 12)$.  Correspondingly, there is a 3-parameter family of R-charge assignments 
 \es{RChargeAssignments}{
  r_{{\cal Z}_1} &= \frac 12 \left( 1+ t_1 + t_2 + t_3 \right) \,, \\
  r_{{\cal Z}_2} &= \frac 12 \left( 1+ t_1 - t_2 - t_3 \right) \,, \\
  r_{{\cal W}_1} &= \frac 12 \left( 1- t_1 + t_2 - t_3 \right) \,, \\
  r_{{\cal W}_2} &= \frac 12 \left( 1- t_1 - t_2 + t_3 \right) \,,
 } 
that can be used to couple the theory to $S^3$ and compute the 2- and 3-point functions of the canonically normalized operators $J_{(\alpha)}$ in the same multiplet as $j^\mu_{(\alpha)}$ using \eqref{JJTwo} and \eqref{JJJAlternative}.

For $N=1$, it is straightforward to apply the formulas in \cite{Jafferis:2010un} to obtain
 \es{FFree}{
  F_\text{free} = -\ell\left(1 - r_{{\cal Z}_1}   \right) -\ell\left(1 - r_{{\cal Z}_2}   \right)-\ell\left(1 - r_{{\cal W}_1}   \right)-\ell\left(1 - r_{{\cal W}_2}   \right) \,,
 }
where $\ell(z)$ is a function defined in \cite{Jafferis:2010un} obeying $\ell'(z) = -\pi z \cot (\pi z)$ and $\ell(0) = 0$.  An expansion at small $t_\alpha$ gives
 \es{FFreeExpansion}{
  F_\text{free} =  2 \log 2  - \frac{\pi^2}{4} \left( t_1^2 + t_2^2 + t_3^2 \right) + \pi^2 t_1 t_2 t_3 + O(t^4) \,.
 }
From \eqref{JJTwo} and \eqref{JJJAlternative} we obtain
 \es{JCorrFree}{
  \langle J_{(\alpha)}(\vec{x}_1) J_{(\beta)}(\vec{x}_2) \rangle_\text{free} 
   &=  \frac{\delta_{\alpha\beta}}{(4 \pi)^2 \abs{\vec{x}_1 - \vec{x}_2}^2} \,, \\
   \langle J_{(1)}(\vec{x}_1) J_{(2)}(\vec{x}_2) J_{(3)}(\vec{x}_3) \rangle_\text{free} 
   &= \frac{1}{(4 \pi)^3 \abs{\vec{x}_1 - \vec{x}_2}
    \abs{\vec{x}_1 - \vec{x}_3} \abs{\vec{x}_2 - \vec{x}_3}}   \,.
 }
This result agrees with \eqref{TwoSummary} when using the matrices in \eqref{MiDef} and $c_T = 16$.

At large $N$, it was shown in \cite{Jafferis:2011zi} that 
 \es{FLargeN}{
  F = \frac{4 \sqrt{2} \pi N^{3/2}}{3} \sqrt{r_{{\cal Z}_1}  r_{{\cal Z}_2}  r_{{\cal W}_1}  r_{{\cal W}_2} } + O(N^{1/2}) \,.
 }
Expanding at small $t_\alpha$, we have
 \es{FLargeNExpansion}{
  F = \frac{4 \sqrt{2} \pi N^{3/2}}{3} \left[ \frac 14 - \frac 14 (t_1^2 + t_2^2 + t_3^2) + t_1 t_2 t_3 + O(t^4) \right] + O(N^{1/2}) \,.
 }
From \eqref{JJTwo} and \eqref{JJJAlternative} we extract
 \es{JCorrLargeN}{
  \langle J_{(\alpha)}(\vec{x}_1) J_{(\beta)}(\vec{x}_2) \rangle
   &=   \frac{4 \sqrt{2} N^{3/2}}{3 \pi}   \frac{\delta_{\alpha\beta}}{(4 \pi)^2 \abs{\vec{x}_1 - \vec{x}_2}^2} + O(N^{1/2})  \,, \\
   \langle J_{(1)}(\vec{x}_1) J_{(2)}(\vec{x}_2) J_{(3)}(\vec{x}_3) \rangle 
   &= \frac{4 \sqrt{2} N^{3/2}}{3 \pi}  \frac{1}{(4 \pi)^3 \abs{\vec{x}_1 - \vec{x}_2}
    \abs{\vec{x}_1 - \vec{x}_3} \abs{\vec{x}_2 - \vec{x}_3}}  + O(N^{1/2})  \,.
 }
Using 
\begin{equation}
 \frac{4 \sqrt{2} N^{3/2}}{3 \pi} ~\approx ~\frac{c_T}{16} ~\approx ~\frac{2L^2}{\pi G_4}\,,
\end{equation}
(see, for example, \cite{Chester:2014fya}) we see that these expressions agree with \eqref{ParticularTwoPoint}--\eqref{ParticularThreePoint}.

\section{Some details of the derivation of \eqref{dskinfinal}}
\label{appendixSkin}

Below, the Killing spinor is assumed to be Majorana.  We start by writing all terms in $\delta S_{kin}$ involving the $P_R\chi$ projection of the spinor field and then add the conjugate terms.
\begin{equation}\label{details1}
\begin{split}
\d S_{\text{kin}~P_R\chi} \eql \int d^4x\sqrt{-g}\bigg[& -\pa_\m(\bar\eps P_R\chi)\pa^\m z -\frac12\bar\chi\gamma^\mu \nabla_\mu(P_L(\slashed{\nabla} z+F)\eps) \\
&+ \frac12\bar\eps(\slashed{\nabla} z-F)\g^\m \nabla_\m P_R\chi+(\bar\eps\g^\m \nabla_\m P_R\chi)(F+ {z\over L}) \\
& +2(\bar\eps P_R\chi) {z\over L^2}\bigg]  \,.
\end{split}
\end{equation}
The 3 terms involving $F$ are
\be
-\frac12 \bar\chi\g^\m \nabla_\mu(P_LF\eps)-\frac12\bar\eps F \g^\m \nabla_\mu P_R\chi +
(\bar\eps\g^\m \nabla_\mu P_R\chi)F\,.
\ee
After a Majorana flip of the first term and adding the last two terms we recognize the total derivative
\be
\frac12 \nabla_\m(\bar \eps F \g^\m P_R \chi) \,.
\ee
This becomes the $P_R\chi$ part of the last term in \eqref{dskinfinal}.
Next use
\be
\g^\m\g^\n \nabla_\mu(\pa_\n z\eps) \eql \Box z\eps + \g^\m\g^\n\pa_\n z \nabla_\mu\eps \eql \Box z\eps + (1/L)\slashed{\nabla} z\eps \,,
\ee
in which the Killing spinor equation and a standard $\g$-matrix identity are used to write the last term.   This relation is used in the second term of \eqref{details1} and, after partial integration, in the third term also.  After partial integration in the first term, one see that the 3  terms containing $\Box z$ cancel.  One is left with the two total derivatives from the partial integrations plus terms in $1/L$ and $1/L^2$, namely   

\begin{equation}\label{}
\begin{split}
\d S_{kin~P_R\chi} \eql \int d^4x\sqrt{-g}\bigg[&-\nabla_\m(\bar\eps P_R\chi \pa^\m z) +\frac12\nabla_\m(\bar\eps \slashed{\nabla}z\g^\m  P_R\chi)\\
&+ {1\over L}\, \bar\eps\,\left[\slashed{\nabla}z P_R\chi +z \g^\m \nabla_\mu P_R\chi\right]+  2\,{z\over L^2}\,\bar\eps P_R\chi\bigg]\,.
\end{split}
\end{equation}
The terms inside the square bracket add to the derivative of the product $z P_R\chi$. This is  partially integrated giving another total derivative plus terms that vanish by Killing spinor equation.

\section{Truncating the $\cn=8$ theory} 
\label{Appendix:Trunc}
\subsection{Truncations and flows}
\label{ss:TandF}

There are many important instances in which the full $\cn=8$ theory is truncated to a subsector with a reduced amount of supersymmetry.   To define the reduced, or truncated, theory we  introduce a  projection matrix, $\Pi^{i}{}_j$, whose task will be to project onto the supersymmetries of interest. Specifically, the supersymmetries in the truncation are given by: 
\begin{equation}
\Pi^{i}{}_j \, \varepsilon^j = \varepsilon^i \,, \qquad \Pi^{j}{}_i  \, \varepsilon_j = \varepsilon_i \,, \qquad  \Pi^{i}{}_j \, \Pi^{j}{}_k ~=~ \Pi^{i}{}_k \,, \qquad {\rm with} \qquad  p \equiv \Pi^{i}{}_i = Tr(\Pi) \,.
 \label{Proj1}
\end{equation}
We are thus truncating to a theory with $p$ supersymmetries.

In the second part of this appendix we show, in particular, that if  $ \Pi^{i}{}_j$  is a projector acting on the supersymmetries in such a way that it reduces their number to $p$ in a manner consistent  with  (\ref{deltagravitino}) and (\ref{deltagaugino}) then the boundary counterterm action is simply:
\begin{equation}\label{bctrred1}
S_\text{b,truncated}  \eql -{2\over p \, L}\int d^3x\,e^{3r_0/L}\,\mathop{\rm Tr} \sqrt{\smash[b]{ \Pi A_1A_1^\dagger}} \,.
\end{equation}
Indeed, in many instances,\footnote{For early examples, see  \cite{Ahn:2000mf,Corrado:2001nv,Ahn:2001by,Ahn:2001kw,Ahn:2002qga}.} $ \sqrt{\smash[b]{ \Pi A_1A_1^\dagger}}$  is simply diagonal on the relevant subspace and has eigenvalues $e^{\cals K/2} W$.  Thus (\ref{bctrred1}) yields the same  result as in (\ref{cutoff}).  

It is also important to note that (\ref{bctrred1}) represents a sum over  a subset of $p$ of the eigenvalues of $A_1A_1^\dagger$.  From (\ref{Aexp1}) one sees that, at quadratic order in $\phi$, the eigenvalues are all the same while at cubic order they will depend on details of the truncation.  Thus, as one would expect, to quadratic (divergent) order, the counterterms are universal\footnote{The factors of $p$ cancel between the coefficient of (\ref{bctrred1}) and the sum in the trace.} but the finite counterterms depend upon the details of the supersymmetry of the truncated theory. In particular, the truncation will generically break $\grSO(8)$ to $\grSO(p)$, or perhaps even some subgroup of $\grSO(p)$.  Thus the form of the finite counterterms is no longer bound by $\grSO(8)$ invariance, and it is quite possible that the truncated analog of (\ref{cubicct1}) might allow some $\alpha (\beta)^2$ terms. Indeed, we encountered precisely such terms in 
Sections~\ref{sec:countN1}-\ref{AB2counter}.  

We would be remiss if we did not mention flow in the context of the Bogomolny factorization. Flows  are solutions that depended solely on $r$ and are thus independent of the boundary directions.  Supersymmetric flow solutions preserve some subset of the supersymmetries and the BPS equations can typically be obtained by requiring each squared term in the Bogomolny action to vanish independently.  This means that $\cA_a{}_{ijkl} = 0$ and, from, (\ref{compsqsub1}):
\begin{equation}\label{floweq1}
\cals A' \delta_{ij} ~=~   \pm \sqrt{2} \, g \,\Pi^{k}{}_i \, D_{kj} \,,  \qquad  \Pi^{i}{}_m  \, \cA_r{}^{mjkl}   ~=~   \mp 2\, g \, \Pi^{i}{}_p \Mmat^{p m}  A_{2m}{}^{jkl} \,.\end{equation}
This means that the eigenvalues of $\Pi \, D$ must all be the same and reduce to essentially a single superpotential, while the second equation in (\ref{floweq1}) takes the form of steepest descents on that superpotential, exactly as in Section \ref{ss:N1bog}.
Also see, for example, \cite{Pope:2003jp,Bobev:2013yra,Pilch:2015dwa}.

\subsection{Calculation of the counterterms}

Consistency with (\ref{deltagravitino}) requires:
\begin{equation}
D_\mu \Pi^{i}{}_j  ~=~  0 \,, \qquad    \Pi^{i}{}_k \, A_1{}^{kj} ~=~  \Pi^{j}{}_k \, A_1{}^{ik} \quad \Rightarrow \quad \Pi^{i}{}_k \,  \Pi^{j}{}_m \, A_1{}^{km} ~=~  \Pi^{i}{}_k \, A_1{}^{kj}  \,.
 \label{Trunc2}
\end{equation}
It follows that our truncation must reduce the $A_1$ tensor to a $p \times p$ sub-matrix and the gravitino variations are restricted to the components of (\ref{deltagravitino}) along $\Pi^{i}{}_k \Pi^{j}{}_m A_1{}^{km}$. 

To perform the Bogomolny trick in the truncated theory we need two identities involving the tensors associated with the scalars.
First, consider the partial contraction: 
\begin{align}
\cA_\mu{}^{iklm}\,\cA^\mu{}_{jklm} ~=~ & \coeff{1}{576} \, \epsilon^{iklmpqrs} \epsilon_{jklmtuvw}\,\cA_\mu{}_{pqrs}\,\cA^\mu{}^{tuvw}~=~  \coeff{5}{4} \, \delta^{[i}_{j} \delta^{p}_{t}  \delta^{q}_{u}\delta^{r}_{v}\delta^{s]}_{w}  ,\cA_\mu{}_{pqrs}\,\cA^\mu{}^{tuvw}   \nonumber \\
~=~ &  \coeff{1}{4} \,  \delta^{i}_{j} \, \cA_\mu{}_{pqrst}\,\cA^\mu{}^{pqrst}  ~-~ \cA^\mu{}_{jklm}  \,\cA_\mu{}^{iklm} \,.
 \label{kinid1}
\end{align}
It follows that the self-duality of the kinetic term implies that one has:
\begin{equation}
\cA_\mu{}^{iklm}\,\cA^\mu{}_{jklm} ~=~  \coeff{1}{8} \,  \delta^{i}_{j} \, \cA_\mu{}^{pqrst}\,\cA^\mu{}_{pqrst}  \label{kinid2} \,.
\end{equation}
There is also a very similar identity in \cite{deWit:1982bul} for the A-tensors:
\begin{equation}
 - \coeff{3}{4}\, A_1{}^{ik} A_1{}_{kj} ~+~ \coeff{1}{24} A_{2}{}^i{}_{klm} A_{2j}{}^{klm} \eql  \coeff{1}{8} \,\delta^i_j \,\big( - \coeff{3}{4}\left|A_1{}^{ij}\right|^2+ \coeff{1}{24}\left|A_{2i}{}^{jkl}\right|^2 \big) ~=~\coeff{1}{8} \, \cP  \,  \delta^i_j \,.
 \label{potid1}
\end{equation}
Contracting  (\ref{kinid2}) and  (\ref{potid1})  with $\Pi^{i}{}_j$ gives
\begin{align}
& \cA_\mu{}^{ijkl}\,\cA^\mu{}_{ijkl} ~=~ \coeff{8}{p}\,\Pi^{j}{}_i \,\cA_\mu{}^{iklm}\,\cA^\mu{}_{jklm}   \label{kinid3} \\ 
& \cP ~=~ \coeff{8}{p}\, ( - \coeff{3}{4}\,\Pi^{j}{}_i \, A_1{}^{ik} A_1{}_{kj} ~+~ \coeff{1}{24}\,\Pi^{j}{}_i \, A_{2}{}^i{}_{klm} A_{2j}{}^{klm}) 
 \,, \label{potid3} 
\end{align}

One can now complete the square, exactly as in Section~\ref{Sect:BogN8}, but now on the truncated subsystem:
\begin{align}
 S_B =  \int d^4 x\, e^{3A} \, \Big[ 3 (\cals A')^2  ~+~ &  \coeff{3}{4}\,g^2 \,\left|A_1{}^{ij}\right|^2~-~ \coeff{1}{96}\, \cals A_r{}^{ijkl}\cals A_r{}_{\,ijkl}  ~-~ \coeff{1}{24}\,g^2 \,\left|A_{2i}{}^{jkl}\right|^2 \,\Big]  \nonumber \\
=   \int d^4 x\, e^{3A} \, \Big[ 3 (\cals A')^2  ~+~   &   \coeff{8}{p}\, \Big( \coeff{3}{4}\,g^2 \,\Pi^{j}{}_i \, A_1{}^{ik} A_1{}_{kj} ~-~ \coeff{1}{96}\, \Pi^{j}{}_i \,\cals A_r{}^{iklm}\,\cals A_r{}_{jklm} \nonumber \\  
  & \qquad\qquad\qquad\qquad\qquad ~+~ \coeff{1}{24}\,g^2\, \Pi^{j}{}_i \, A_{2}{}^i{}_{klm} A_{2j}{}^{klm}  \Big)\,\Big]  \nonumber \\
   =   \coeff{1}{p}\,  \int d^4 x\, e^{3A} \, \Big[   3\,\Big |  \cals A' \,  \Mmat_{ij} & ~\mp~   \sqrt{2} \, g \,\Pi^{k}{}_i \, A_1{}_{kj} \Big|^2 ~-~ \coeff{1}{12}\, \Big |  \Pi^{i}{}_m  \, \cA_r{}^{mjkl}   ~\pm~   2\, g \, \Mmat^{i m}  A_{2m}{}^{jkl} \Big|^2     \nonumber \\
  & \qquad  \qquad  \pm  \sqrt{2}\,g \,   \Big(\Mmat^{i j}  D_r  (e^{3A} \,A_{1\, ij})  ~+~    \Mmat_{i j} \, D_r (e^{3A} \, A_1{}^{ij}) \Big)  \,\Big]
  \,.
\label{compsqsub1}
\end{align}

Here the matrices, $\Mmat_{ij} = (\Mmat^{ij})^*$,  are again allowed to be dynamical but satisfy:
\begin{equation}
\Mmat_{ij}  ~=~ \Mmat_{ji} \,, \qquad \Mmat_{ij} \, \Mmat^{kj} ~=~ \Pi^{k}{}_i \,, \qquad  \Pi^{i}{}_k  \Mmat^{kj}  = \Mmat^{ij} \,, \quad \Pi^{j}{}_k  \Mmat^{ik}  = \Mmat^{ij}  \,.
\label{MMmatadj}
\end{equation}
That is, it is an $\grSU(p)$  matrix on the  remaining supersymmetries.   As in Section~\ref{Sect:BogN8}, we choose $\Mmat$ so as to diagonalize $A_1$ on the subspace defined by $\Pi$, and the same arguments lead to a counterterm action:
\begin{equation}\label{bctrred1a}
S_\text{s-ct,truncated}  \eql -{2\over p \, L}\int d^3x\,e^{3r_0/L}\,\mathop{\rm Tr} \sqrt{\smash[b]{ \Pi A_1A_1^\dagger}} \,.
\end{equation}
In particular, for truncations to $\cn=1$ or $\cn=2$ supersymmetric theories, the superpotential emerges as one or two, respectively, of the eigenvalues of $A_1A_1^\dagger$ while the other eigenvalues of this matrix play no role in the supersymmetry of the theory. (These other eigenvalues give mass to the gravitini for the broken supersymmetries.)  Thus the projection by $\Pi$  in (\ref{bctrred1a})   onto the subspace of residual supersymmetries is an essential part of getting the correct supersymmetric boundary terms.  Indeed, for such truncations, this projection extracts the superpotential terms and thus generates boundary terms exactly of the form (\ref{bpsct}).

\section{Some identities for SO(8) (anti-)self-dual tensors}
\label{app:so8tens}

Let  $\alpha^{ijkl}=\alpha^{[ijkl]_+}$ be a self-dual and  $\beta^{ijkl}=\beta^{[ijkl]_-}$ an anti-self-dual real SO(8) tensor, 
\begin{equation}\label{defself}\begin{split}
\alpha^{ijkl}    \eql {1\over 24} \eta^{ijklmnpq}\alpha^{mnpq}\,,\qquad 
\beta^{ijkl} & \eql -{1\over 24} \eta^{ijklmnpq}\beta^{mnpq}\,.
\end{split}\end{equation}
By a repeated use of \eqref{defself} together with the contraction identities for the completely antisymmetric symbol, $\eta^{ijklmnpq}$, one can prove the following identities (see, e.g., \cite{deWit:1978sh}):
\begin{align}\label{cid1}
\alpha^{ijkl}\beta^{ijkl} & \eql 0\,,\\[6 pt]\label{cid2}
\alpha^{iklm}\alpha^{jklm} & \eql {1\over 8}\,\delta ^{ij}\,\alpha^{klmn}\alpha^{klmn}\,,\\[6 pt]\label{cid3}
\beta^{iklm}\beta^{jklm} & \eql {1\over 8}\,\delta ^{ij}\,\beta^{klmn}\beta^{klmn}\,,\\[6 pt]\label{cid4}
\alpha^{iklm}\beta^{jklm} & \eql\beta^{iklm}\alpha^{jklm}\,,\\[6 pt]\label{cid5}
\alpha^{mn[ij}\alpha^{k]lmn} & \eql \alpha^{mn[ij}\alpha^{kl]mn}\qquad \text{self-dual}\,,\\[6 pt]\label{cid6}
\beta^{mn[ij}\beta^{k]lmn} & \eql \beta^{mn[ij}\beta^{kl]mn}\qquad \text{anti-self-dual}\,,\\[6 pt]\label{cid7}
\alpha^{mn[ij}\beta^{k]lmn} & \eql -\beta^{mn[ij}\alpha^{k]lmn}\,, 
\end{align}
and 
\begin{equation}\label{epsalal}
\begin{split}
\eta_{klmnpqrs} \alpha_{(1)}{}^{iklm}\alpha_{(1)}{}^{jnpq}\eql  & 18\,\alpha_{(1)}{}^{ijmn}\alpha_{(1)}{}^{rsmn}\\ & +6\, \delta_{si}\alpha_{(1)}{}^{jmnp}\alpha_{(1)}{}^{rmnp}-6\, \delta_{ri}\alpha_{(1)}{}^{jmnp}\alpha_{(1)}{}^{smnp}\\
\eql & 18\,\alpha_{(1)}{}^{ijmn}\alpha_{(1)}{}^{rsmn}+{3\over 4}(\delta_{ir}\delta_{js}-\delta_{ir}\delta_{js})\alpha_{(1)}{}^{mnpq}\alpha_{(1)}{}^{mnpq}\,,
\end{split}\end{equation}
\begin{equation}\label{epsalbe}
\begin{split}
\eta_{klmnpqrs} \alpha_{(1)}{}^{iklm}\beta_{(1)}{}^{jnpq}\eql  &- 18\,\alpha_{(1)}{}^{ijmn}\beta_{(1)}{}^{rsmn}\\ & +6\, \delta_{sj}\alpha_{(1)}{}^{imnp}\beta_{(1)}{}^{rmnp}-6\, \delta_{rj}\alpha_{(1)}{}^{imnp}\beta_{(1)}{}^{smnp}\\
\eql & 18\,\alpha_{(1)}{}^{rsmn}\beta_{(1)}{}^{ijmn}\\& + 6\,\delta_{si}\alpha_{(1)}{}^{rmnp}\beta_{(1)}{}^{jmnp}-
6\,\delta_{ri}\alpha_{(1)}{}^{smnp}\beta_{(1)}{}^{jmnp}\,,
\end{split}\end{equation}
which are used in Sections~\ref{Sect:BogN8} and \ref{N8sugrsusy}.

\section{The U(1)$^\mathbf{3}$-invariant truncation in \cite{Freedman:2013ryh}}
\label{app:FP}

The scalar sector of the $\rm U(1)^3$-invariant truncation of $\cals N=8$ supergravity\footnote{For an early work on this truncation, see  \cite{Cvetic:1999xp}.} studied  in \cite{Freedman:2013ryh}, in the notation of the present paper, is given by
\begin{equation}\label{}
\begin{split}
\alpha^{1234}&\eql \alpha^{5678}\eql \rho_1\cos\theta_1\,,\\
\alpha^{1256}&\eql \alpha^{3478}\eql \rho_2\cos\theta_2\,,\\
\alpha^{3456} & \eql \alpha^{1278}\eql\rho_3\cos\theta_3\,,
\end{split}
\end{equation}
where
\begin{equation}\label{}
z_\alpha\eql\tanh\rho_\alpha\,e^{i\theta_\alpha}\,,\qquad \alpha=1,2,3\,.
\end{equation}
After the change from the $\rm SU(8)$ to the $\rm SL(8,\mathbb{R})$ basis, only the diagonal fields, $A^{II}$, are nonzero and are given by\footnote{The particular arrangement of the signs in \eqref{AIIs} depends on a  representation of $\Gamma$-matrices of $\rm SO(8)$.} 
\begin{equation}\label{AIIs}
\begin{split}
A^{11}&\eql A^{77}\eql {1\over 2}\,(\rho_1\cos\theta_1-\rho_2\cos\theta_2-\rho_3\cos\theta_3)\,,\\
A^{22} & \eql A^{88}\eql {1\over 2}\,(\rho_1\cos\theta_1+\rho_2\cos\theta_2+\rho_3\cos\theta_3)\,,\\
A^{33}& \eql A^{66}\eql {1\over 2}\,(-\rho_1\cos\theta_1-\rho_2\cos\theta_2+\rho_3\cos\theta_3)\,,\\
A^{44}& \eql A^{55}\eql {1\over 2}\,(-\rho_1\cos\theta_1+\rho_2\cos\theta_2-\rho_3\cos\theta_3)\,.
\end{split}
\end{equation}
The qudartic and cubic counterterms are then
\begin{equation}\label{}
\begin{split}
-{1\over 4L}\,A^{IJ}A^{IJ} & \eql -{1\over 2L}\,(\rho_1^2\cos^2\theta_1+\rho_2^2\cos^2\theta_2+\rho_3^2\cos^2\theta_3)\\[6 pt] &\eql -{1\over 2L}\,(z_1\bar z_1+z_2\bar z_2+z_3\bar z_3)+\ldots\,,
\end{split}
\end{equation}
\begin{equation}\label{}
\begin{split}
{1\over 6\sqrt 2\,L}\,A^{IJ}A^{JK}A^{KI}  & \eql {1\over \sqrt 2\,L}\,\rho_1\rho_2\rho_3\cos\theta_1\cos\theta_2\cos\theta_3\\
& \eql {1\over 2\sqrt 2L}(z_1z_2z_3+\bar z_1\bar z_2\bar z_3)+\ldots\,.
\end{split}
\end{equation}
where one must set the pseudoscalars to zero. The $\ldots$ stand for  terms quartic in the fields due to the expansion $\tanh\rho_\alpha\eql \rho_\alpha+\ldots$.



\end{document}